\documentclass[a4paper,11pt]{article}
\pdfoutput=1 

\usepackage{jheppub} 
\usepackage{etoolbox}                     
\makeatletter
\patchcmd{\maketitle}{\@fpheader}{Prepared for submission to JHEP}{}{}
\makeatother

 \usepackage[titles]{tocloft}
 
\renewcommand\afterTocRuleSpace{\flushbottom\pagebreak}

\usepackage[T1]{fontenc} 
\usepackage{float}
\usepackage{cprotect}
\usepackage{caption,subcaption}
\usepackage[utf8]{inputenc}
\usepackage{slashed}
\usepackage{epstopdf}
\usepackage{booktabs}
\usepackage{multirow}
\usepackage{float}
\usepackage[printonlyused]{acronym}
\usepackage[T1]{fontenc}
\usepackage{array}
\usepackage{booktabs}
\usepackage{rotating}
\usepackage{tablefootnote}
\usepackage{diagbox}
\usepackage{hhline}
\usepackage{enumerate}

\hypersetup{
    linktocpage,
     colorlinks,
     citecolor=darkgreen,
     linkcolor= darkgreen,
     urlcolor=darkgreen
}
\definecolor{darkred}{rgb}{0.6,0.0,0.0}
\definecolor{darkerblue}{rgb}{0.0,0.0,0.5}
\definecolor{purple}{rgb}{0.0,0.4,0.2}
\definecolor{darkgreen}{rgb}{.0,0.5,0.0}
\definecolor{black}{rgb}{0.0,0.0,0.0}
\definecolor{brown}{rgb}{0.6,0.4,0.2}
\definecolor{newpurple}{rgb}{0.65, 0.38, 0.61}
\definecolor{newyellow}{rgb}{0.9718, 0.6093, 0.0759}
\definecolor{amber}{rgb}{1.0, 0.75, 0.0}
\definecolor{newblue}{rgb}{0.4, 0.52, 0.85}
\definecolor{newred}{rgb}{0.8524, 0.2595, 0.3294}
\definecolor{newgreen}{rgb}{0.2, 0.8, 0.2}
\definecolor{SMgreen}{rgb}{0.56, 0.69, 0.19}
\definecolor{neworange}{rgb}{0.94, 0.462, 0.162}

\definecolor{BrickRed}{rgb}{0.9,0.1,0}

\def\LO{\text{LO}}
\def\NLO{\text{NLO}}
\newcommand{\fs}{\,\textrm{fs}}
\newcommand{\pico}{\,\textrm{ps}}
\newcommand{\GeV}{\,\textrm{GeV}}
\newcommand{\MeV}{\,\textrm{MeV}}

\acrodef{CKM}[CKM]{Cabibbo-Kobayashi-Maskawa}
\acrodef{HQE}[HQE]{heavy quark expansion}
\acrodef{HQET}[HQET]{Heavy Quark Effective Theory}
\acrodef{LO}[LO]{leading order}
\acrodef{NL}[NL]{nonleptonic}
\acrodef{NLO}[NLO]{next-to-leading order}
\acrodef{NRCQM}[NRCQM]{nonrelativistic constituent quark model}
\acrodef{OPE}[OPE]{Operator Product Expansion}
\acrodef{PI}[PI]{Pauli interference}
\acrodef{QCD}[QCD]{Quantum Chromodynamics}
\acrodef{SL}[SL]{semileptonic}
\acrodef{VIA}[VIA]{vacuum insertion approximation}
\acrodef{VSA}[VSA]{vacuum saturization approximation}
\acrodef{WA}[WA]{weak annihilation}
\acrodef{WE}[WE]{weak exchange}

\newcommand{\App}{appendix~}
\newcommand{\Fig}{figure~}
\newcommand{\Figs}{figures~}
\newcommand{\Sec}{section~}
\newcommand{\Secs}{sections~}
\newcommand{\Tab}{table~}
\newcommand{\Tabs}{tables~}
\newcommand{\Eqn}{eq.~}
\newcommand{\Eqns}{eqs.~}
\renewcommand{\arraystretch}{1.5}

\newcommand{\al}{\alpha}
\newcommand{\be}{\beta}
\newcommand{\ga}{\gamma}
\newcommand{\Ga}{\Gamma}
\newcommand{\de}{\delta}

\newcommand{\eps}{\epsilon}

\newcommand{\la}{\lambda}
\newcommand{\La}{\Lambda}

\newcommand{\Om}{\Omega}
\newcommand{\sig}{\sigma}

 \renewcommand{\Im}{\textrm{Im}}

\definecolor{Ivn}{rgb}{0.6, 0.0, 0.0}
\definecolor{Jms}{rgb}{0.2, 0.5, 0.9}

\newcommand{\als}{\al_s}
\newcommand{\NC}{N_C}
\newcommand{\MSbar}[1]{\overline{#1}}
\newcommand{\msbar}{{\overline{\rm MS}}}

\newcommand{\mukin}{\mu_\pi^2}
\newcommand{\muG}{\mu_G^2}
\newcommand{\darwin}{\rho_D^3}

\newcommand{\Dleft}{\overset{\leftarrow}{D}}

\newcommand{\cdarwinc}[4]{\mathcal{K}_{\rho,#1 #2}^{(#3\bar{#4})}}
\newcommand{\cdarwinlepc}[2]{\mathcal{K}_{\rho}^{(#1 #2)}}

\newcommand{\Opsix}[2]{O_{#1}^{#2}}
\newcommand{\Opsixt}[2]{\tilde{O}_{#1}^{#2}}
\newcommand{\OpsixT}[2]{T_{#1}^{#2}}
\newcommand{\OpsevenP}[2]{P_{#1}^{#2}}
\newcommand{\OpsevenPt}[2]{\tilde{P}_{#1}^{#2}}
\newcommand{\OpsevenS}[2]{S_{#1}^{#2}}
\newcommand{\Opsixprime}[2]{\mathcal{O}_{#1}^{#2}}

\newcommand{\OpsixTprime}[2]{\mathcal{T}_{#1}^{#2}}
\newcommand{\OpsevenPprime}[2]{\mathcal{P}_{#1}^{#2}}
\newcommand{\OpsevenRprime}[2]{\mathcal{R}_{#1}^{#2}}

\newcommand{\OpsevenSprime}[2]{\mathcal{S}_{#1}^{#2}}
\newcommand{\OpsevenUprime}[2]{\mathcal{U}_{#1}^{#2}}
\newcommand{\tGamma}{\tilde{\Gamma}}

\newcommand{\Bary}{\mathcal{B}}
\newcommand{\Had}{H}
\newcommand{\Mes}{M}
\newcommand{\Lac}{\La_c^+}
\newcommand{\Xicp}{\Xi_c^+}
\newcommand{\Xico}{\Xi_c^0}
\newcommand{\Omc}{\Om_c^0}
\newcommand{\Lab}{\La_b^0}

\newcommand{\Omb}{\Om_b^-}
\newcommand{\mQheavy}{m_Q}
\newcommand{\Lifetime}[1]{\tau \left( #1 \right)}

\newcommand{\intm}{\textrm{int}^{-}}
\newcommand{\intp}{\textrm{int}^{+}}
\newcommand{\exc}{\textrm{exc}}

\title{\boldmath Lifetimes of singly charmed hadrons
}

\author[a]{James Gratrex,}
\author[a]{Bla\v zenka Meli\' c,}
\author[a]{and Ivan  Ni\v sand\v zi\'c}

\affiliation[a]{Division of Theoretical Physics, Ru\dj er Bo\v skovi\'c Institute, Bijeni\v cka cesta 54, 10000, Zagreb, Croatia.}

\emailAdd{jgratrex@irb.hr}
\emailAdd{blazenka.melic@irb.hr}
\emailAdd{ivan.nisandzic@irb.hr}

\abstract{
We provide an extensive study of the lifetimes of singly charmed baryons and mesons, within the heavy quark expansion with all known corrections included. A special attention is devoted to the choice of the charm mass and wavefunctions of heavy baryons.
We give our predictions for lifetimes, lifetime ratios, and semileptonic branching ratios of singly charmed baryons. Our results accommodate the experimentally-favoured hierarchy of singly charmed baryon lifetimes 
\begin{eqnarray*}
\Lifetime{\Xico} < \Lifetime{\Lac}< \Lifetime{\Omc} < \Lifetime{\Xicp}\,
\end{eqnarray*}
in contrast to earlier theoretical findings. 
Predictions for charmed meson lifetimes and semileptonic decay rates are in agreement with a recent comprehensive study and experimental results within uncertainties.
}

\begin{document} 
\preprint{RBI-ThPhys-2022-8}
\maketitle
\flushbottom

\section{Introduction}

The recent measurements of charmed baryon lifetimes by the LHCb Collaboration \cite{LHCbOmegac2018,LHCbcharmedLifetimes2019,LHCb2021Omega0} stand in marked contrast to earlier determinations. Whereas the lifetimes of the $\Lac$ and $\Xicp$ are compatible with previous experiments \cite{SELEXLambdac,CLEOLambdac,FOCUSLambdac2002,CLEOXic,FOCUSXicp2001}, that of the $\Xico$ is in roughly $3\sig$ tension with the older measurement \cite{FOCUSXic02002,PDG2018}. Even more dramatically, the measured lifetime of the $\Omc$ of $274.5 \fs$ is four times larger than, and wholly inconsistent with, earlier results \cite{FOCUSOmegac2003,SELEXOmegac}. Moreover, the newly-established hierarchy of experimental lifetimes,
\begin{eqnarray}
\Lifetime{\Xico} < \Lifetime{\Lac}< \Lifetime{\Omc} < \Lifetime{\Xicp}\,,
\end{eqnarray}
is in conflict with earlier theoretical predictions \cite{Melic97c,Cheng1997c}, where in particular the $\Omc$ was expected to be the shortest-lived among the singly charmed baryons.  

On the theoretical side, the approach to calculating lifetimes proceeds via the \ac{HQE}, which is an expansion of the inclusive decay width in inverse powers of the heavy quark mass, developed in the 1980s and early 1990s, eg \cite{SV1985,Chay:1990da,BUV1992HQE,BBSUV92HQE,BSUV1993I,BSUV1993II}. This was motivated by the experimental observation
that the lifetime ratio of $D$ mesons \cite{PDG2020},
\begin{equation}
    \frac{\Lifetime{D^+}}{\Lifetime{D^0}} = 2.54(2) \, ,
\end{equation}
is significantly different from the naive prediction of unity, based upon the assumption that the charm quark decay is the dominant contribution. Once contributions sensitive to the flavour of the light valence quark were taken into account, the then-experimental hierarchy could be reproduced \cite{GNPR1979,Cortes1980,KS1983,BGT1984,GRT1986,SV1986}.\footnote{See \cite{LenzRauh2013} and \cite{Lenz2014} for further details about the history of the \ac{HQE}, as well as further references.} In response to the temporary $\tau(\La_b)/\tau(B)$ lifetime puzzle (discussed for example in \cite{Lenz2014} and references therein), the focus of the \ac{HQE} turned to $b$-quark hadrons, where the  much-improved convergence of the $1/m_Q$ series motivated the analysis of higher-order terms \cite{FLMT2002,BBGLN2002,GOP2004,BUZ2005,DMT2006,MTU2010,GHT2016}. 

The applicability of the \ac{HQE} to charm decays, as well as the correct way to perform the expansion, is an open question, with some alternative approaches appearing recently in \cite{FMV2019,MMP2021}. In the most recent study of $D$ meson lifetimes within the \ac{HQE}, including the most complete set of contributions \cite{LenzNote:2021}, the central value of the decay width of the $D^+$ was found to be extremely small or even negative, driven by a large Pauli interference contribution, an observation also made in \cite{Cheng18c}. On the other hand, the uncertainties in such predictions are sizeable, due to large hadronic and scale uncertainties. In the same paper \cite{LenzNote:2021}, the \ac{HQE} predictions for ratios of decay widths and of the semileptonic branching fraction of mesons were found to be compatible with the experimental values, notwithstanding the ratio of the lifetimes of $D_s^+$ and $D^0$ that remained in a slight tension with experiment. This seems to support the possibility that the \ac{HQE} is a successful approach in understanding at least the qualitative nature of charm physics in the meson sector.

The most recent update to the theoretical prediction of singly charmed baryon lifetimes was made in 2018 \cite{Cheng18c}. That calculation considered the effects of subleading spectator corrections, but neglected the Darwin term and \acs{QCD} corrections, which in the case of charm physics are sizeable \cite{LenzNote:2021}, and did not provide an error analysis. Moreover, whilst the prediction in \cite{Cheng18c} managed to accommodate the new experimental lifetime hierarchy implied by \cite{LHCbOmegac2018}, it was only able to do so by introducing an arbitrary factor suppressing certain contributions to the $\Omc$ decay width. Such a resolution is hardly satisfying.

In this paper, we extend the analysis of \cite{LenzNote:2021} by revisiting the inclusive lifetime predictions for the baryon sector. As compared with previous studies \cite{Cheng1997c,Cheng18c,Melic97c}, we include the Darwin contributions, recently computed in \cite{MMP2020,Moreno2020,LPR2020} and extended to charm hadrons in \cite{LenzNote:2021}, and dimension-seven four-quark operator contributions, first considered in the context of $B$ hadrons in \cite{GOP2003,GOP2004} and subsequently in the charm sector in \cite{LenzRauh2013,Cheng18c}.  In addition, we include existing \acf{NLO} contributions to the Wilson coefficients of two-quark operators at dimension-three and four-quark operators \cite{CFLM2001,FLMT2002} at dimension-six. We also repeat the computations of $D$ meson lifetimes performed in \cite{LenzNote:2021}, verifying their results, with minor differences originating from different estimates of some of the hadronic parameters.

The paper is organized as follows. In \Sec\ref{sec:background}, we briefly outline the \ac{HQE}, defining our notation and the contributions to be included, with current  experimental results presented in \Sec\ref{sec:experimentalstatus} for ease of comparison. Some comments on the charm mass schemes used are presented in \Sec\ref{sec:charmmassscheme}. In \Sec\ref{sec:twoquarkmatel}, we discuss the two-quark contributions and values for the matrix elements for all hadrons of interest to the paper, presenting the numerical results for these ``non-spectator'' contributions. In \Sec\ref{sec:mesonsresultsmain}, we present results for inclusive observables for charmed mesons, and in \Sec\ref{sec:baryonsresultsmain}, we do likewise for the baryons, after  an extensive discussion of the baryon wavefunctions in \Sec\ref{sec:matrix_baryons}. The paper ends with conclusions in \Sec\ref{sec:conclusion}. Appendix~\ref{app:parameterinputs} collects numerical inputs used in this work, and \App\ref{app:c356} collates various useful analytic expressions. In \App\ref{app:HQEtoQCD}  we give some technical details on properly relating HQET and QCD four-quark matrix elements. Finally, in \App\ref{app:detailedtables} we provide supplementary tables with a detailed breakdown of contributions to meson and baryon observables.

\section{Theoretical and experimental background}
\label{sec:background}
\acresetall
\subsection{Experimental status}
\label{sec:experimentalstatus}
Before proceeding to a discussion of the theoretical approach to predicting lifetimes, we briefly review the present status of experimental measurements of inclusive charmed hadron lifetimes.
\subsubsection{Charmed mesons}\label{subsec:charmed_mesons}
\begin{table}[ht]
\centering
\begin{tabular}{|c||c|c|c|c|}
\hline
Quantity & $D^0$ & $D^+$ & $D_s^+$ \\ \hline
$\tau\,[ps]$& $ 0.4101 \pm 0.0015$ & $1.040 \pm 0.007$ & $ 0.504\pm 0.004$\\
\hline
$\Gamma\,[ps^{-1}]$& $ 2.438 \pm 0.009$ & $0.962 \pm 0.006$ & $ 1.984\pm 0.0016$\\
\hline
$BR(D_i\to X e\nu)\,[\%]$ & $6.49\pm 0.16$ & $16.07\pm 0.30$ & $6.30\pm 0.16$\\
\hline
$\Gamma(D_i\to X e\nu)\,[ps^{-1}]$ & $0.158\pm 0.004$ & $0.155\pm 0.003$ & $0.125\pm 0.003$\\
\hline
\end{tabular}
\caption{\small Summary of measured values for the lifetimes and semileptonic branching fractions of charmed mesons. 
The quoted values are the latest PDG averages \cite{PDG2020}, with the exception of $BR(D_s\to X e\nu)$, for which we show the recent result by BESIII \cite{BESIII2021}. We combine the statistical and systematic errors in quadrature in cases where both are given by experimental collaborations. 
A recent Belle II measurement \cite{BelleII2021} is compatible with the world averages.}
\label{tab:lMesonLifetimesexpt}
\end{table}
The current experimental values of the lifetimes (largely unchanged since the early 2000s) and the semileptonic branching fractions of charmed mesons are summarized in \Tab\ref{tab:lMesonLifetimesexpt}. The experimental values for the lifetime ratios are therefore
\begin{equation}
\frac{\tau(D^+)}{\tau(D^0)}=2.54\pm 0.02\,,\qquad\qquad \frac{\tau(D_s^+)}{\tau(D^0)}=1.23\pm 0.01.\label{eq:patern-ratios}
\end{equation}
Since, however, the \acl{HQE} does not account for the pure leptonic decay of $D_s \to \tau\nu_\tau$, one usually defines the modified width 
\begin{equation}
    \tilde{\Gamma}(D_s^+)=\Gamma(D_s)\big(1-BR(D_s^+\to\tau\nu)\big)\,.\label{eq:ModifiedDs0}
\end{equation}
Using the PDG average value $BR(D_s^+\to\tau\nu)=(5.48\pm 0.23)\%$ \cite{PDG2020} and the above value for $\tau(D_s)$, we obtain
\begin{equation}
   \tilde{\tau}(D_s^+)= 0.533\pm 0.004\,\text{ps}\,,\label{eq:ModifiedDs1}
\end{equation}
with the corresponding ratio, to be compared to theoretical estimates, of
\begin{equation}
    \frac{\tilde{\tau}(D_s^+)}{\tau(D^0)}=1.30\pm 0.01.\label{eq:ModifiedDs2}
\end{equation}
As can be seen, the experimental precision for charmed meson measurements is now at a sub-\% level. In particular, the most recent measurement, from Belle II \cite{BelleII2021}, is compatible with earlier values, indicating that the lifetime measurements of charmed mesons are robust. 

We will also consider the ratios of the inclusive semileptonic decay widths involving the electrons in the final states. Denoting $\Gamma(D\to X e\nu)\equiv \Gamma^{(e)}(D)$, and combining the experimental results from \Tab\ref{tab:lMesonLifetimesexpt}, we have:
\begin{equation}
\frac{\Gamma^{(e)}(D^+)}{\Gamma^{(e)}(D^0)}=0.977\pm 0.031\,,
\end{equation}
while for the remaining ratio we adopt the value given by the BESIII Collaboration \cite{BESIII2021},
    \begin{equation}
\frac{\Gamma^{(e)}(D_s^+)}{\Gamma^{(e)}(D^0)}=0.790\pm 0.026\,,
\end{equation}
with the statistic and systematic uncertainties combined in quadrature.

\subsubsection{Singly charmed baryons}

\begin{table}[ht]
\centering
\begin{tabular}{|c||c|c|c|c|}
\hline
Collaboration & $\Lifetime{\Lac}$/fs & $\Lifetime{\Xicp}$/fs & $\Lifetime{\Xico}$/fs & $\Lifetime{\Omc}$/fs \\ \hline
CLEO \cite{CLEOLambdac,CLEOXic} & $179.6 \pm 8.2$ & $503\pm 50$ & N/A & N/A \\
FOCUS \cite{FOCUSLambdac2002,FOCUSXic02002,FOCUSXicp2001,FOCUSOmegac2003} & $203.5 \pm 4.2$  & $439 \pm 24$ & $118^{+15}_{-13} $ & $72 \pm 16$\\
SELEX  \cite{SELEXOmegac,SELEXLambdac} & $198.1 \pm 9.0 $ & N/A & N/A & $65 \pm 16$ \tablefootnote{Unpublished except in a preprint, not cited in PDG.}   \\
LHCb \cite{LHCbOmegac2018,LHCbcharmedLifetimes2019} & $203.5 \pm 2.2$  & $457 \pm 6 $ & $154.5 \pm 2.6$ & $268 \pm 26 $ \\ 
LHCb 2021 \cite{LHCb2021Omega0} & N/A  & N/A & $148.0 \pm 3.2$ & $276.5 \pm 14.1$ \\ \hline
PDG 2018 \cite{PDG2018} & $200 \pm 6$ & $442 \pm 26$ & $112^{+13}_{-10}$ & $69 \pm12$ \\
PDG 2020 \cite{PDG2020} & $ 202.4 \pm 3.1$ & $456 \pm 5$ & $153 \pm 6$ & $268 \pm 24 \pm 10$ \\ 
\hhline{|=||=|=|=|=|}
Reference values & $202.4\pm 3.1$~\cite{PDG2020} & $456  \pm 5$~\cite{PDG2020} & $152.0 \pm 2.0$ ~\cite{LHCb2021Omega0} & $274.5 \pm 12.4$ ~ \cite{LHCb2021Omega0}\\
\hline
\end{tabular}
\caption{\small Summary of lifetime measurements of singly charmed baryons. All results are expressed in femtoseconds. As in \Tab\ref{tab:lMesonLifetimesexpt}, we combine statistical and systematic errors in quadrature. The PDG world-average (as of 2018 and 2020, showing the changes due to the LHCb results) is also included. Lifetime measurements that are not available (due to never being performed at the given experiment, or at least with no existing reference) are marked ``N/A''. The most recent LHCb determination, of the $\Omc$ and $\Xico$ lifetimes, is given in a separate row to highlight the two separate measurements. In the last row we list the most precise current results used as the reference values for the comparisons to our theoretical predictions, taken from PDG \cite{PDG2020} for the $\Lac$ and $\Xicp$, and from the LHCb averages given in \cite{LHCb2021Omega0} for the $\Xico$ and $\Omc$. }
\label{tab:lifetimesexpt}
\end{table}
 For baryons, experimental results are listed in \Tab\ref{tab:lifetimesexpt}. Given the new LHCb results, which significantly differ from previous measurements, it is useful to be more comprehensive about the lifetime measurements.
As compared with the meson lifetimes, there are some tensions in the available data. In particular:
\begin{enumerate}
\item The lifetime measurement for $\Xico$ has significantly shifted between the two eras, representing a $\sim3\sig$ tension. The most recent PDG update includes LHCb's earlier results \cite{LHCbcharmedLifetimes2019}.
\item Likewise, the $\Omc$ lifetime has shifted even more dramatically. PDG has in fact abandoned all earlier measurements, given that the LHCb data set is approximately five times larger than from all previous experiments.
\item We also note that the CLEO measurements are in some tension with other results, in particular in the case of $\Lifetime{\Lac}$. The most recent LHCb measurement is in good agreement with other results, supporting the conclusion that $\Lifetime{\Lac}$ is close to $200 \fs$.
\end{enumerate}
The most notable of these is undoubtedly the shift in the $\Omc$ lifetime, which  is now almost four times longer, but was previously found to be the shortest-lived charmed baryon. This prompts a new experimental hierarchy of charmed baryon lifetimes, 
\begin{eqnarray}
\text{exp:} \qquad\Lifetime{\Xico} < \Lifetime{\Lac} < \Lifetime{\Omc} < \Lifetime{\Xicp}\,.
\end{eqnarray} 

Using the reference values of the lifetimes shown in the last row of \Tab\ref{tab:lifetimesexpt} we obtain the lifetime ratios
\begin{equation}
    \frac{\tau(\Xi_c^+)}{\tau(\Lac)}=2.25 \pm 0.04\,,\qquad \frac{\tau(\Xi_c^0)}{\tau(\Lac)}=0.75 \pm 0.02\,,\qquad\frac{\tau(\Omega_c^0)}{\tau(\Lac)}=1.36 \pm 0.06\,,
\end{equation}
to which we compare our theoretical predictions.

Finally, the inclusive semileptonic branching fraction of $\Lac \to X e \nu$ has been measured experimentally as \cite{BESIII:2018mug}
\begin{equation}
    \text{BR}(\Lac\to Xe\nu)=(3.95\pm 0.35)\%\,.
\end{equation}
The remaining three semileptonic branching fractions ($\text{BR}(\Xicp \to Xe \nu)$, etc), have not yet been measured. Such measurements would provide further important checks of any specific theoretical approach. We provide our predictions of these branching fractions in \Sec\ref{sec:results_baryons}.

\subsection{The heavy quark expansion and inclusive decays}
In this section, we briefly overview the \acf{HQE}, and refer the reader to \cite{Lenz2014} for a more detailed review.

 Via the optical theorem, the total decay width can be related to the imaginary part of the forward transition operator:
\begin{equation}
\frac{1}{\Lifetime{H}}= \Ga(H) = \frac{1}{2m_H} \langle H| \mathcal{T} |H\rangle\,,\qquad \mathcal{T} = \Im\,\,i\int d^4x\, T\left[\mathcal{H}_{eff}(x)\mathcal{H}_{eff}(0)\right]\,,\label{eq:OpticalTheorem}
\end{equation}
where $\mathcal{H}_{eff}$ is the effective Hamiltonian describing the charged current interactions of the charm quark (eg \cite{BBL1995})
\begin{equation}
    \begin{split}
    \mathcal{H} &=\frac{G_F}{\sqrt{2}}\bigg[\sum_{q,q'=d,s} V_{cq}V^\ast_{uq'}\big(C_1(\mu) Q_1^{(qq')}+C_2(\mu) Q_2^{(qq')}\big)-V_{ub}V^\ast_{cb}\sum_{k=3}^{6}C_k(\mu)Q_k\\
    &+\sum\limits_{\substack{q=d,s \\ \ell=e,\mu}}V_{cq}Q^{(q\ell)}\bigg]\,,
    \label{eq:HeffSM}
    \end{split}
\end{equation}
where $G_F$ is the Fermi constant, $V_{ab}$ are \ac{CKM} matrix elements, and
\begin{equation}
\begin{split}
    Q_1^{(qq')}&=(\bar{c}^i\gamma_\mu(1-\gamma_5)q^j)(\bar{q}'^j\gamma^\mu(1-\gamma_5)u^i)\,,\\
    Q_2^{(qq')}&=(\bar{c}^i\gamma_\mu(1-\gamma_5)q^i)(\bar{q}'^j\gamma^\mu(1-\gamma_5)u^j)\,,\\
    Q_{\text{SL}}^{(q\ell)}&=(\bar{c}\gamma_\mu(1-\gamma_5)q)(\bar{\ell}\gamma^\mu(1-\gamma_5)\nu_\ell)\,,
    \end{split}
\end{equation}
where $i,j$ are colour indices. The remaining operators $Q_{3\text{-}6}$ denote the penguin operators, which are suppressed by the \ac{CKM} factor $V_{ub}V^\ast_{cb}$. Since the Wilson coefficients $C_3\text{-}C_6$ are also numerically small (eg \cite{BBL1995,LenzNote:2021}), we will neglect these contributions in the present paper. Note that $Q_2^{q q'}$ denotes the colour-singlet operator in our convention, following \cite{BBL1995} but opposite to the choice by some other authors, eg \cite{LenzNote:2021,Cheng18c,Melic97c}, where $Q_1$ is the colour-singlet. 

\begin{figure}
\centering
\includegraphics[scale=0.33,clip=true,trim = 110 320 140 60]{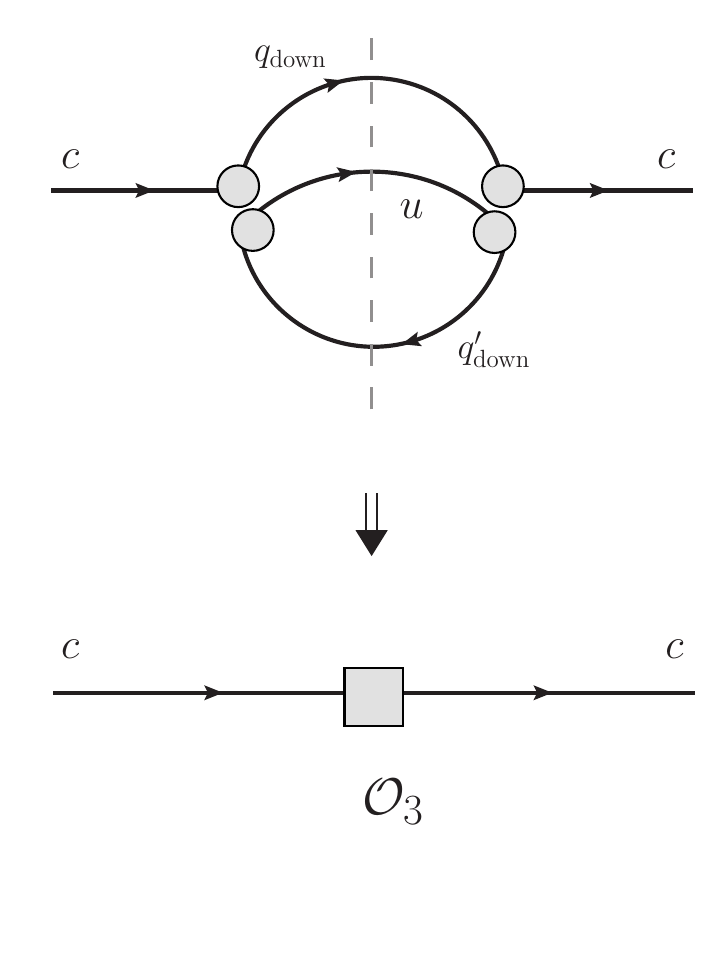}
\includegraphics[scale=0.33,clip=true,trim = 110 290 140 60]{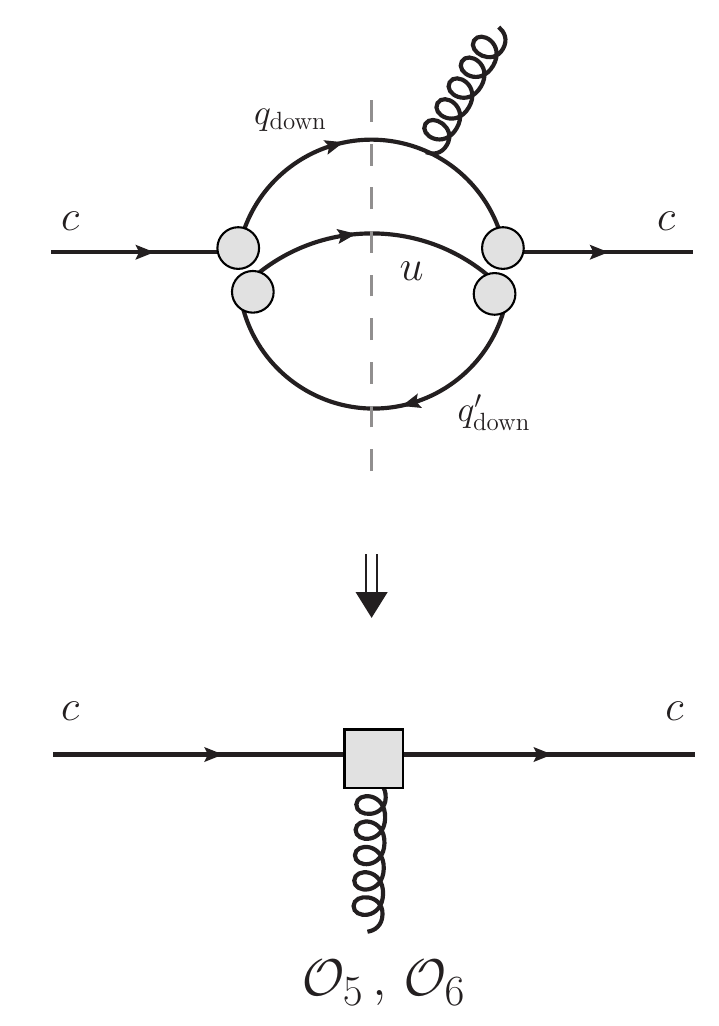}
\includegraphics[scale=0.33,clip=true,trim = 110 313 140 60]{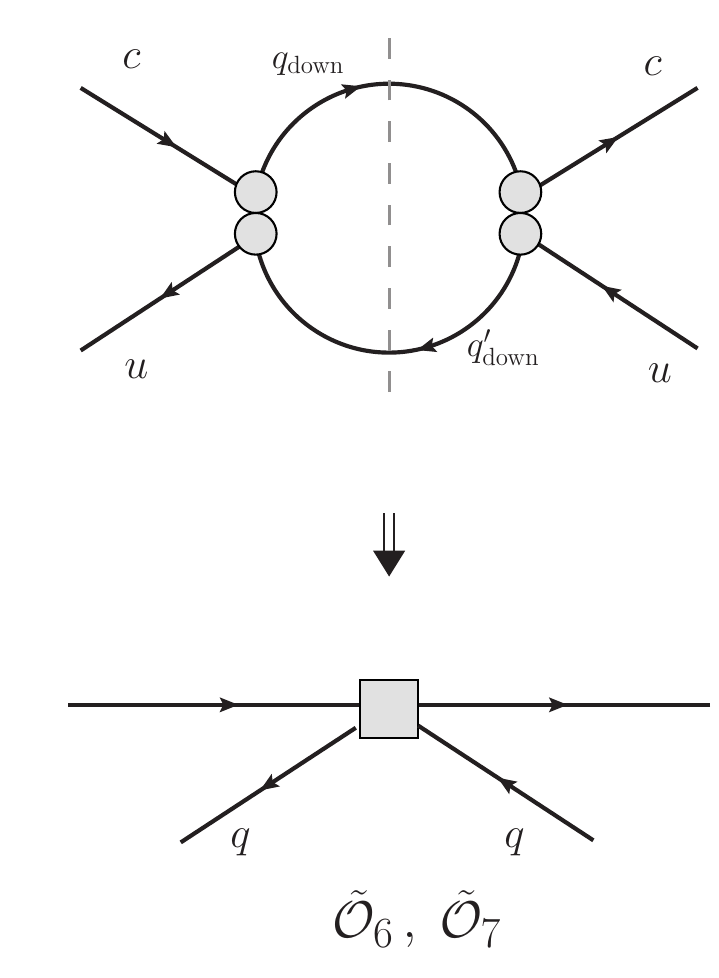}
\caption{Representative diagrams illustrating various contributions to the inclusive width of charmed hadrons. (a) The leading contribution $\mathcal{O}_3$. (b) higher-order ``non-spectator'' terms in the first series, generated by insertion of covariant derivatives with respect to the background gluon field. (c) An example of a four-quark ``spectator'' contribution, in this case Weak~exchange.\label{fig:HQE-1}}
\end{figure}

The right-hand side of \eqref{eq:OpticalTheorem} can then be expanded, using the \ac{HQE}, in powers of $\Lambda_{QCD}/m_Q$ and $\als$, where $m_Q$ is the heavy-quark mass and $\Lambda_{QCD}$ is the \acs{QCD} scale \cite{Chay:1990da,BSUV1993I}. 
 This yields a tower of local operators $\mathcal{O}_i$, ordered by increasing powers of the inverse heavy quark mass $m_Q$,
\begin{equation}
\mathcal{T} = \bigg(\mathcal{C}_3\mathcal{O}_3 +\frac{\mathcal{C}_5}{m_Q^2}\mathcal{O}_5 +\frac{\mathcal{C}_6}{m_Q^3}\mathcal{O}_6+\dots\bigg)+ 16\pi^2\bigg(\frac{\tilde{\mathcal{C}}_6}{m_Q^3}\tilde{\mathcal{O}}_6 + \frac{\tilde{\mathcal{C}}_7}{m_Q^4}\tilde{\mathcal{O}}_7 + \dots\bigg)\,,
\label{eq:HQEsystematic}
\end{equation}
where the Wilson coefficients $\mathcal{C}_i$ contain the short-distance physics, analogously to the $C_i$ in \eqref{eq:HeffSM}.\footnote{The absence of the dimension-four operator, suppressed by $\Lambda_{\text{QCD}}/m_{Q}$, was demonstrated in \cite{Chay:1990da,LukeThm}.} 
The operators within the first bracket are each composed of heavy-quark field bilinears, with operators of increasing dimension generated by insertion of covariant derivatives, and will be referred to below as the ``non-spectator'' contributions. The leading term $\mathcal{O}_3$ is represented by diagram (a) in \Fig\ref{fig:HQE-1}, while $\mathcal{O}_{5,6}$ is represented by diagrams similar to (b) in \Fig\ref{fig:HQE-1}. The terms within the second bracket involve the contributions of four-quark operators, with one example given by diagram (c) in \Fig\ref{fig:HQE-1}. These ``spectator contributions'' are sensitive to the flavour of the light quark in the hadron, and are one-loop enhanced relative to the non-spectator contributions by the factor $16\pi^2$. Therefore, they can result in significant lifetime splitting effects.

The $\mathcal{C}_i$ can be calculated perturbatively in powers of the strong coupling constant $\als$,
\begin{equation}
\mathcal{C}_i= \mathcal{C}_i^{(0)}(\mu,\mu_0) + \mathcal{C}_i^{(1)}(\mu,\mu_0) \als(\mu) + \mathcal{C}_i^{(2)}(\mu,\mu_0) \als(\mu)^2 + \dots ,
\label{eq:WilsonCoeffsSchematic}
\end{equation}
where $\mu$ is the renormalization scale arising from evolution of the weak Hamiltonian. Both the Wilson coefficients and the operators further depend on the operator factorization scale $\mu_0$. As will be made clear later, only a few of the $\mathcal{C}_i$ in \eqref{eq:HQEsystematic} are known beyond leading order.

Having summarized some of the key ideas, we now express the decay width more explicitly, in the form
\begin{align}
\Ga(H) = &~\Ga_0 \bigg[c_3 + \frac{c_\pi\mukin+c_G\muG}{\mQheavy^2} +  \frac{c_\rho\darwin}{\mQheavy^3} + \dots + \frac{16 \pi^2}{2m_H} \bigg( \sum\limits_{i,q}\frac{c_{6,i}^q\langle H| \Opsix{i}{q}| H\rangle}{\mQheavy^3} \nonumber \\
&{}+\sum\limits_{i}\dfrac{c_{7,i}^q\langle H| \OpsevenP{i}{q}| H\rangle}{\mQheavy^4}+ \dots \bigg) \bigg]\,,
\label{eq:DecayRateExpansion}
\end{align}
with the sum over $q=u,d,s$, where $\mQheavy = m_{c}$ is the pole mass of the charm quark and $m_H$ the mass of the heavy hadron. The normalization factor is
\begin{equation}
\Ga_0 = \frac{G_F^2 \mQheavy^5}{192\pi^3} \,, \label{eq:Ga0}
\end{equation}
and all \ac{CKM} contributions are included implicitly in the  coefficients $c_i$ in the equation above. The definitions of the non-perturbative parameters $\mukin,\,\muG,\, \darwin$, and the description of the spectator contributions, are given in the next section. 
The overall decay width \eqref{eq:DecayRateExpansion} can be further split into contributions from \ac{SL} decays and \ac{NL} decays, where the \ac{SL} decays can also be measured separately in experiments. Note that the above form, and in particular the coefficients $c_i$, applies to all hadrons considered in this paper, and can also be applied to hadrons containing a $b$ quark upon appropriate replacement of the quark/lepton masses, quark fields in the matrix elements, and CKM factors.\footnote{In \cite{FMV2019}, it was argued that, when applying the \ac{HQE} to inclusive charm decays, the strange quark mass should also be considered as an expansion parameter with the same status as $\La_{QCD}/m_Q$. We return to this point in \Sec\ref{sec:nonspeccontgen}, but note here that in the present approach we follow eg \cite{LenzNote:2021,LenzRauh2013,Cheng18c} in handling $m_s$ contributions.} Given that the $c_i$ are therefore universal, we now turn to briefly discussing their present status, as well as defining the operator basis of \eqref{eq:DecayRateExpansion}.

\subsection{Contributions to the decay width and operator basis}
\label{sec:contgeneral}
To calculate the lifetimes, we apply the following expansions, given schematically as:
\begin{align}
\Gamma = & \Gamma^{\rm NL} +  \Ga^{\textrm{SL}} \,, \nonumber \\
    \Gamma^{\rm NL} &= g_{3}^{(0)} + \als g_{3}^{(1)} + \frac{1}{m_c^2} \left( g_{\pi}^{(0)} + g_G^{(0)} \right) + \frac{1}{m_c^3} g_{\text{Darwin}}^{(0)} + \frac{16\pi^2}{m_c^3} \left( \tilde{g}_6^{(0)} + \als \tilde{g}_{6}^{(1)} +\frac{1}{m_c}\tilde{g}_7^{(0)} \right) \,, \nonumber \\
    \Gamma^{\textrm{SL}} &= g_{3}^{(0)} + \als g_{3}^{(1)} + \frac{1}{m_c^2} \left( g_{\pi}^{(0)} + \als g_{\pi}^{(1)} + g_G^{(0)} + \als g_G^{(1)} \right) + \frac{1}{m_c^3} g_{\text{Darwin}}^{(0)}  \nonumber \\
    & \quad + \frac{16\pi^2}{m_c^3} \left( \tilde{g}_6^{(0)} + \als \tilde{g}_{6}^{(1)} +\frac{1}{m_c}\tilde{g}_7^{(0)} \right) \,,
    \label{eq:decaywidthcontribs}
\end{align}
where the individual contributions will be described in this section. 
$g_{3}$ is the leading non-spectator contribution, while $g_G, g_{\pi}$ and $g_{\text{Darwin}}$ are $1/m_c^n$  suppressed contributions; $\tilde{g}_{6,7}$ label the four-quark  spectator  contributions. Superscripts $(0)$ and $(1)$ denote \ac{LO} and \ac{NLO} contributions respectively.

\subsubsection{Non-spectator contributions}
\label{sec:nonspeccontgen}

The non-spectator terms are given by the first series in \eqref{eq:DecayRateExpansion}:
\begin{equation}
    \Ga(H) = \Ga_0 \bigg[c_3 + \frac{c_\pi\mukin+c_G\muG}{\mQheavy^2} +  \frac{c_\rho\darwin}{\mQheavy^3} +\dots \bigg]
\end{equation}
where the matrix elements \cite{DMT2006,LenzNote:2021,Mannel1994,BSUV1993II,BSUV1994}
\begin{align}
\mukin(H) &= \frac{-1}{2m_H} \langle H |\bar{c}_v (iD)^2 c_v | H \rangle \,, \nonumber \\
\muG(H)&=\frac{1}{2m_H} \langle H |\bar{c}_v \frac{1}{2} \sig \cdot (g_s G) c_v  |H \rangle\,, \nonumber \\
\darwin(H) &= \frac{1}{2m_H}\langle H |\bar{c}_v (i D_\mu) (i v \cdot D) (i D^\mu) c_v | H \rangle \,,
\label{eq:nonspecdefns}
\end{align} 
are the kinetic, chromomagnetic, and Darwin operators respectively ($\sig_{\mu\nu} = \frac{i}{2}[\ga_\mu,\ga_\nu]$).\footnote{In some early literature, eg \cite{Mannel1994}, the alternative notation $\la_{1,2}$ was employed for the first two matrix elements, with $\mukin = -\la_1$ and $\mu_G = d_H \la_2$, where $d_H$ is a spin factor given explicitly in \eqref{eq:dHdef}.} In some conventions, an additional operator arises at dimension-six, the spin-orbit term $\rho^3_{LS}$, but in the basis above its coefficient in the total decay rate vanishes. The above expressions are defined in terms of the heavy charm \acs{QCD} field, with the large momentum fraction removed, ie \cite{Neubert1993,Manohar:2000dt}
\begin{equation}
    c_v(x) = e^{i m_c v\cdot x}c(x) \,,
\end{equation}
where $v^\mu$ is the four-velocity of the hadron. The leading term $c_3$ is the coefficient of the matrix element $\langle H| \bar{c}_vc_v| H \rangle/(2m_H)$, normalized to 1 at leading order in the $1/m_c$ expansion. In fact, at dimension-five, diagrams of the type exemplified by \Fig\ref{fig:HQE-1}(b) do not contribute to  $c_\pi$, meaning that it is only non-zero owing to the expansion of the dimension-three matrix element \cite{BSUV1994II,DMT2006}
\begin{equation}
   \frac{1}{2m_H}\langle H| \bar{c}_v c_v| H \rangle = 1-\frac{\mukin-\muG}{2m_Q^2}+\mathcal{O}(1/m_Q^4)\,.\label{eq:HQE1}
\end{equation}
Consequently, $c_\pi = -c_3/2$. On the other hand, \Fig\ref{fig:HQE-1}(b) does contribute to  $c_G$, alongside the contribution generated by the expansion \eqref{eq:HQE1}.  

To \ac{LO}, the general form of the coefficients is \begin{equation}
c_n^{(0)} = N_C C_1^2 \mathcal{K}^{(0)}_{n,11}  + 2 C_1 C_2 \mathcal{K}^{(0)}_{n,12}  + N_C C_2^2 \mathcal{K}^{(0)}_{n,22}  +\mathcal{K}^{(0)}_{n,\textrm{SL}} \, ,
\label{eq:decompcns}
\end{equation}
where $N_C=3$ is the number of colours, and we remind the reader that in our conventions $C_2$ is the Wilson coefficient for the colour-singlet operator. 
Defining the mass ratios as 
\begin{equation}
x_s = \frac{m_s^2}{m_c^2} \,, \qquad  x_\mu = \frac{m_\mu^2}{m_c^2}\,,
\end{equation} 
then the \ac{LO} results for $c_{3}$ are \cite{Cheng18c,Lenz2014,CPT1982,Koyrakh1993}
\begin{align}
\mathcal{K}^{(0)}_{3,ij} &= |V_{cs}|^2 \left( |V_{ud}|^2 I_0(x_s,0,0) + |V_{us}|^2 I_0(x_s,x_s,0) \right) + |V_{cd}|^2 |V_{ud}|^2 I_0(0,0,0)  \, , \nonumber \\
\mathcal{K}^{(0)}_{3,\textrm{SL}} &= |V_{cs}|^2 \left(  I_0(x_s,0,0) +  I_0(x_s,x_\mu,0) \right) + |V_{cd}|^2 \left(  I_0(0,0,0) + I_0(0,x_\mu,0) \right) \, , 
\label{eq:c3atLO}
\end{align}
where we have included the singly \ac{CKM}-suppressed 
contributions, while $I_0(x,y,z)$ is a phase-space function defined explicitly in \App\ref{app:c356}. At \ac{LO}, all the Wilson coefficient structures in $c_3$ are identical, while the \ac{SL} part can be recovered from the $C_2^2$ coefficient by applying the replacements $C_2 \to 1$, $N_C \to 1$, $|V_{ud}|^2 \to 1$, and by appropriate redefinition of masses. For $c_G$ we have
\begin{align}
\mathcal{K}^{(0)}_{G,ii} &= -|V_{cs}|^2 \left( |V_{ud}|^2 I_1(x_s,0,0) + |V_{us}|^2 I_1(x_s,x_s,0) \right) -|V_{cd}|^2  |V_{ud}|^2 I_1(0,0,0)  \, , \nonumber \\
\mathcal{K}^{(0)}_{G,12} &= |V_{cs}|^2 \left( |V_{ud}|^2 \left(4I_2(x_s,0,0) - I_1(x_s,0,0) \right)+ |V_{us}|^2 \left(4I_2(x_s,x_s,0) - I_1(x_s,x_s,0) \right) \right) \nonumber \\
& \quad {} + |V_{cd}|^2 |V_{ud}|^2 \left(4I_2(0,0,0) - I_1(0,0,0) \right)  \, , \nonumber \\
\mathcal{K}^{(0)}_{G,SL} &= -|V_{cs}|^2 \left( I_1(x_s,0,0) + I_1(x_s,x_\mu,0) \right) -|V_{cd}|^2 \left( I_1(0,0,0) + I_1(0,x_\mu,0) \right) \, , 
\label{eq:c5atLO}
\end{align}
where $I_{1,2}(x,y,z)$ are additional phase-space functions given in \App\ref{app:c356} \cite{BUV1992HQE,BS1992I,BS1992II,BBSUV92HQE}. It is worth noting that the scale dependence of $c_G$ at \ac{LO} is quite significant, and can even cause $c_G$ to change sign, going negative at larger values of $\mu$ \cite{LenzNote:2021}. 

As for $c_\rho$, no such compact expressions are available, but we provide the analytic expressions, taken from \cite{LenzNote:2021} (see also \cite{GK1996,MRS2017,MMP2020,LPR2020,Moreno2020}), in \App\ref{app:c356}. The \ac{SL} parts of $c_{G,\rho}$ can again be recovered from the $C_2^2$ coefficient by applying the same replacements listed below \Eqn\eqref{eq:c3atLO}. 

In terms of the $\als$ expansion, $c_3$ has been computed to \ac{NLO} for \ac{NL} decays in \cite{HokimPham84,BBBG1994,BBFG1995,KLR2013}, with a partial result at N\ac{NLO} in \cite{CST2005}. For \ac{SL} decays, results are available at \ac{NLO} in \cite{CJK1994}, N\ac{NLO} in \cite{LSW1994,Ritbergen1999,PC2008I,PC2008II,BM2009}, and recently to N$^3$LO in \cite{FSS2020,CCD2021}. However, since many results beyond \ac{NLO} are partial, and in view of the divergent nature of the $c_3$ series, we restrict our analysis to include only the \ac{NLO} contributions in the present work, and will treat the $\als$ series as asymptotic.\footnote{Some further comments on the $c_3$ series can be found in \Sec\ref{sec:alsseriescomments}.}

For $c_G$, results are available at \ac{NLO} \cite{AGN2013,MPR2014,MPR2015} only for \ac{SL} decays. We include these results in our analysis.  For the $c_\rho$ contribution, the \ac{NLO} result was computed, for $b \to c \tau \nu_\tau $ decays, in \cite{MP2019,MMP2021II}. It is worth stressing, however, that in handling the $\darwin$ contributions, care must be taken in handling the mixing effects with the other dimension-six operators \cite{FMV2019,LenzNote:2021}, with a different treatment required for $c \to s $ decays as compared with $b \to c$ decays. As a result, the results of \cite{MP2019,MMP2021II} cannot be naively applied to $c \to s$ decays, and so we do not include them in our analysis.  

In summary, we include the available \ac{NLO} contributions for all non-spectator terms apart from in the Darwin contribution, but do not include contributions beyond \ac{NLO}. A list of relevant contributions and references is given in \Tab\ref{tab:pertcorr}.

Before proceeding to discuss the contributions arising from four-quark operators, we wish to discuss two alternative approaches to the \ac{HQE} in the charm sector advanced recently, specifically those in \cite{FMV2019} and \cite{MMP2021}. The work of the former is particularly interesting, and may not yet have received enough attention in the inclusive \ac{HQE} literature. The principal idea in that paper is that, for consistency, the parameter $m_s/m_c$ should be treated as an expansion parameter in the \ac{HQE}, on the same footing as $\La_{\text{QCD}}/m_c$. This differs from the approach of \cite{Melic97c,Cheng18c,LenzNote:2021}, whereby results obtained traditionally in the setting of inclusive $b\to c$ decays are assumed to apply to $c \to s$ decays, with appropriate replacement of quark masses and \ac{CKM} factors, eg \cite{Cheng18c,Melic97c}. In the context of the Darwin contribution, the work of \cite{LenzNote:2021} partially confirms this (compare with \cite{MMP2020,Moreno2020,LPR2020} computing the same terms for $b \to c$ decays), but in fact the modified \ac{HQE} in the charm sector goes beyond just the Darwin contribution, as can be confirmed by comparing the expressions in \Eqns(5.6) and (A.1) in \cite{FMV2019}. To what extent this represents a genuine difference that cannot be reconciled with the standard approach, as opposed to merely a re-ordering of the same expression, remains to be seen. Since, however, the factor $m_s^2/m_c^2$ is fairly small, it is reasonable to assume that any errors in the handling of the strange quark in the present approach are negligible compared with other uncertainties. Further studies of this point would be welcome, particularly if determinations of other parameters in the \ac{HQE} come with reduced uncertainties. 

In \cite{MMP2021}, it was argued that the contribution to inclusive decays of four-quark operators should be re-summed, and considered part of the leading term, in order to render the \ac{HQE} a true expansion with parameters of order unity. This, however, presents the difficulty of evaluating non-local matrix elements, and while this could perhaps be done on the lattice in the future, the approach has thus far only been tested in an extremely simplified setting. Further work exploring these questions would also be welcome, especially as an alternative to merely evaluating further terms in the slowly-converging $1/m_c$ series. 

In any case, the validity of the current approach, or equivalently the urgency of re-formulating the \ac{HQE} for charm decays, can be assessed most strongly by comparing the results obtained with experimental data. The debate over the proper application of the \ac{HQE} to charm decays is hardly new, eg \cite{BlokShifman93Review,BSU97ReviewHQE,LenzRauh2013,Lenz2014}, and is likely to continue for some time. Applying the approaches advocated in \cite{FMV2019,MMP2021} of tailoring the \ac{HQE} more suitably for charm decays in a more concrete setting may serve to clarify the issue.

\subsubsection{Spectator contributions}
\label{sec:speccontgen}

\begin{figure}[t]
    \centering
\includegraphics[scale=0.33,clip=true,trim = 110 550 140 60]{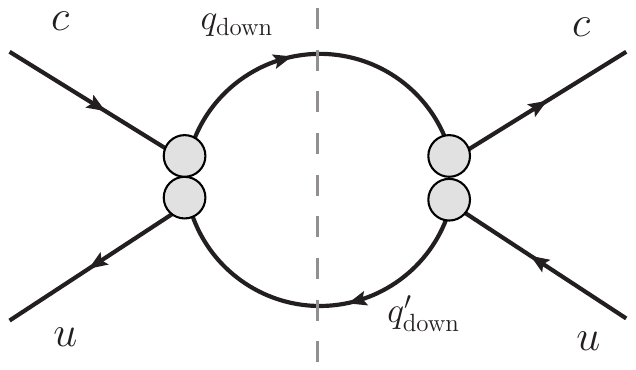}
   \includegraphics[scale=0.33,clip=true,trim = 110 550 140 60]{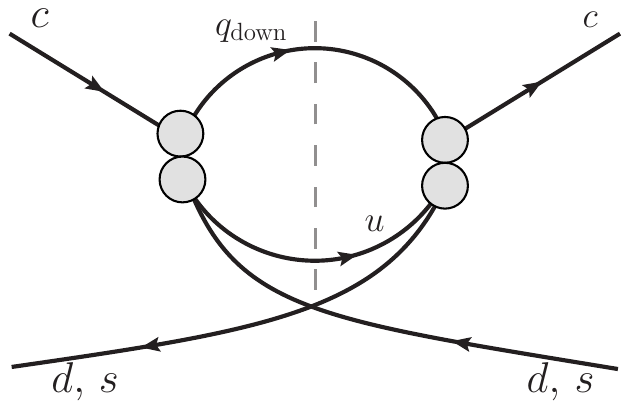}
   \includegraphics[scale=0.33,clip=true,trim = 110 550 140 60]{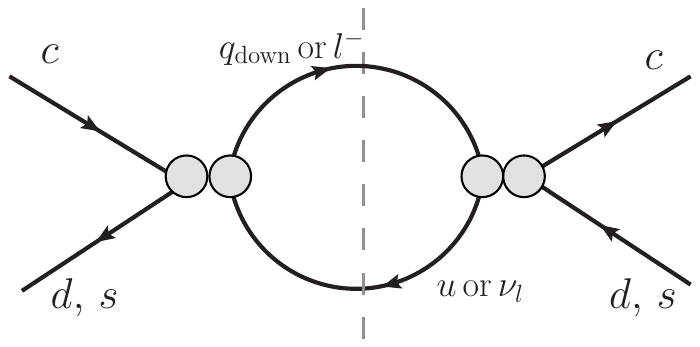}
       \caption{\small Diagrammatic representations of the spectator contributions resulting in the four-quark operators in {\it mesons}. From left to right: (a) \acf{WE}, (b) \acf{PI}, (c) \acf{WA}.}
       \label{fig:1-DiagramsMesons}
\end{figure}

\begin{figure}[t]
    \centering
   \includegraphics[scale=0.33,clip=true,trim = 110 511 140 60]{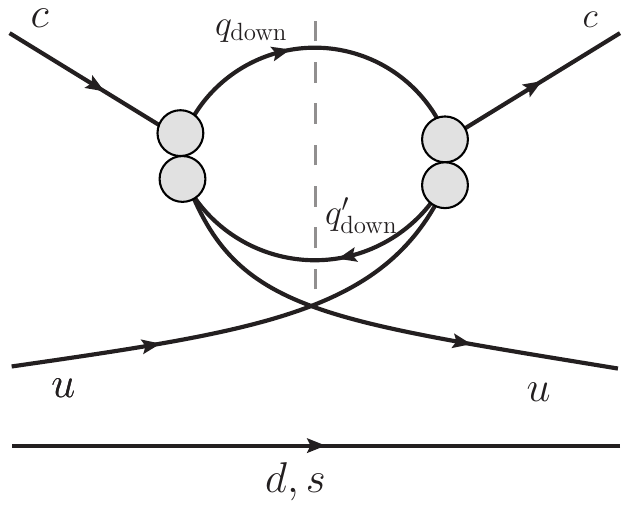}
   \includegraphics[scale=0.33,clip=true,trim = 110 540 140 60]{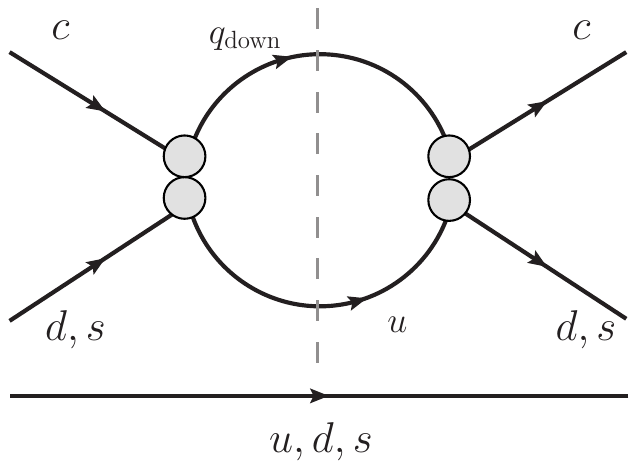}
   \includegraphics[scale=0.33,clip=true,trim = 110 520 140 60]{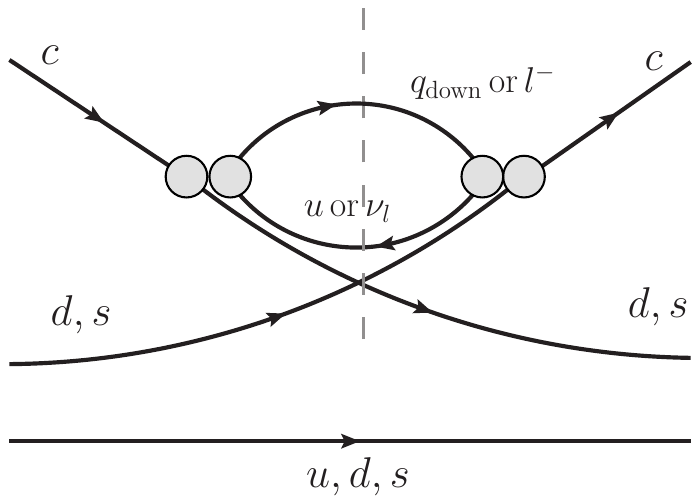}
       \caption{\small Diagrammatic representations of the spectator contributions resulting in the four-quark operators in {\it baryons}. From left to right: (a) destructive Pauli interference, labelled `$\intm$'; (b) Weak exchange, labelled `$\exc$'; (c) constructive Pauli interference, labelled `$\intp$'. The non-participating light quark is also indicated. The correspondence to the equivalent meson diagrams is clearly visible by comparison with \Fig\ref{fig:1-DiagramsMesons}.}
     \label{fig:1-DiagramsBaryons}
\end{figure}

The remaining contributions to the decay width \eqref{eq:DecayRateExpansion} arise from four-quark operators, and can be described, for mesons, by the topologies in \Fig\ref{fig:1-DiagramsMesons}. The three topologies are typically referred to as \acf{WE}, \acf{PI}, and \acf{WA}. As compared with the terms discussed in the previous section, these are enhanced by the factor $16\pi^2$, being one-loop effects, and are primarily responsible for the lifetime splitting between heavy hadrons. For baryons, the equivalent topologies are represented in \Fig\ref{fig:1-DiagramsBaryons}, and are referred to as weak exchange ($\exc$), constructive Pauli interference ($\intp$), and destructive  Pauli interference ($\intm$). The correspondence to the
equivalent meson contributions (WE $\leftrightarrow \intm$, PI $\leftrightarrow \exc$, WA $\leftrightarrow \intp$) is clearly visible by comparing \Fig\ref{fig:1-DiagramsMesons} and \Fig\ref{fig:1-DiagramsBaryons}. Practically, this means that the expressions are the same at the operator level, although they differ at the level  of the resulting matrix elements, as discussed in \Secs\ref{sec:mesonsresultsmain} and \ref{sec:baryonsresultsmain}. 
\begin{figure}[t]
    \centering
  \includegraphics[scale=0.35,clip=true,trim = 110 510 100 60]{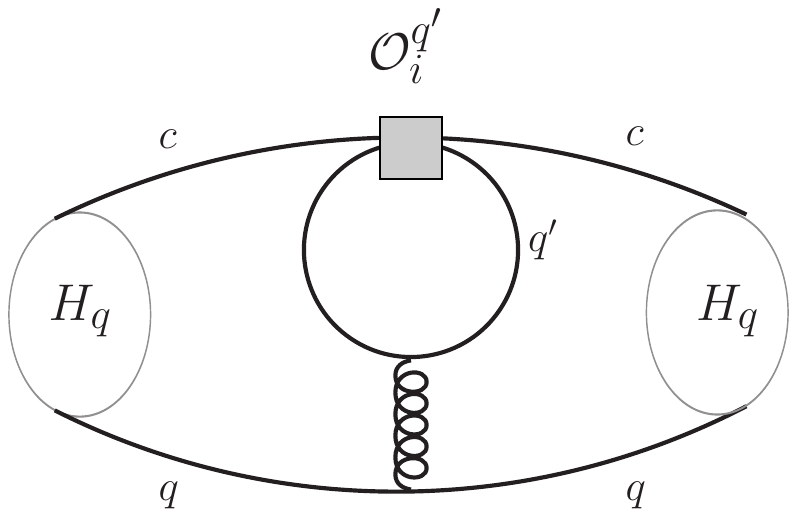}
       \caption{\small Diagrammatic representation of ``eye contractions'', which serve as a further contribution relevant in the four-quark operators \cite{Voloshin2001,LenzNote:2021,KLR2021}. In particular, the quark running in the loop $q'$ need not match the valence quark in the hadron, $q$, leading to further non-valence contributions to the decay widths. Such contributions are numerically small \cite{LenzNote:2021,KLR2021}, but allow, for example, for a \ac{PI} contribution to the $D^0$ lifetime.}
     \label{fig:1-DiagramsEye}
\end{figure}
Spectator contributions are also present in semileptonic decays of both mesons and baryons, as indicated in \Figs\ref{fig:1-DiagramsMesons}(c) and \ref{fig:1-DiagramsBaryons}(c) respectively, with the identical topology to \ac{WA}/$\intp$. For mesons, such contributions are helicity-suppressed, but this does not apply in baryon decays, where these contributions are  relevant. The effect of these contributions in semileptonic decays was first noticed by Voloshin in \cite{Voloshin96}.

A further relevant topology, known as the ``eye contraction'' \cite{Voloshin2001,LenzNote:2021,KLR2021}, is represented in \Fig\ref{fig:1-DiagramsEye}. The eye contractions allow for contributions where the light valence quark in the hadron does not necessarily match the quark involved in the short-distance interaction.

The relevant operator basis varies depending on whether we are considering mesons or baryons, although, as can be seen by comparing \Fig\ref{fig:1-DiagramsMesons} and \Fig\ref{fig:1-DiagramsBaryons}, the different bases are related. Beginning with the mesons, the dimension-six operators are, following \cite{NS1996,LenzRauh2013,LenzNote:2021},\footnote{A useful dictionary relating this parametrization to other historical parametrizations of the four-quark matrix elements is given in appendix C of \cite{PU1998}.}
\begin{alignat}{3}
\Opsix{1}{q} &= (\bar{c}_i \ga_\mu(1-\ga_5) q_i) (\bar{q}_j \ga^\mu(1-\ga_5) c_j)\,, \quad
& \Opsix{2}{q} &= (\bar{c}_i (1-\ga_5) q_i) (\bar{q}_j (1+\ga_5) c_j) \,, \nonumber \\
\OpsixT{1}{q} &= (\bar{c}_i\ga_\mu(1-\ga_5)  t^a_{ij} q_j) (\bar{q}_k \ga^\mu(1-\ga_5) t^a_{kl}  c_l)\,, \quad
&\OpsixT{2}{q} &= (\bar{c}_i(1-\ga_5)  t^a_{ij} q_j) (\bar{q}_k (1+\ga_5) t^a_{kl}  c_l)\;,\label{eq:Dim6MesonBasis}
\end{alignat}
where colour indices are denoted by $i,j$, the $t^a$ are colour matrices, and $q=u,d,s$ are light-quark flavours. This basis is most suited for mesons, because within the \ac{VIA} the matrix elements $\langle D | \OpsixT{i}{q} | D \rangle$ vanish \cite{NS1996}. Note that these operators are defined with full QCD fields. To obtain a consistent expansion in $1/m_Q$, one uses instead the basis
\begin{alignat}{3}
\Opsixprime{1}{q} &= (\bar{h}_{v,i} \ga_\mu(1-\ga_5) q_i) (\bar{q}_j \ga^\mu(1-\ga_5) h_{v,j})\,, \quad
& \Opsixprime{2}{q} &= (\bar{h}_{v,i} (1-\ga_5) q_i) (\bar{q}_j (1+\ga_5) h_{v,j}) \,, \nonumber \\
\OpsixTprime{1}{q} &= (\bar{h}_{v,i}\ga_\mu(1-\ga_5)  t^a_{ij} q_j) (\bar{q}_k \ga^\mu(1-\ga_5) t^a_{kl}  h_{v,l})\,, \quad
&\OpsixTprime{2}{q} &= (\bar{h}_{v,i}(1-\ga_5)  t^a_{ij} q_j) (\bar{q}_k (1+\ga_5) t^a_{kl}  h_{v,l})\;,\label{eq:Dim6MesonBasisHQE}
\end{alignat}
where $h_v$ is the heavy-quark field, and again the colour-octet contributions $\langle D|\OpsixTprime{i}{q}| D \rangle$ vanish in the \ac{VIA}. This basis will be referred to as the \ac{HQET} basis. 

The corresponding \ac{LO} coefficients $c_{6,i}^q$ in \eqref{eq:DecayRateExpansion} have been known for some time \cite{SV1985,SV1986,GRT1986,Voloshin96,NS1996}, while the \ac{NLO} contributions were first reported in \cite{ CFLM2001,BBGLN2002,FLMT2002,LenzRauh2013}. We adapt the \ac{NLO} results from \cite{CFLM2001}, which were computed for $b$ decays, to the charm sector. The explicit results in that reference correspond to the basis
\begin{alignat}{3}
&\bar{\mathcal{O}}^q_1=(\bar{h}_{v,i}\gamma_\mu(1-\gamma_5)h_i)(\bar{q}_j\gamma^\mu(1-\gamma_5)q_j)\,, &&\bar{\mathcal{O}}^q_2=(\bar{h}_{v,i}\gamma_\mu(1+\gamma_5)h_{v,i})(\bar{q}_j\gamma^\mu(1-\gamma_5)q_j) \,,\nonumber\\
&\bar{\mathcal{O}}^q_3=(\bar{h}_{v,i}\gamma_\mu(1-\gamma_5)t^a_{ij}h _{v,j})(\bar{q}_i\gamma^\mu(1-\gamma_5)t^a_{ij} q_j)\,,\quad  &&\bar{\mathcal{O}}^q_4=(\bar{h}_{v,i}\gamma_\mu(1+\gamma_5)t^a_{ij} h_{v,j})(\bar{q}_i\gamma^\mu(1-\gamma_5)t^a_{ij} q_j)\,,
\end{alignat}
with the operators denoted by bars, in order to avoid overlap with the notation in \Eqn\eqref{eq:Dim6MesonBasisHQE}. The two bases are related by the Fierz transformation
\begin{equation}
\vec{\mathcal{O}}=F\vec{\mathcal{Q}}\,,
\end{equation}
where $\vec{\mathcal{O}}=(\bar{\mathcal{O}}^q_1,\bar{\mathcal{O}}^q_2,\bar{\mathcal{O}}^q_3,\bar{\mathcal{O}}^q_4)$ and $\vec{\mathcal{Q}}=(\Opsixprime{1}{q}, \Opsixprime{2}{q}, \OpsixTprime{1}{q}, \OpsixTprime{2}{q})$, with the transformation matrix  
\begin{equation}
F=
\begin{pmatrix}
\frac{1}{3} && 0 && 2 && 0 \\
0 && -\frac{2}{3} && 0 && -4 \\
\frac{4}{9} && 0 && -\frac{1}{3} && 0 \\
0 && -\frac{8}{9} && 0 && \frac{2}{3}
\end{pmatrix}\,.
\end{equation}
The matching between the flavour nonsinglet dimension-six four quark operators defined in HQET $\vec{\mathcal{Q}}$ and the corresponding QCD operators $\vec{Q}$ is performed using the relation
\begin{equation}
    \vec{\mathcal{Q}}(\mu_0)=F^{-1}\widehat{C}^{-1}(\mu_{h},\mu_0,m_c)F\vec{Q}(\mu_h)\,,\label{eq: C-matrix1}
\end{equation}
with the matrix \cite{CFLM2001}
\begin{equation}
    \widehat{C}(\mu_h,\mu_0,m_c)=(W^T(\mu_h, m_c))^{-1}\hat{C}(m_c)\widetilde{W}^T(\mu_0, m_c) \,,
    \label{eq: C-matrix2}
\end{equation}
where $\widetilde{W}(\mu_0,m_c)$ and $W(\mu_h,m_c)$ denote the renormalization group evolution matrices within HQET and QCD, respectively, and $\hat{C}(m_c)$ is the matrix used to match between the two frameworks at the scale $m_c$. The scale $\mu_h\sim 1\GeV$ denotes the low hadronic scale. If one remains within the HQET framework, the matrix $\widehat{C}(\mu_{h},\mu_0,m_c)$ reduces to the usual HQET evolution matrix implementing the so-called hybrid renormalization (eg \cite{NS1996}). 
In general, this evolution to the lower scale $\mu_{h}$ will also bring penguin operators into play \cite{SV1986,BlokShifman93Review}.
There are two types of penguin contributions: those from the penguin operators themselves, and those arising from penguin-like diagrams. Both of these can be safely neglected, since on the one hand the penguin diagrams are neglible, while on the other hand the penguin operator contributions bring at most a few \% corrections, which are not relevant compared to other uncertainties in the computation. 

For our numerical evaluation, we use the fixed value $\mu_0=1.5\GeV$, and we also set $m_c=1.5\GeV$ in the formulas in \Eqns\eqref{eq: C-matrix1} and \eqref{eq: C-matrix2}. For the case of charmed mesons, we use the explicit numerical results for the corresponding matrix elements of the four-quark operators evaluated within the framework of HQET sum rules \cite{LenzNote:2021, KLR2021}. Therefore, in this case we have $\widehat{C}$ equal to the identity matrix. For the case of baryons, we have assumed that the matrix elements of the four-quark operators evaluated within the \ac{NRCQM} correspond to the matrix elements of the QCD operators renormalized at the low hadronic scale, which we set to $\mu_h=1\GeV$. Note that the above running and matching matrices are currently known only for the nonsinglet flavour operators, eg $\mathcal{O}_i^u-\mathcal{O}_i^d$, that do not mix under renormalization with flavour singlet operators involving penguin contractions. The matrix elements of flavour nonsinglet operators appear in the differences between the baryon lifetimes  within a given $SU(3)_F$ multiplet. However, we have applied these results for evaluations of the absolute values of the decay widths, due to the expectation that the neglected missing contributions are small compared with other uncertainties in the matrix elements.


At the next order in the $1/m_c$ expansion, the dimension-seven operators in the meson basis are
\begin{align}
\OpsevenP{1}{q} &=  m_q (\bar{c}_i (1-\ga_5) q_i) (\bar{q}_j (1-\ga_5) c_j) \,, \nonumber \\   
\OpsevenP{2}{q} &= \frac{1}{\mQheavy}(\bar{c}_i \Dleft_\rho \ga_\mu(1-\ga_5) D^\rho q_i) (\bar{q}_j  \ga^\mu(1-\ga_5) c_j) \,, \nonumber \\
\OpsevenP{3}{q} &= \frac{1}{\mQheavy}(\bar{c}_i \Dleft_\rho (1-\ga_5) D^\rho q_i) (\bar{q}_j  (1+\ga_5) c_j) \,, \nonumber \\
\OpsevenS{1}{q} &=  m_q (\bar{c}_i (1-\ga_5)t^a_{ij} q_j) (\bar{q}_k (1-\ga_5)t^a_{kl} c_l) \,, \nonumber \\ 
\OpsevenS{2}{q}&=\frac{1}{\mQheavy}(\bar{c}_i \Dleft_\rho\ga_\mu(1-\ga_5) t^a_{ij} D^\rho q_j) (\bar{q}_k \ga^\mu(1-\ga_5) t^a_{kl} c_l) \,, \nonumber \\
\OpsevenS{3}{q} &=\frac{1}{\mQheavy}(\bar{c}_i \Dleft_\rho (1-\ga_5) t^a_{ij} D^\rho q_j) (\bar{q}_k  (1+\ga_5)t^a_{kl} c_l) \,.
\label{eq:dim7matelmes}
\end{align}
Note that an earlier basis for the dimension-seven operators, used in \cite{LenzRauh2013}, contained an additional operator $
{\OpsevenP{2}{q}}^{\textrm{\cite{LenzRauh2013}}} =  m_q (\bar{c}_i (1+\ga_5) q_i) (\bar{q}_j (1+\ga_5) c_j)  $, but this operator can be omitted, since it is related to $\OpsevenP{1}{q}$ above by hermitian conjugation \cite{LenzNote:2021}. Again, within this basis the matrix elements $\langle D| S_i^q|D \rangle$ can be expected to vanish within the \ac{VIA}. Likewise, for a consistent $1/m_Q$ expansion, one should re-express the operators in terms of heavy-quark fields. However, unlike at dimension-six, this time new operators emerge:
\begin{align}
\OpsevenPprime{1}{q} &=  m_q (\bar{h}_{v}^i (1-\ga_5) q^i) (\bar{q}^j (1-\ga_5) h_v^j) \,, \nonumber  \\  
\OpsevenPprime{2}{q} &= (\bar{h}_v^i \ga_\mu(1-\ga_5) iv \cdot D q_i) (\bar{q}_j  \ga^\mu(1-\ga_5) h_v^j) \,, \nonumber \\
\OpsevenPprime{3}{q} &= (\bar{h}_v^i (1-\ga_5) iv \cdot D q^i) (\bar{q}^j (1+\ga_5) h_v^j) \,, \nonumber   \\
\OpsevenRprime{1}{q}&= (\bar{h}_v^i\ga^\mu (1-\ga_5)q^i) (\bar{q}^j \ga_\mu(1-\ga_5)\slashed{D}h_v^j) \,, \nonumber  \\ 
\OpsevenRprime{2}{q} &=(\bar{h}_v^i (1-\ga_5)q^i) (\bar{q}^j (1-\ga_5)\slashed{D}h_v^j)\,, 
\label{eq:dim7matelmesHQE}
\end{align}
supplemented with the colour-octet operators, $\mathcal{S}^{q}_i$ corresponding to $\mathcal{P}^{q}_i$, and $\mathcal{U}^q_{1,2}$ corresponding to $\mathcal{R}^q_{1,2}$:
\begin{align}
\OpsevenSprime{1}{q} &=  m_q (\bar{h}_{v}^i (1-\ga_5) t^a_{ij} q_j) (\bar{q}^k (1-\ga_5)t^a_{kl} h_v^l) \,, \nonumber  \\  
\OpsevenSprime{2}{q} &= (\bar{h}_v^i \ga_\mu(1-\ga_5) iv \cdot D t^a_{ij} q_j) (\bar{q}_k  \ga^\mu(1-\ga_5) t^a_{kl} h_v^l) \,, \nonumber \\
\OpsevenSprime{3}{q} &= (\bar{h}_v^i (1-\ga_5) iv \cdot D t^a_{ij} q_j) (\bar{q}^k (1+\ga_5) t^a_{kl} h_v^l) \,, \nonumber   \\
\OpsevenUprime{1}{q}&= (\bar{h}_v^i\ga^\mu (1-\ga_5)t^a_{ij} q_j) (\bar{q}^k \ga_\mu(1-\ga_5)\slashed{D}h_v^jt^a_{kl} h_v^l) \,, \nonumber  \\ 
\OpsevenUprime{2}{q} &=(\bar{h}_v^i (1-\ga_5)t^a_{ij} q_j) (\bar{q}^k (1-\ga_5)\slashed{D}t^a_{kl} h_v^l)\,, 
\label{eq:dim7matelmesHQEoctet}
\end{align}
Finally, for a complete dimension-seven \ac{HQE} basis, one should include non-local operators, defined explicitly in \Eqns(2.66)-(2.69) of \cite{LenzNote:2021} (see also \Sec3.6 in \cite{Neubert1993} and \Eqns(30), (31) in \cite{KM1992}). However, as was shown in \cite{BlokShifman93Review,Neubert1993,KM1992} (see also \cite{LenzNote:2021} and \App\ref{app:HQEtoQCD}), in meson decay widths the effect of the non-local dimension-seven matrix elements can be re-absorbed into the dimension-six matrix elements, up to higher-order corrections in $1/m_c$ and $\als$. For our purposes, therefore, we do not quote the definitions of the non-local elements.

For baryons, the equivalent dimension-six basis is 
\begin{alignat}{3}
\Opsix{1}{q} &= (\bar{c}_i \ga_\mu(1-\ga_5) q_i) (\bar{q}_j \ga^\mu(1-\ga_5) c_j) \,, \quad
& \Opsix{2}{q} &= (\bar{c}_i (1-\ga_5) q_i) (\bar{q}_j (1+\ga_5) c_j) \,,\nonumber \\
\Opsixt{1}{q} &= (\bar{c}_i \ga_\mu(1-\ga_5) q_j) (\bar{q}_j \ga^\mu(1-\ga_5) c_i) \;,\quad
&\Opsixt{2}{q} &= (\bar{c}_i (1-\ga_5) q_j) (\bar{q}_j (1+\ga_5) c_i) \,, \label{eq:Dim6BaryonBasis}
\end{alignat}
which is related to the meson basis, at the operator level, owing to the colour identity
\begin{equation}
 t^a_{ij}t^a_{kl}  = \frac{1}{2}\de_{il}\de_{jk} - \frac{1}{2\NC}\de_{ij}\de_{kl} \, , \label{eq:colourIdentity}
\end{equation}
which implies
\begin{equation}
\OpsixT{1,2}{q} = \frac{1}{2} \Opsixt{1,2}{q} - \frac{1}{2\NC} \Opsix{1,2}{q} \,, 
\label{eq:colourid}
\end{equation}
and similarly for all colour-octet operators. The reason for choosing a different basis for baryons is that, in this case, the totally antisymmetric colour wavefunction imposes the relation between the matrix elements
\begin{equation}
\langle\mathcal{B}_c|\Opsixt{i}{q}|\mathcal{B}_c\rangle = - \tilde{B} \langle\mathcal{B}_c| \Opsix{i}{q} |\mathcal{B}_c\rangle \,, \label{eq:BaryonMERelation1}
\end{equation}
where $\mathcal{B}_c$ denotes a singly charmed baryon, and $\tilde{B} \simeq 1$, with equality in the valence quark approximation \cite{NS1996}. 
\begin{table}
\setlength{\tabcolsep}{2pt}
\small
    \centering
    \begin{tabular}{|c|c|c|c|c||c|c|}
    \hline  
 \multirow{2}{*}{\diagbox[width=5em,height=4em]{$\phantom{H_c}H_c$}{decay}}  & CE NL & SCS-s NL & SCS-d NL & DCS NL & CE SL & CS SL \\
  &$c \to s \bar{d}u$& $c \to s \bar{s}u$ &$c \to d \bar{d}u$&$c \to d \bar{s}u$&$c\to s \bar{l}\nu_l$ &$c \to d\bar{l}\nu_l$\\ 
        \hhline{|=|=|=|=|=||=|=|}
     $D^0(\bar{u}c)$ & $\tGamma_{\text{WE}}$& $\tGamma_{\text{WE}}$&$\tGamma_{\text{WE}}$&$\tGamma_{\text{WE}}$ &-&- 
      \\
      $D^+(\bar{d}c)$&$\tGamma_{\text{PI}}$&-& $\tGamma_{\text{PI}} +\tGamma_{\text{WA}}$&$\tGamma_{\text{WA}}$&-& $\tGamma_{\text{WA}}^{\text{SL}}$ 
      \\ 
      $D_s^+(\bar{s}c)$&$\tGamma_{\text{WA}}$&$\tGamma_{\text{PI}} +\tGamma_{\text{WA}}$&-&$\tGamma_{\text{PI}}$&  $\tGamma_{\text{WA}}^{\text{SL}}$& -
      \\
       \hhline{|=|=|=|=|=||=|=|}
      $\Lac(udc)$&$\tGamma_{\exc}+\tGamma_{\intm}$& $\tGamma_{\intm}$&$\tGamma_{\exc}+\tGamma_{\intm} + \tGamma_{\intp}$&$\tGamma_{\intm} + \tGamma_{\intp}$&-& $\tGamma_{\intp}^{\text{SL}}$ 
      \\
      $\Xi_c^+(usc)$&$\tGamma_{\intm} + \tGamma_{\intp}$& $\tGamma_{\exc}+\tGamma_{\intm} + \tGamma_{\intp}$&  $\tGamma_{\intm}$ &$\tGamma_{\exc} + \tGamma_{\intm} $  
      & $\tGamma_{\intp}^{\text{SL}}$ & -
      \\
      $\Xi_c^0(dsc)$&$\tGamma_{\exc}+\tGamma_{\intp}$& $\tGamma_{\exc}+\tGamma_{\intp}$&$\tGamma_{\exc}+ \tGamma_{\intp}$&$\tGamma_{\exc}+\Gamma_{\intp}$
      & $\tGamma_{\intp}^{\text{SL}}$ & $\tGamma_{\intp}^{\text{SL}}$
      \\
      $\Omega_c^0(ssc)$&$\tGamma_{\intp}$&$\tGamma_{\exc}+ \tGamma_{\intp}$&-&$\tGamma_{\exc}$&
      $\tGamma_{\intp}^{\text{SL}}$&-\\
      \hline
\end{tabular}
\caption{\small Synopsis of various contributions of the four-quark operators to the lifetimes of singly charmed hadrons in the valence approximation,
whereby the hierarchies stemming from various CKM scalings are denoted by CE (Cabibbo enhanced), SCS (singly Cabibbo suppressed), DCS (doubly Cabibbo suppressed). As before, NL and SL respectively denote nonleptonic and semileptonic contributions. The decay processes in the first row schematically denote the kinds of weak transitions appearing in the diagrams in \Fig\ref{fig:1-DiagramsMesons}, and are to be rearranged according to the quark involved in the decay.}
\label{tab:FourQuarkContribs}
\end{table}%
The dimension-seven basis for baryons is
\begin{align}
\OpsevenP{1}{q} &=  m_q (\bar{c}_i (1-\ga_5) q_i) (\bar{q}_j (1-\ga_5) c_j) \,, \nonumber \\ 
\OpsevenP{2}{q} &= \frac{1}{\mQheavy}(\bar{c}_i \Dleft_\rho \ga_\mu(1-\ga_5) D^\rho q_i) (\bar{q}_j  \ga^\mu(1-\ga_5) c_j) \,, \nonumber \\
\OpsevenP{3}{q} &= \frac{1}{\mQheavy}(\bar{c}_i \Dleft_\rho (1-\ga_5) D^\rho q_i) (\bar{q}_j  (1+\ga_5) c_j) \,, \nonumber \\
\tilde{P}_1^q &=  m_q (\bar{c}_i (1-\ga_5) q_j) (\bar{q}_j (1-\ga_5) c_i) \,, \nonumber \\
\tilde{P}_{2}^{q} &= \frac{1}{\mQheavy}(\bar{c}_i \Dleft_\rho \ga_\mu(1-\ga_5) D^\rho q_j) (\bar{q}_j  \ga^\mu(1-\ga_5) c_i) \,, \nonumber \\
\tilde{P}_{3}^{q} &= \frac{1}{\mQheavy}(\bar{c}_i \Dleft_\rho (1-\ga_5) D^\rho q_j) (\bar{q}_j  (1+\ga_5) c_i) \,, \label{eq:Dim-7Baryon}
\end{align}
which is again related to the meson basis \eqref{eq:dim7matelmes} by the colour identity \eqref{eq:colourid}. As before, to obtain a consistent $1/m_c$ expansion one should use a basis with the heavy-quark field, equivalent to that in \Eqns\eqref{eq:Dim6MesonBasisHQE} and \eqref{eq:dim7matelmesHQE}, supplemented by the non-local operators. However, there is no currently-available parametrization for the non-local baryon matrix elements. Furthermore, since the re-absorption of the non-local dimension-seven contributions, following the arguments of \cite{Neubert1993,KM1992,BlokShifman93Review,LenzNote:2021}, cannot be readily applied to baryons, we prefer to use QCD matrix elements throughout in the analysis of baryon lifetimes.

Having defined the operator basis, we are now in a position to quote expressions for the spectator contributions to decay width. At leading order, the expressions for mesons have been computed in \cite{SV1985,SV1986,GRT1986,Voloshin96,NS1996,LenzRauh2013,Cheng18c,LenzNote:2021}, and are provided in \App\ref{app:c356}. The contributions to the lifetimes of specific charmed mesons are then obtained from the general expressions in \Eqn\eqref{eq:explicitDim6Mes} by inserting the appropriate \ac{CKM} factors according to \Fig\ref{fig:1-DiagramsMesons}, and evaluation of the appropriate matrix elements. In practice, several simplifications could be invoked. In the valence approximation, only those contributions in which the light quark $q'$ within the operator coincides with the meson's valence quark $q$ are included. Within this approximation, the contributions to the decay widths of $D$ mesons from the four-quark operators at dimension-six are
\begin{align}
     \tilde{\Ga}_{6, D^0}&=|V_{cs}|^2 |V_{ud}|^2\langle \widehat{\Ga}^u_{\text{6,WE}}(x_s,0)\rangle_{D^0}+| V_{cs}|^2 | V_{us}|^2\langle \widehat{\Ga}^u_{\text{6,WE}}(x_s,x_s)\rangle_{D_0}\nonumber\\
     &{}+| V_{cd}|^2 | V_{ud}|^2\langle \widehat{\Ga}^u_{\text{6,WE}}(0,0)\rangle_{D^0}\,,\nonumber\\[8pt]
     \tilde{\Ga}_{6, D^+}&=| V_{cs}|^2 | V_{ud}|^2\langle \widehat{\Ga}^d_{6, \text{PI}}(x_s,0)\rangle_{D_+}+| V_{cd}|^2 | V_{ud}|^2\langle \widehat{\Ga}^d_{\text{6,WA}}(0,0)\rangle_{D^+}\nonumber\\
     &{}+| V_{cd}|^2 | V_{ud}|^2\langle \widehat{\Ga}^d_{6,\text{PI}}(0,0)\rangle_{D^+} +| V_{cd}|^2\sum\limits_{\ell=e,\mu}\langle\widehat{\Ga}^{d,\text{SL}}_{6,\text{WA}}(x_\ell,0)\rangle_{D^+}\,,\nonumber\\[5pt]
      \tilde{\Ga}_{6, D_s}&=| V_{cs}|^2 | V_{ud}|^2\langle \widehat{\Ga}^s_{6, \text{WA}}(0,0)\rangle_{D_s}+| V_{cs}|^2 | V_{us}|^2\langle \widehat{\Ga}^s_{6, \text{WA}}(x_s,0)\rangle_{D_s}\nonumber\\
     &{}+| V_{cs}|^2 | V_{us}|^2\langle \widehat{\Ga}^s_{6, \text{PI}}(x_s,0)\rangle_{D_s}\,+| V_{cs}|^2\sum\limits_{\ell=e,\mu}\langle\widehat{\Ga}^{s,\text{SL}}_{6,\text{WA}}(x_\ell,0)\rangle_{D_s} \,,
     \label{eq:mesonDim6contribs}
\end{align}
where $\langle\widehat{\Ga}^q(x_1,x_2)\rangle_{M_q}$ is a shorthand for $\langle M_q|\widehat{\Ga}^q(x_1,x_2)|M_q \rangle$, and we have neglected the doubly Cabibbo-suppressed terms. The contributions for dimension-seven, using the analytic expressions in \Eqn\eqref{eq:explicitDim7Mes}, are exactly analogous. To include non-valence contributions, we insert terms arising from $\langle\widehat{\Ga}^{q'}(x_1,x_2)\rangle_{M_q}$, where $q'\neq q$, which would for example generate \ac{WA} and \ac{PI} contributions to $\tilde{\Ga}_{6, D^0}$. The resulting modification to \eqref{eq:mesonDim6contribs} is systematic, since, by including non-valence terms, all possible topologies contribute to any given meson \cite{LenzNote:2021,KLR2021}. The expressions above reflect the clear hierarchy of contributions, presented in \Tab\ref{tab:FourQuarkContribs}: for example, the $D^+$ width receives a large \ac{PI} contribution and a small, Cabibbo-suppressed, semileptonic contribution, whereas the $D_s$ receives a smaller, Cabibbo-suppressed \ac{PI} contribution alongside the semileptonic and \ac{WA} contributions.

\pagebreak
Considering only the valence contributions, and neglecting doubly Cabibbo-suppressed terms,  the analogous expressions to \eqref{eq:mesonDim6contribs} for baryons are
\begin{align}
     \tilde{\Ga}_{6, \Lac}&=| V_{cs}|^2 | V_{ud}|^2 \left(\langle  \widehat{\Ga}^u_{6,\intm}(x_s,0)\rangle_{\Lac} + \langle \widehat{\Ga}^d_{6,\exc}(x_s,0)\rangle_{\Lac} \right)+| V_{cs}|^2 | V_{us}|^2\langle \widehat{\Ga}^u_{6,\intm}(x_s,x_s)\rangle_{\Lac} \nonumber \\ & {}
      + | V_{cd}|^2 | V_{ud}|^2\left( \langle \widehat{\Ga}^u_{6,\intm}(0,0)\rangle_{\Lac} +\langle \widehat{\Ga}^d_{6, \intp}(0,0)\rangle_{\Lac}+\langle \widehat{\Ga}^d_{6,\exc}(0,0)\rangle_{\Lac} \right) \nonumber\\
     & {} +| V_{cd}|^2\sum_{\ell=e,\mu}\langle \widehat{\Ga}^{d,\text{SL}}_{6,\intp}(x_\ell,0)\rangle_{\Lac} \,, \nonumber\\
     \tilde{\Ga}_{6, \Xicp}&=| V_{cs}|^2 | V_{ud}|^2 \left(\langle  \widehat{\Ga}^u_{6,\intm}(x_s,0)\rangle_{\Xicp} + \langle \widehat{\Ga}^s_{6,\intp}(0,0)\rangle_{\Xicp} \right) + | V_{cd}|^2 | V_{ud}|^2\langle \widehat{\Ga}^u_{6,\intm}(0,0)\rangle_{\Xicp} \nonumber \\ & {}+| V_{cs}|^2 | V_{us}|^2\left( \langle \widehat{\Ga}^s_{6,\exc}(x_s,0)\rangle_{\Xicp} + \langle \widehat{\Ga}^s_{6,\intp}(x_s,0)\rangle_{\Xicp}+ \langle \widehat{\Ga}^u_{6,\intm}(x_s,x_s)\rangle_{\Xicp}\right)
       \nonumber\\
     & {} +| V_{cs}|^2\sum_{\ell=e,\mu}\langle \widehat{\Ga}^{s,\text{SL}}_{6,\intp}(x_\ell,0)\rangle_{\Xicp} \,, \nonumber\\
     \tilde{\Ga}_{6, \Xico}&= | V_{cs}|^2 | V_{ud}|^2 \left( \langle \widehat{\Ga}^s_{6,\intp}(0,0)\rangle_{\Xico}+ \langle \widehat{\Ga}^d_{6,\exc}(x_s,0)\rangle_{\Xico} \right)
       \nonumber\\
       & {} + | V_{cd}|^2 | V_{ud}|^2 \left(\langle \widehat{\Ga}^d_{6,\intp}(0,0)\rangle_{\Xico} + \langle \widehat{\Ga}^d_{6,\exc}(0,0)\rangle_{\Xico}  \right) \nonumber \\
       & {} + | V_{cs}|^2 | V_{us}|^2 \left(\langle  \widehat{\Ga}^s_{6,\intp}(x_s,0)\rangle_{\Xico} + \langle \widehat{\Ga}^s_{6,\exc}(x_s,0)\rangle_{\Xico}  \right) \nonumber \\
     & {} +| V_{cs}|^2\sum_{\ell=e,\mu}\langle \widehat{\Ga}^{s,\text{SL}}_{6,\intp}(x_\ell,0)\rangle_{\Xico} +| V_{cd}|^2\sum_{\ell=e,\mu}\langle \widehat{\Ga}^{d,\text{SL}}_{6,\intp}(x_\ell,0)\rangle_{\Xico} \,, \nonumber\\
     \tilde{\Ga}_{6, \Omc}&= | V_{cs}|^2 | V_{ud}|^2 \langle \widehat{\Ga}^s_{6,\intp}(0,0)\rangle_{\Omc} +  | V_{cs}|^2 | V_{us}|^2 \left( \langle \widehat{\Ga}^s_{6,\intp}(x_s,0)\rangle_{\Omc} + \langle \widehat{\Ga}^s_{6,\exc}(x_s,0)\rangle_{\Omc}  \right)\nonumber \\
     & {} + | V_{cs}|^2\sum_{\ell=e,\mu}\langle \widehat{\Ga}^{s,\text{SL}}_{6,\intp}(x_\ell,0)\rangle_{\Omc} \,, 
     \label{eq:baryonDim6contribs}
\end{align}
and likewise for dimension-seven, where analytic forms at tree level for the various contributions are given in \Eqns\eqref{eq:explicitDim6Bar} and \eqref{eq:explicitDim7Bar}. Again, one could in principle insert non-valence contributions, resulting in longer expressions. Unlike in mesons, however, no reliable estimate of the matrix elements for non-valence contributions exist, and we therefore do not include these contributions in our analysis of the baryon lifetimes. Given that non-valence contributions in mesons are small, albeit with significant uncertainties \cite{KLR2021}, this is justified to the present degree of accuracy. As is the case with mesons, the contributions exhibit a clear hierarchy, also presented in \Tab\ref{tab:FourQuarkContribs}, although for baryons there are more relevant contributions, owing to the additional spectator quark, than there are for mesons. It is apparent from \eqref{eq:baryonDim6contribs} that four-quark operator contributions in semileptonic decays, which always accompany the nonleptonic $\Gamma_{\intp}$ contributions, are pronounced in all singly charmed baryon decays, except in $\Lac$. In particular, the semileptonic decay of $\Omc$ is expected to get a significant four-quark contribution due to the interference of two $s$ quarks \cite{Voloshin96}.\footnote{Although we do not include them in our analysis, penguin contributions in baryons accompany $\Gamma_{\intm}$, meaning that one could only observe such contributions in \Eqn\eqref{eq:baryonDim6contribs} in $\Lac$ and $\Xicp$.}
\begin{table}[ht]\small
\centering
\begin{tabular}{|c c ||c|c|c|c|}
\hline
& & $\als^0$ (LO)  & $\als^1$ (NLO) & $\als^2$ (NNLO) & $\als^3$ (N$^3$LO) \\ \hline
\multirow{2}{*}{$c_3$} & SL & & \cite{CJK1994} & (\cite{LSW1994,Ritbergen1999,PC2008I,PC2008II,BM2009}) & (\cite{FSS2020,CCD2021}) \\
                                           & NL & & \cite{HokimPham84,BBBG1994,BBFG1995,KLR2013} & (\cite{CST2005}) & \\ \hline
\multirow{2}{*}{$c_{\pi,G}$} & SL & \multirow{2}{*}{\cite{BUV1992HQE,BS1992I,BS1992II,BBSUV92HQE}} & \cite{AGN2013,MPR2014,MPR2015} & & \\
                                           & NL & & & & \\ \hline
\multirow{2}{*}{$c_\rho$} & SL & \cite{GK1996,MRS2017} &(\cite{MP2019,MMP2021II}) & & \\
                                           & NL & \cite{MMP2020,LPR2020,Moreno2020,LenzNote:2021} & & & \\ \hline
\multirow{2}{*}{$c_{6,i}^q$} & SL & \multirow{2}{*}{\cite{SV1985,SV1986,GRT1986,Voloshin96,NS1996}} &  \multirow{2}{*}{\cite{CFLM2001,BBGLN2002,FLMT2002,LenzRauh2013}} & & \\
                                           & NL & & & & \\ \hline
\multirow{2}{*}{$c_{7,i}^q$} & SL &\multirow{2}{*}{\cite{GOP2004,LenzRauh2013,Cheng18c,LenzNote:2021}} & & & \\
                                           & NL & & & & \\ \hline
\end{tabular}
\caption{\small Summary of main references for contributions to the heavy hadron decay rates, arranged by order in the $1/m_c$ and $\als$ expansions, with the upper three rows referring to non-spectator contributions (ie two-quark matrix elements), and the lower two rows referring to spectator contributions (ie four-quark matrix elements). The notation in the leftmost column corresponds to \Eqn\eqref{eq:DecayRateExpansion}. If a contribution is not available, the cell is left empty (the exception is the leading $\Ga_3$ term, which in effect has been known since the 1950s). Contributions that have not been included in our computations, but may be useful for future studies, are indicated in brackets (). More details, and other key references, are also provided in the main text. Some exploratory work on higher-order non-spectator contributions in semileptonic channels is available in, for example, \cite{DMT2006,MTU2010,GHT2016,FMV2018,FMV2019}, but since little is known about matrix elements at this order, we do not include such contributions in our analysis.}
\label{tab:pertcorr}
\end{table}

A summary of all the contributions considered in this paper, along with useful references, is available in \Tab\ref{tab:pertcorr}. 

\subsection{Charm quark mass}
\label{sec:charmmassscheme}
The major input parameter influencing the decay rate is the mass of the charm quark itself, $m_c$, since in the leading term it enters the expression for the decay width as $m_c^5$. Consequently, it is important to define this mass precisely. 

 Typically, the starting point for all analytic expressions derived earlier is to treat $m_c$ as the pole mass, which is defined as a pole in the complex $p^2$ plane of the heavy-quark propagator and can be perturbatively related to other renormalized masses order-by-order. The pole mass is suitable for processes with nearly on-shell heavy quarks, such as heavy hadron decays. However, the pole mass of a quark is not a physical parameter. Moreover, any perturbative definition of the pole mass suffers from a divergence due to infrared (IR) renormalons, which impose a minimal uncertainty of $\mathcal{O}(\La_{\rm QCD})$ (see, for example, \cite{Beneke1998Renormalon,Beneke2021} and references therein for a detailed discussion).\footnote{The authors of \cite{BSUV1994II} were the first to point out this important drawback in the concept of a pole mass, which becomes apparent as soon as one addresses leading non-perturbative corrections to order
$1/m_Q$.} This can be seen explicitly in the relation between the $\msbar$ and pole masses, which to third order reads \cite{CS2000,MR2000}
\begin{equation}
\begin{split}
m_{c}^{\rm pole}&=\overline{m}_c(\overline{m}_c)\bigg[1+\frac{4}{3}\frac{\als(\overline{m}_c)}{\pi}+ 10.3 \bigg(\frac{\als(\overline{m}_c)}{\pi}\bigg)^2+116.5 \bigg(\frac{\als(\overline{m}_c)}{\pi}\bigg)^3+\ldots\bigg]\\
&= \overline{m}_c(\overline{m}_c)(1+0.16+0.15+0.21+\ldots) \,,
\label{eq:RelationOS-MS}
\end{split}
\end{equation}
where we have used $\alpha_s(\MSbar{m}_c) = 0.38 $. Assuming that the value of $\overline{m}_c$ can be extracted without ambiguities from lattice QCD, as has been done in \cite{ETM2014I,ETM2014II,HPQCD2014I,MILC2018,HPQCD2018}, this demonstrates an instability in the determination of $m_c^{\rm pole}$ starting already from two-loop corrections.\footnote{As expected, heavier quark pole masses are less influenced by the problem of the non-convergence of the $\alpha_s$ series. Thus, the series for the $b$-quark pole mass starts to diverge from fourth order, and for the top quark mass at the eighth order in the expansion \cite{Beneke2021} 
(note that there are some numerical inconsistencies in tables 2 and 3 therein, confirmed by the author in a private communication).}


To circumvent this issue, various alternative renormalon-free mass definitions have been proposed. A common alternative definition for the heavy quark mass is the above-mentioned $\msbar$ mass ($m^{\rm \msbar} \equiv \overline{m}$), which is the short-distance, perturbatively-defined mass appearing as a parameter in the Lagrangian. Being the running mass defined at the scale $\mu$, it includes only effects of momenta higher than $\mu$ \cite{Beneke2021}. The scale at which the $\msbar$ mass is evaluated is usually taken to be the mass itself, $\overline{m}_c(\mu_m) = \overline{m}_c(\overline{m}_c)$, the default value which is used throughout this paper. 
Although  $\msbar$ is a good scheme for quantities that involve energies much larger than $m$, it is not an appropriate choice for physical processes such as heavy quark (practically on-shell) decays, since it is a strictly defined short-distance object and is therefore quite far away from the on-shell pole mass by $O(\overline{m}\alpha_s)$. On the other hand, it was shown in \cite{BSUV1996} that the characteristic normalization scale for the mass in inclusive heavy quark decays is $m_Q/5$. Whereas, for the $b$-quark decay, this becomes $m_b/5 \sim 1 \GeV$, for inclusive $c$-quark decays, the equivalent scale $m_c/5 \sim 0.3\GeV$ is nonperturbative. 

There are several other renormalon-free short-distance masses, $m_{c}^{X}(\mu_f)$ (also called low-scale short-distance masses, since $\mu_f \ll m$), each constructed in a way to be more appropriate for a particular process. Here, $X$ labels the given scheme. All of them use a perturbative relation to the $\msbar$ mass and are defined such that the leading IR renormalon divergence is subtracted by construction \cite{Beneke2021}:
\begin{eqnarray}
m_{c}^X(\mu_f) &=& m_c^{\rm pole}  - \delta m_{c}^{X}(\mu_f)
\nonumber \\
&=& \overline{m}_c(\overline{m}_c) + \overline{m}_c(\overline{m}_c)    \sum_{n=1}^{\infty} \left [  c_n(\mu,\overline{m}_c(\overline{m}_c))   - \frac{\mu_f}{\overline{m}_c(\overline{m}_c) } s_{n}^X (\mu/\mu_f) \right ]  \alpha_s^n(\mu) \,,
\nonumber \\
\label{eq:mCX}
\end{eqnarray}
where both coefficients diverge as $c_n, s_n^X \sim ( -2 \beta_0)^n n! n^a$; $\be_0$ is the leading coefficient of the QCD $\be$-function. The $s_N^X$ are constructed so that the leading IR renormalon divergence cancels in the bracket. This introduces a new, arbitrary, mass scale $\mu_f$
in the range $\Lambda_{QCD} \ll \mu_f \ll m$. To have a perturbative expansion the scale $\mu_f$ has to be somewhat large, but on the other hand, it must also be sufficiently small so as not to run into the same problem faced by the $\msbar$ mass in the description of the heavy hadron decays, discussed above. In practice, we should construct renormalon-free masses which are numerically close to the pole mass. Since the difference between the pole mass and an arbitrary renormalon-free mass is 
$\delta m^X \sim \mu_f \alpha_s(\mu_f)$, the arbitrary parameter $\mu_f$ is typically chosen to minimize this relation and to ensure the convergence of the expansion (\ref{eq:mCX}).\footnote{For the $\msbar$ mass, $\mu_f = \overline{m}$.}

Theoretically, the most appropriate mass in semileptonic heavy hadron decays is the kinetic mass \cite{BSUV1994,BSUV1996}, defined from the relation between the heavy quark and heavy meson masses
\begin{eqnarray}
m_Q(\mu) = \overline{M}_H - \overline{\Lambda}(\mu) - \frac{\mu_{\pi}^2(\mu)}{2 m_Q(\mu)} +...
\label{eq:mM_mass1}
\end{eqnarray}
where $\overline{M}_H= (M_H + 3 M_{H^{\ast}})/4$, the  spin-averaged mass of $H$ and $H^\ast$ mesons, is introduced to cancel $\mu_G$ dependence that would otherwise enter (\ref{eq:mM_mass1}). The short-distance, renormalon-free, mass $m_c^{\rm kin}$ is then defined by perturbative loop-momentum calculations below the scale $\mu_f$:
\begin{eqnarray}
m_c^{\rm kin}(\mu_f) = m_c^{\rm pole} - \overline{\Lambda}(\mu_f)_{\rm pert} - \frac{\mu_{\pi}^2(\mu_f)_{\rm pert}}{2 m_c^{\rm kin}(\mu_f)} + \dots \,,
\label{eq:mM_mass2}
\end{eqnarray}
and in such a way the leading IR renormalon of the pole mass is subtracted order-by-order in $\alpha_s$ and $\mu_f/m_Q$.
The kinetic mass scheme was defined for the description of the inclusive $b$-quark decays and therefore, again, it might not be fully appropriate for the $c$-quark decays, although there is a good convergence of the perturbative series $m^{\rm kin} (\overline{m})$ for both heavy quark masses \cite{FSS2020I,FSS2020II}.  

In addition, we will consider the MSR mass scheme \cite{HJS2008,HJLMPSS2017}, 
 which uses $\mu_f$ to interpolate between the $\msbar$ mass ($\mu_f = \overline{m}$) and the pole mass ($\mu_f = 0)$, and avoids the drawback (see discussion in \cite{HJLMPSS2017}, also in \cite{Beneke2021}) of the  Renormalon-Subtracted (RS) mass scheme, the simplest renormalon-free mass scheme \cite{Pineda2001}. There are two versions of the MSR mass, the ``natural'' and ``practical'', as defined in \cite{HJLMPSS2017}, which differ conceptually, but are numerically close; we will use the practical definition. 
Its simplicity is in the chosen scales, such that $\mu = \mu_f = \overline{m}$, and therefore the subtraction coefficients $s_n^X$ in \eqref{eq:mCX} are simply the same coefficients of the pole-$\msbar$ mass conversion,  
\begin{eqnarray}
{s_n^{\rm MSR}}_{|\, \mu = \mu_f} = c_n( \overline{m}_c, \overline{m}_c(\overline{m}_c)) \,.
\end{eqnarray}
\\
\subsubsection{Application of mass schemes in inclusive decays}
\label{sec:massschemesexplicit}

Since the analytic results of \Sec\ref{sec:contgeneral} are expressed initially in terms of the pole mass, they must then be rearranged to match the new definition. The mass in the new scheme is then treated as an input parameter, determined as precisely as possible. 
To leading order in $\alpha_s$, this has little consequence beyond changing the value of the mass, but when including \ac{NLO} contributions, one must also take into account the $\alpha_s$ relation between $m_c^{\text{pole}}$ and $m_{c}^{X}$ for a given choice of scheme. Specifically, supposing that a mass scheme at NLO can be related to the pole scheme as
\begin{equation}
    m_c^{\text{pole}} = m_c^X (1 + \frac{\als}{\pi}a_X^{(1)} ) \,,
\end{equation}
then the leading dimension-three term can be rearranged as
\begin{equation}
    \Ga_3^{X} = \frac{G_F^2}{192\pi^3}(m_c^X)^5\left(c_3^{(0)}+\left(c_3^{(1)}+5\, a_X^{(1)} \right)\frac{\als}{\pi} \right) \, ,
\end{equation}
where $X$ denotes the scheme of interest, and similar relations hold for all other terms in the decay width. 

We now spell out the details of our usage of different mass schemes for the charm quark mass. 
For definiteness, we consider the dimension-three contribution with the Wilson coefficient $c_3$ evaluated to the first order in $\als$. 
\begin{enumerate}

\item \emph{Pole scheme}. For the presentation of the results in the pole scheme for the charm quark mass we adopt the one-loop result for the pole mass
\begin{equation}
    m_c^{\text{pole}}= 1.49\GeV\,.
\end{equation}

\item \emph{$\msbar$ scheme}. The expression,
\begin{equation}
\Gamma_3^{\text{pole}}=\frac{G_F^2}{192\pi^3}m_{c\,\text{pole}}^5\big(c_3^{(0)}+c_3^{(1)}\als(\mu)+\ldots\big)\,,
\end{equation}
is translated to the $\msbar$ scheme using the relation between the masses to the same order in $\als$,
\begin{equation}
m_{c}^{\text{pole}}=\overline{m}_c(\mu_m)\bigg[1+\frac{\als(\mu_m)}{\pi}\bigg(\frac{4}{3}+\log \frac{\mu_m^2}{\overline{m}_c(\mu_m)^2}\bigg)+\ldots\bigg]\,.\label{eq:PoleMS-relation}
\end{equation}
Our strategy is to set $\mu_m = \overline{m}_c$ as already discussed at the beginning of this section, and to expand $\alpha_s$ about $\mu = \overline{m}_c$. Expanding the fifth power of the mass to first order in $\als$ we have
\begin{equation}
  (m_{c}^{\,\text{pole}})^5=\overline{m}_c(\mu_m)^5\bigg[1+5\,\frac{\als(\mu)}{\pi}\bigg(\frac{4}{3}\bigg)\bigg]\,,
  \label{eq:PoleMS5-relation}
\end{equation}
resulting in 
\begin{equation}
    \Gamma_3^{\msbar}=\frac{G_F^2}{192\pi^3}\overline{m}_c^5\Bigg[c_3^{(0)}\ + \, \frac{\als(\mu)}{\pi}\Bigg(5\cdot \frac{4}{3}+c_3^{(1)}\Bigg)+\ldots  \Bigg] \,.
    \label{eq:Gamma3MSbar}
\end{equation}
For any individual contribution $c_i$ corresponding to a given order in $m_c^{-1}$, evaluated to a specific order in $\als$, we use the relation \eqref{eq:PoleMS-relation} to this same order of $\als$. Hence, for the coefficients $c_i$ known only to leading order in $\als$, we simply perform the replacement $m_{c}^{\text{pole}}\to\overline{m}_c(\overline{m}_c)$.
The numerical value of the $\msbar$ mass is \cite{FLAG2019,FLAG2021}
\begin{equation}
\overline{m}_c(\overline{m}_c) = 1.28 \GeV\,.\label{eq:mmMSbar}
\end{equation}

\item \emph{Kinetic scheme}.  
Expressing the results in terms of kinetic scheme proceeds in analogy to the above case of $\msbar$. Pole mass is expressed in terms of the kinetic mass with the $\mathcal{O}(\als)$ relation 
\begin{equation}
m_{c}^{\text{pole}} = m_c^{\text{kin}}(\mu_f)\left[1 + \frac{\als(\mu)}{\pi} \left(\frac{16}{9}\frac{\mu_f}{m_c^{\text{kin}}} + \frac{2}{3}\frac{\mu_f^2}{{m_c^{\text{kin}}}^2} \right) \right] \,.\label{eq:PoletoKin}
\end{equation}
The relation between the kinetic and $\msbar$ mass is known to three loops \cite{FSS2020I,FSS2020II}. We evaluate the numerical three-loop value for $m_c^{\text{kin}}$ at $\mu_f=0.5\GeV$ using \verb+RunDec+ \cite{RunDec, RunDecv3},
\begin{equation}
    m_c^{\text{kin}}=1.40\GeV\,,\label{eq:Kin3loops}
\end{equation}
with the input $\MSbar{m}_c(\MSbar{m}_c)=1.28\GeV$, see \Tab\ref{Tab:cMasses}. 
The scale $\mu_f = 0.5$ GeV is chosen to be as close to the expected value of $\mu_f = m_c/5$, without going below the scale $\Lambda_{QCD}$ \cite{BSUV1996}. 

\item \emph{MSR scheme}.  
For the MSR scheme, we have \cite{HJLMPSS2017}
\begin{equation}
  m_{c}^{\textrm{pole}}= m_{c}^{\text{MSR}}(\mu_f)\bigg(1+\frac{4}{3}\frac{\als(\mu)}{\pi} \frac{\mu_f}{m_{c}^{\text{MSR}}(\mu_f)}+\ldots\bigg).
  \label{eq:OS-MSRMass}
\end{equation}

For the numerical value of the MSR mass we use the four-loop result \cite{Beneke2021}
\begin{equation}
    m_c^{\text{MSR}}=1.36\GeV\,.\label{eq:MSR4loops}
\end{equation}
at the scale $\mu_f = 1$ GeV. 

\end{enumerate}
In \Tab\ref{Tab:cMasses} we illustrate the behaviour of higher-order $\als$ contributions for the charm mass schemes used in the paper.\footnote{Another, physically-motivated, mass definition is  the $1S$ scheme, where the mass of the $c$ ($b$) quark is extracted from the $J/\psi$ ($\Upsilon$) resonance \cite{HLM1999I,HLM1999II,BS1999}. This has, however, received criticisms in its applicability to 
heavy-light systems due to an inconsistent $\als$ expansion \cite{Uraltsev2004,Bigi2006}, and also a large non-perturbative correction from the gluon condensate, calculated in \cite{BS1999} for the $m_b^{\Upsilon}$ mass, which could be even more significant in the charm sector. Therefore, the $1S$ mass should  be seen as unsuitable for such $D$ (or $B$) decays, despite its use in recent papers on charm lifetimes \cite{LenzNote:2021}. However, it could still be relevant to inclusive decays of the  $B_c$ meson, which is more 'quarkonium-like' \cite{AG2021}. }
\begin{table}[ht]
	\centering
\begin{tabular}{|c||c|c|c|c|}
 \hline
$\overline{m}_c(\overline{m}_c)$ = $1.28$ GeV & 1-loop & 2-loop & 3-loop & 4-loop \\
 \hline
$ m_c^{\rm pole}$& $1.49 $ &  $1.68$ & $1.95$ & $2.43$   \\
\hline
 $m_c^{\rm kin} $& $1.36$ &  $1.39$ & $1.40$ & -\\
 \hline
$ m_c^{\rm MSR}$ & $1.33$ &  $1.35$ & $1.36$ & $1.36$\\
   \hline
\end{tabular}
\cprotect\caption{\small Illustration of convergences of the charm quark masses for different mass schemes, expressed in units of $\GeV$. In all cases the masses are expanded around the $\msbar$ mass $\overline{m}_c(\overline{m}_c)=1.28\GeV$ \cite{FLAG2019,FLAG2021}. The scale $\mu_f$, defined in \eqref{eq:mCX}, is $\mu_f=0.5\GeV$ for the kinetic mass, and $\mu_f=1\GeV$ for the  MSR mass, with the values taken from \cite{Beneke2021}. The numerical values for $m_c^{\text{pole}}$ and $m_c^{\text{kin}}$ are evaluated using the Mathematica package \verb+RunDec+ \cite{RunDec,RunDecv3}.}
\label{Tab:cMasses}
\end{table}

For the light masses, which enter phase space mass corrections as $(m_q/m_c)^n$, we will use their $\msbar$ masses, since these corrections are relatively small and change of mass schemes there bring numerically no difference to the results. We neglect the $m_u,m_d$ masses, and for the strange quark mass we use $\overline{m}_s(\mu = 1.5 \GeV)=0.1 \GeV$.

Finally, it is important to emphasize that various renormalon-free mass schemes yield different, renormalon-free definitions of the $\overline{\Lambda}$ parameter that appear in the HQET expression of the heavy hadron masses (see also \Eqn\eqref{eq:mHexpand}):
\begin{eqnarray}
M_H &=& m_c^{\text{pole}} + \overline{\Lambda} \nonumber \, + \cdots\\
&=& \left [ m_c^{\text{pole}} - \delta m_c^X(\mu_f) \right  ] + \left [  \overline{\Lambda} + \delta m_c^X(\mu_f) \right  ] \, + \cdots \nonumber \\
&=& m_c^X(\mu_f) + \overline{\Lambda}^X(\mu_f) \, + \cdots
\end{eqnarray}
which then becomes a scale-dependent quantity. 
In our calculation, $\overline{\Lambda}$ parametrizes meson matrix elements of the four-quark operators in dimension-seven contributions, cf. \Sec\ref{sec:mesonmateldefs}. Since the estimation of these contributions is anyhow plagued by large uncertainties, we refrain from using different values of $\overline{\Lambda}^X$ when applying different mass schemes, but instead associate the $\overline{\Lambda}$ parameter with a large uncertainty, made explicit below in  \Eqn\eqref{eq:labarvals}. 
%
\subsubsection{Effect of different mass schemes on the $\alpha_s$ convergence of the results}
\label{sec:alsseriescomments}
We also wish to comment on the $\als$ series, which is (partially) known for semileptonic decays up to N$^3$LO in the leading $\Ga_3$ term \cite{FSS2020}. Numerically, one finds in the pole scheme\footnote{To reproduce these results, one needs to set $n_b (N_H) =1$, $n_c (N_C) = 0$, and $n_l (N_L)= 3$ in \cite{FSS2020} (\cite{PC2008II}), as well as $\delta=1$. Since the results in \cite{FSS2020} are presented as a series expansion in $\delta = 1- m_c/m_b$, the value of the $\als^3$ coefficient is not exact, although the behaviour of the series suggests that higher-order corrections are of order no more than a few \%. }
\begin{equation}
     \frac{c_3^{\text{pole}}(X_c \to X_d e \nu_e)}{|V_{cd}|^2\Ga_0} = \begin{cases} 1 - 0.25 - 0.26 - 0.37 \,, & m_c = 1.68 \GeV \\
     1 - 0.27 - 0.28 -0.43 \,, & m_c = 1.49 \GeV
     \end{cases} \,,
\end{equation}
where the respective corrections on the right-hand side are the first-, second-, and third-order $\als(m_c)$ contributions respectively. It can be seen that the series is divergent, with the divergent behaviour already emerging at second order. This again reflects the unsuitability of the pole mass, and shows that care must be taken to define an appropriate scheme. Likewise, in the $\msbar$ scheme (at $\MSbar{m}_c(\MSbar{m}_c) = 1.28 \GeV$), we obtain
\begin{equation}
     \frac{c_3^{\msbar}(X_c \to X_d e \nu_e)}{|V_{cd}|^2\Ga_0} = 1 + 0.52 + 0.44 + 0.45 \,,  
     \end{equation}
which also shows signs of divergence, and represents only a mild improvement on the pole scheme result. Note that the definition of $\Ga_0$ here is adapted to the $\msbar$ scheme.

As a further example, in the  kinetic scheme, the behaviour of the perturbative series is heavily dependent on the choice of the scale $\mu_f$. For illustration, we present the behaviour of the series for values of $\mu_f = (0.3,0.5,1) \GeV$, with $\als$ evaluated at the scale $m_c^{\text{kin}}$ (with $\Ga_0$ adapted to the kinetic scheme):
\begin{equation}
    \frac{c_3^{\text{kin}}(X_c \to X_d e \nu_e)}{|V_{cd}|^2\Ga_0} = \begin{cases} 1 - 0.06 + 0.00 + 0.06 \,, \quad \mu_f = 0.3 \GeV \\
    1 + 0.13 + 0.19 + 0.24 \,, \quad \mu_f = 0.5 \GeV \\
    1 + 1.08 + 1.26 + 1.30 \,, \quad \mu_f = 1 \GeV
    \end{cases}\,,
\end{equation}
where the perturbative nature of the series is particularly poor for larger values of $\mu_f$. On the other hand, the excellent behaviour of the $\als$ series for $\mu_f = 0.3 \GeV$ is artificial, since this does not reflect the limiting value of $\mu_f \gg \La_{QCD}$ \cite{FaelComms}.\footnote{It should be noted that the value of $c_3$ in the kinetic scheme is quite stable at N$^3$LO, with little dependence on the value of the mass, and so on the value of $\mu_f$. However, this is likely to be an accident, and cannot be expected to hold at higher orders in the $\als$ series.} However, merely improving the perturbative behaviour of $c_3$ is not sufficient to motivate a mass scheme, as the entire \ac{HQE} should be considered: to some extent, the renormalon divergence in $c_3$ motivates the presence of the non-perturbative contributions from $\mukin$, $\muG$, $\darwin$ etc \cite{BSUV1994II}. One could expect similarly divergent behaviour in all the coefficients that have yet to be computed, such as $c_{G,\rho}$ and the coefficients of four-quark operators. It is therefore important to consider the series as a whole, rather than the behaviour of a single coefficient in the \ac{HQE}.\footnote{The point can be illustrated by considering the behaviour of the dimension-eight contribution: since the leading $m_c$ dependence cancels here, owing to $\Ga_8 = \Ga_0 c_8 O_8/m_c^5 \sim m_c^5 c_8 O_8/m_c^5 = c_8 O_8$, every mass scheme is equivalent in this contribution, and any divergent behaviour therefore cannot be addressed separately by a given mass scheme.} 

In our work, we circumvent these problems by neglecting the available higher-order contributions in the semileptonic $c_3$ coefficient, preferring to treat the $\als$ expansion as asymptotic and keeping only the NLO contribution.


\section{Matrix elements for the non-spectator parameters}
\label{sec:twoquarkmatel}
The three inputs relevant to the non-spectator series are the kinetic, chromomagnetic and Darwin parameters $\mukin$, $\muG$, and $\darwin$, arising from the matrix elements defined in \Eqn\eqref{eq:nonspecdefns}. The first two can be defined in terms of the heavy-quark expansion of the hadron mass \cite{FN92I,FN92II,Neubert1996},
\begin{equation}
m_H = m_c + \bar{\La} + \frac{\mukin(H)}{2m_c} - \frac{\muG(H)}{2m_c} + \mathcal{O}\left(\frac{1}{m_c^2}\right) \,.
\label{eq:mHexpand}
\end{equation}
where $\bar{\La} \sim 0.5 \GeV$ for charmed hadrons, and all parameters in the expansion are formally independent of the heavy quark mass. 

The $\darwin$ parameter enters the series \eqref{eq:mHexpand} at the next order in the $1/m_c$ expansion. However, it does so alongside other, non-local, parameters, and cannot be disentangled from them. As discussed in \Sec\ref{sec:darwin}, we will instead extract $\darwin$ by applying the equation of motion, and so do not give the explicit dependence of $m_H$ on $\darwin$ in \eqref{eq:mHexpand}. 
\subsection{Chromomagnetic parameter $\muG(H)$}
\label{sec:muG}
By also considering resonances $H^{*}$, it is possible to extract the value of $\muG$ for a given hadron, with the result
\begin{equation}
\label{eq:la2def1}
\muG(H) = d_H 2 m_c \frac{m_{H^*} - m_H}{d_H - d_{H^*}}\,, 
\end{equation}
where $d_H$ denotes the spin factor
\begin{equation}
\label{eq:dHdef}
d_H = -2 \left(S_H (S_H+1) - S_h (S_h +1 ) - S_l(S_l+1)\right) \,,
\end{equation}
which follows from the fact that the chromomagnetic operator \label{eq:muGdef} describes the interaction of the spin between the heavy quark and light quark(s) in the hadron. $S_H$ is the spin of the heavy hadron, $S_h$ that of the heavy quark, and $S_l$ that of the light quark system. $H^{*}$ is an excited state with analogous definition of $d_{H^*}$ to \eqref{eq:dHdef}. In mesons, $d_H= 3$ and $d_{H^*}=-1$, while in baryons, only $d_{\Omc}$ is non-zero, with $d_{\Omc}=4$ and $d_{{\Omc}^*}=-2$ (see also \Tab\ref{tab:dH}). Here, it has been assumed that the other parameters in the expansion \eqref{eq:mHexpand}, $\bar{\La}$ and $\mukin$, are identical for a given hadron $H$ and the excited state $H^{*}$. Formally, $\muG$ and all other parameters in \eqref{eq:mHexpand} are independent of the heavy quark mass, and to reflect this it is common to apply the relation $2 m_c \approx m_{H^{*}} + m_H$, so that
\begin{equation}
\label{eq:la2def2}
\muG(H) \equiv d_H  \la_2 = d_H \frac{m_{H^{*}}^2 - m_H^2}{d_H - d_{H^{*}}} \,.
\end{equation}
\begin{table}[ht]
    \centering
    \begin{tabular}{|c||c|c|c|c|c|} \hline
     $H$    & $D$ & $D^*$ & $\Lac,\, \Xicp, \,\Xico$  & $\Omc$ &${\Omc}^*$ \\ \hline
    $S_H$     & $0$ & $1$ & $1/2$ & $1/2$  & $3/2$  \\
    $S_l$     & $1/2$ & $1/2$ & $0$ & $1$ & $1$  \\ \hline
    $d_H$ & 3 & $-1$ & 0  & 4 & $-2$ \\\hline
    \end{tabular}
    \caption{\small Values of spins and of the parameter $d_H$ for all relevant particles. In all cases $S_h = 1/2$, as this is merely the spin of a single quark, and is therefore not included in the table. Values of $S_H$, $S_l$ and $d_{H^*}$ for the triplet baryons ${\Lac}$, $\Xico$, and $\Xicp$ are not indicated, since $d_H$ is zero for all three of these baryons, making $\muG$ identically zero.    }
    \label{tab:dH}
\end{table}
The two determinations \eqref{eq:la2def1} and \eqref{eq:la2def2} differ by $1/m_Q$ 
corrections, which for charm quarks could be significant. We will use the relation \eqref{eq:la2def2} to determine values of $\muG$. Using the latest values for the masses in PDG \cite{PDG2020}, which are also quoted in \App\ref{app:parameterinputs}, we obtain the values in \Tab\ref{tab:nonpertNScharm}. Note that the chromomagnetic operator obtains a non-vanishing anomalous dimension, known to three loops \cite{Grozin:2007fh}. Since, however, we assume that $\mu_G^2$ for the charm hadrons is renormalized at the charm mass scale, the running of $\mu_G^2$ does not play any role. 

\subsection{Kinetic parameter $\mukin(H)$}
\label{sec:mukin}
Unlike $\muG(\Had)$, the value of $\mukin$ cannot be so easily extracted in terms of known hadron masses, and is therefore less well-determined than $\muG$. Because of this, various attempts have been made to evaluate $\mukin$ using sum rules, lattice QCD, or experimental fits, with several different values available throughout the literature, (see eg \Tab I of \cite{Neubert1997Btheory}, \Eqn(3.133) of \cite{Lenz2014}, and \Tab 10 from \cite{LenzNote:2021}). Some experimental fits, in the case of $B$ mesons, give $\mukin = 0.465(68)\GeV^2$ \cite{AGHN2014} and $\mukin = 0.432(68)\GeV^2$ \cite{GHT2016}.  For the charm sector, no such analyses have been performed. Hence, previous studies have tended to assume that $\mukin(D) = \mukin(B) = \mukin(\Lac)$, etc, as used in \cite{Cheng18c,LenzNote:2021}. In addition, there have been attempts to estimate $SU(3)_F$-breaking corrections, finding them to be no more than 20\% \cite{LenzRauh2013,BMU2011}. 

There does exist the theoretical lower bound, derived in \cite{Voloshinmukin,BSUV1994}, that
\begin{equation}
\label{eq:mukinlower}
\mukin \geq \muG \,,
\end{equation}
which in principle provides a constraint on its value.
 Some other estimates can be derived by imposing the heavy-quark symmetry relation $\mukin(D^{(*)}) = \mukin(B^{(*)})$, and repeated application of \eqref{eq:mHexpand}. In \cite{Neubert1997Btheory,BU1993,BUV1992HQE}, for example, $\mukin$ was related to the pole mass difference $m_b - m_c$ as
\begin{equation}
    m_b - m_c \simeq \left(\frac{m_B + 3m_{B^*}}{4} - \frac{m_D + 3m_{D^*}}{4} \right) + \mukin(D) \left(\frac{1}{2m_c } - \frac{1}{2m_b} \right) \,.
    \label{eq:Neubertmukin}
\end{equation}
In \cite{Neubert1997Btheory}, the further replacement on the right-hand side was made of $m_c \to (m_D + 3m_{D^*})/4$, and $m_b \to (m_B + 3m_{B^*})/4$, the spin-averaged masses. This is not the only possible choice, and one can also approximate $m_c \to m_D$, $m_b \to m_B$, which is valid to leading order in the heavy quark mass when taking the difference $m_b - m_c$. Since there is a further uncertainty in the value of the pole masses, we also replace $m_b - m_c \to m_B - m_D$ on the left-hand side. Regardless of the approach, one must accept uncertainties due to neglected higher-order terms in the $1/m_c$ expansion. With this replacement, and using the equation as originally presented in \cite{Neubert1997Btheory}, we find a central value of $\mukin(D) = 0.45 \GeV^2$. On the other hand, making the approximations $m_c \to m_D$, $m_b \to m_B$ throughout, we obtain a central value $\mukin(B) = 0.42 \GeV^2$. Remarkably, both are compatible with the experimental fits given above, although any estimate for $\mukin$ obtained in this manner is highly sensitive to the choice of how to express $m_b - m_c$ in terms of known quantities. 

In \cite{BUV1992HQE}, it was shown how to extract the difference $\mukin(\Lac) - \mukin(D)$, again assuming that $\mukin(D^{(*)}) = \mukin(B^{(*)})$ and $\mukin(\Lac) = \mukin(\Lab)$. Correcting a typo, the relationship reads
\begin{equation}
\begin{split}
\label{eq:BUVmukin}
    \left(m_D + 3m_{D^{*}} - 4 m_{\Lac} \right) - \left(m_B + 3m_{B^{*}} - 4 m_{\Lab} \right) \simeq{}&2 \left( \frac{1}{m_c} - \frac{1}{m_b} \right) \left( \mukin(D) - \mukin(\Lac) \right)\\ + &\mathcal{O}\left(\frac{1}{m_Q}\right) \,.
\end{split}
\end{equation}
 Using the latest particle mass values from PDG \cite{PDG2020}, we obtain
\begin{equation}
    \mukin(D) - \mukin(\Lac) \approx -0.05 \GeV^2 \,,
\end{equation}
which suggests that, to this order in $1/m_Q$, the kinetic parameter is approximately identical for baryons and mesons. This argument extends to the $\Xico$ and $\Xicp$, with appropriate replacement of masses. For the $\Omc$ and $\Omb$, the contribution from $\muG$ in this case does not cancel in the relation \eqref{eq:BUVmukin}, but one can use the combination
\begin{equation}
\label{eq:mukinspecOm}
    \mukin(\Omc)\left(\frac{1}{2m_b} - \frac{1}{2m_c} \right) = m_c - m_b + \frac{1}{3}\left( \left(m_{\Omb} + 2 m_{{\Omb}^{*}} \right) - \left(m_{\Omc} + 2 m_{{\Omc}^{*}} \right)\right) \,.
\end{equation}

In our numerical analysis, we will use the spectroscopic estimates from above, applying also the replacement
\begin{equation}
    \frac{1}{m_c} - \frac{1}{m_b} \to \frac{4}{m^{\vphantom{*}}_{D^{\vphantom{*}}_{(s)}} +3 m^{\vphantom{*}}_{D_{(s)}^*}} - \frac{4}{m^{\vphantom{*}}_{B^{\vphantom{*}}_{(s)}} +3 m^{\vphantom{*}}_{B_{(s)}^*}} \,.
\end{equation} 
The central values so obtained are compatible with previous studies and the experimental fits \cite{AGHN2014,GHT2016}, and are given in \Tab\ref{tab:nonpertNScharm}. These relations rely on the heavy-quark symmetry limit, are affected by $1/m_Q$ corrections, and are unlikely to capture all $SU(3)_F$-breaking effects. In particular, our central estimate of the $SU(3)_F$-breaking in mesons is
\begin{equation}
    \mukin(D_s)-\mukin(D^0) = 0.03 \GeV^2 \, ,
    \label{eq:SU3mukin}
\end{equation}
which can be compared with previous estimates of up to $ \mukin(D_s)-\mukin(D^0) = 0.10 \GeV^2$ from the literature, eg \cite{LenzRauh2013,BMU2011}. Our estimate of $SU(3)_F$-breaking should not therefore be taken as definitive, and in analysing observables sensitive to the difference \eqref{eq:SU3mukin}, the two parameters will be treated as uncorrelated. In light of these considerations, we assign a 30\% uncertainty to all central values for $\mukin$. A more precise determination of these parameters, either from the lattice or from experimental studies, in the spirit of \cite{AGHN2014,GHT2016} but applied to the charm sector, would serve to clarify these issues. 

\subsection{Darwin parameter $\darwin(H)$}
\label{sec:darwin}
Several attempts have been made to fit to experimental data to extract the Darwin parameter, again for $B$ mesons only, in \cite{GS2013,AGHN2014,GHT2016}. The resulting values fall within the approximate range $0.15\GeV^3 \leq \darwin(B_{(s)}) \leq 0.2\GeV^3$. 

An alternative approach \cite{BSUV1993II} is to relate the Darwin parameter to the matrix elements of the dimension-six spectator contributions by applying the equations of motion for a gluon field, $[D^\mu, G_{\mu\nu}] = - g_s\,t_a\, \bar{q}_f t_a\gamma_\nu q_f$, where all the repeated indices are summed over, including the index $f$ that denotes the light quark flavours, and the gluon field strength is $i g_s G^{\mu\nu}\equiv[iD^\mu,iD^\nu]$. In addition, applying the equation of motion $(i v\cdot D)h_v=0$, one obtains the leading order relation between the matrix elements
\begin{equation}
\begin{split}
    2m_H\rho_D^3&=\langle H| \bar{h}_v(iD_\mu)(iv\cdot D)(iD^\mu)h_v| H\rangle+\mathcal{O}(1/m_c)\\
    =&g_s^2\langle H| \big(-\frac{1}{9}\Opsixprime{1}{q}+\frac{2}{9}\Opsixprime{2}{q}+\frac{1}{12}\OpsixTprime{1}{q}-\frac{1}{6}\OpsixTprime{2}{q}\big)| H\rangle\,+\mathcal{O}(1/m_c),\label{eq:Darwin4q1}
\end{split}
\end{equation}
where $\Opsixprime{i}{q}, \OpsixTprime{i}{q}$ are the HQET four-quark operators defined in \Eqn\eqref{eq:Dim6MesonBasisHQE}. This approach has been recently employed for $D$ mesons beyond the \ac{VIA} in \cite{LenzNote:2021}. Using this approximation for charmed mesons we have  
\begin{equation}
    \rho_D^3(D_q)=\frac{g_s^2}{18}f_{D_{q}}^2 m_{D_q}+\mathcal{O}(1/m_c)\,,
\end{equation}
where $f_{D_q}$ denotes the decay constant of $D_q$, also defined in \Eqn\eqref{eq:fDandFDdefs}. 

For the charmed baryons, the same expression \eqref{eq:Darwin4q1} is rewritten in terms of the operators \eqref{eq:Dim6BaryonBasis} in the baryon basis:
\begin{equation}
    2m_H\rho_D^3=
    g_s^2\langle H| \big(-\frac{1}{8}O^q_1+\frac{1}{24}\tilde{O}^q_1+\frac{1}{4}O^q_2-\frac{1}{12}\tilde{O}^q_2\big)| H\rangle\,+\mathcal{O}(1/m_c)\,.\label{eq:Darwin4qbar}
\end{equation}
In \eqref{eq:Darwin4q1} and \eqref{eq:Darwin4qbar}, $g_s^2 = 4\pi \als$ represents a non-perturbative scale; following eg \cite{BSUV1993II}, we set $\als = 1$. With explicit values of the matrix elements of operators from \eqref{eq:Darwin4q1} given in \Sec\ref{sec:matrix_baryons}, we obtain the Darwin parameters of charmed baryons given in \Tab\ref{tab:nonpertNScharm}. 

\begin{table}[t]
    \centering
    \begin{tabular}{|c|c|c|c||c|c|c|c|} \hline
        & $D^0$ & $D^+$& $D_s^+$ & $\Lac$ & $\Xicp$ & $\Xico$ & $\Omc$  \\ \hline
      $\muG/ \GeV^2$  & $0.41(12) $ & $0.41(12)$ & $0.44(13)$ & 0 & 0 & 0 & $0.26(8)$ \\ 
      $\mukin/ \GeV^2$ & $ 0.45(14) $ & $ 0.45(14) $ & $0.48(14)$ & $0.50(15)$ & $0.55(17)$ & $0.55(17)$ & $0.55(17)$ \\
      $\darwin/ \GeV^3$ &$0.056(12)$ & $0.056(22)$&$0.082(33)$ & $0.04(1)$ & $0.05(2)$ &$0.06(2)$ & $0.06(2)$ \\ \hline
    \end{tabular}
    \caption{\small Non-perturbative parameters in the non-spectator contributions used in our analysis, including assigned $\sim$30\% uncertainties to all the entries, by applying the methods discussed in \Secs\ref{sec:muG}, \ref{sec:mukin} and \ref{sec:darwin}. The values for $\mukin$ for the $\Xi$ baryons are derived by replacing $m_{\Lac}$ in \eqref{eq:BUVmukin} with $m_{\Xicp}$, and similarly for the $\Xico$, which has the effect of lifting the values slightly and accounts for $SU(3)_F$-breaking. The central values of the Darwin parameters for the mesons are adopted from \cite{LenzNote:2021}. We estimate the central values of the Darwin parameters for the baryons using the equation of motion for the gluon field with $\alpha_s=1$, see \Sec\ref{sec:baryonsresultsmain}, and our constituent quark model estimates of the dimension-six four-quark matrix elements in \Eqns\eqref{eq:DarwinBary} and \eqref{eq:EstimateDarwinBary}). 
    }
    \label{tab:nonpertNScharm}
\end{table}
\subsection{Results for non-spectator contributions}

In \Tab\ref{Tab:Gamma-3} we present the numerical results for the contribution $\Gamma_3$ at LO (NLO). In accord with our discussion in \Sec\ref{sec:massschemesexplicit}, the quantity $\Gamma_3^{(1)}$ involves terms proportional to $\als$ that originate from the evaluation of the leading coefficient $c_3^{(0)}$ using $C_{1,2}^{(1)}$, the order-$\als$ term from the conversion formula to a given mass scheme, as well as the genuine $c_3^{(1)}$ contribution evaluated with $C_{1,2}^{(0)}$.

Values of the $\Gamma_3\, [\text{ps}^{-1}]$ contribution evaluated using different schemes for the charm-quark mass are listed in \Tab\ref{Tab:Gamma-3}. For these evaluations, we set $\mu_m=\overline{m}_c$ within the expression in \Eqn\eqref{eq:PoleMS5-relation}. We used the value of the pole mass $m_{c}^{\text{pole}}=1.49\GeV$, found using the relation \eqref{eq:PoleMS-relation} to order $\als$. The uncertainties are estimated by varying the scale $\mu$ in the range $(1\GeV\,, 3\GeV)$, while the central values correspond to fixing $\mu=1.5\GeV$. The scale $\mu_f$ in \Eqn\eqref{eq:OS-MSRMass} is set to $\mu_f=1\GeV$, the same value used for the evaluation of the MSR mass in terms of $\overline{m}_c$. The results in the kinetic scheme are for $\mu_f=0.5\GeV$.

Values of the $\Gamma_5$ contributions $\Gamma_\pi$ and $\Gamma_G$ are listed in \Tab\ref{Tab:Gamma-5}. The procedure for the evaluations follows the description for $\Gamma_3$ shown in the caption in \Tab\ref{Tab:Gamma-3}. For consistency of the power-counting in the perturbative expansion, we keep only the leading part of $\Gamma_3$ in the expansion $\Gamma_\pi^{(0)}=\Gamma_3^{(0)}(-\mu_\pi^2/(2m_Q^2))$, in the absence of complete $\als$-corrections to the operators at the dimension-five. Note, however, that in the case of the semileptonic contribution, the complete $\als$-corrections to the dimension-five operators are retained for our final numerical results presented in \Secs\ref{sec:results_mesons} and \ref{sec:results_baryons}. However, the effects of including these additional corrections turn out to be small. 

One can notice several features of the dimension-five and dimension-six non-spectator contributions, previously described in \cite{LenzNote:2021}. Firstly, the Wilson coefficient of the chromomagnetic operator $c_G$ shows strong dependence on the renormalization scale, but nonetheless stays relatively small. The coefficient of the Darwin operator turns out unexpectedly large. This hints at a bad convergence of the $1/m_c$ expansion in the non-spectator sector. However, this can be determined only with the knowledge of the unknown higher-dimension terms.
We also note that the above estimate of the Darwin matrix element involves unusually large $SU(3)_F$-breaking, ie $\rho(D_s^+)/\rho(D^+)\simeq(f_{D_s}^2 m_{D_s})/(f_{D}^2 m_{D})= 1.5$.
Finally, our results agree with those given in  \cite{LenzNote:2021}.

\begin{table}[ht]
	\centering
	\begin{tabular}{ |c|c|c|c| }
		\hline
		Mass scheme & $\Gamma_3^{(0)}$ & $\Gamma_3^{(1)}$\\
			\hline\hline
		Pole & $1.49^{+0.17}_{-0.14}$&  $1.62^{+0.26}_{-0.22}$\\
		\hline
		$\msbar$ & $0.69^{+0.08}_{-0.07}$ & $1.28^{+0.37}_{-0.29}$\\
		\hline
		Kinetic & $1.10^{+0.13}_{-0.11}$ & $1.65^{+0.41}_{-0.32}$\\
		\hline
		MSR & $0.93^{+0.11}_{-0.09}$& $1.54^{+0.41}_{-0.32}$\\
		\hline
		\end{tabular}
	\caption{\small Values of $\Gamma_3\, [\text{ps}^{-1}]$ contribution to the total decay widths at LO ((0)) and NLO((1)) for various schemes of the charm-quark mass. For the central values, we set $\mu=1.5\GeV$, while the errors result from the variation of the scale $\mu$ in the range $(1\GeV, 3\GeV)$. For the pole scheme, we use $m_{c}^{\text{pole}}=1.49\GeV$, evaluated using the relation \eqref{eq:PoleMS-relation} to order $\als$; see \Tab\ref{Tab:cMasses} for the values of charm quark mass in other schemes. We adopt the values $m^{\text{MSR}}$ and $m^{\text{kin}}$ evaluated to four and three loops, respectively, see \Eqns\eqref{eq:Kin3loops} and \eqref{eq:MSR4loops}.}
	\label{Tab:Gamma-3}
\end{table} 

\begin{table}[ht]
	\centering
	\begin{tabular}{ |c|c|c||c|}
		\hline
		Mass scheme & $\Gamma_\pi^{(0)}$ & $\Gamma_G^{(0)}$ & $\Gamma_\text{Darwin}^{(0)}$\\
			\hline\hline
		Pole & $(-0.17^{+0.02}_{-0.02})\frac{\mukin}{0.5\GeV^2}$ & $(0.01^{+0.07}_{-0.07})\frac{\mu_G^2}{0.25\GeV^2}$ & $(0.44^{+0.09}_{-0.09})\frac{\rho_D^3}{0.1\GeV^3}$\\
		\hline
		$\msbar$ & $(-0.10^{+0.01}_{-0.01})\frac{\mukin}{0.5\GeV^2}$ & $(0.00^{+0.05}_{-0.05})\frac{\mu_G^2}{0.25\GeV^2}$ & $(0.43^{+0.08}_{-0.08})\frac{\rho_D^3}{0.1\GeV^3}$\\
		\hline
		Kinetic & $(-0.14^{+0.01}_{-0.02})\frac{\mukin}{0.5\GeV^2}$ & $(0.01^{+0.06}_{-0.06})\frac{\mu_G^2}{0.25\GeV^2}$ & $(0.44^{+0.09}_{-0.08})\frac{\rho_D^3}{0.1\GeV^3}$\\
		\hline
		MSR & $(-0.13^{+0.01}_{-0.01})\frac{\mukin}{0.5\GeV^2}$ & $(0.01^{+0.06}_{-0.06})\frac{\mu_G^2}{0.25\GeV^2}$ & $(0.44^{+0.09}_{-0.08})\frac{\rho_D^3}{0.1\GeV^3}$\\
		\hline
		\end{tabular}
	\caption{\small LO values of dimension-five and dimension-six non-spectator contributions to the total decay widths, $\Gamma_{\pi, G}$ and $\Gamma_{\text{Darwin}}$, respectively, for several schemes of the charm-quark mass. The numerical coefficients are in units ps$^{-1}$, and the non-perturbative parameters $\mu_\pi^2, \mu_G^2$ and $\rho_D^3$ in units $\GeV^2,\GeV^2$ and $\GeV^3$, respectively. See the caption of \Tab\ref{Tab:Gamma-3}, and \Tab\ref{Tab:cMasses}, for further details about the input parameters. The expressions are valid for both mesons and baryons, after the suitable replacements of the hadronic parameters.}
	\label{Tab:Gamma-5}
\end{table} 

\section{Charmed mesons}
\label{sec:mesonsresultsmain}
In this section we present results for the charmed mesons $D^0$, $D^+$, and $D_s^+$. A similar computation was already presented in \cite{LenzNote:2021}, which serves as a useful cross-check of our method and analytic inputs.

\subsection{Matrix elements of spectator contributions}
\label{sec:mesonmateldefs}

The final set of inputs to the lifetime are the four-quark matrix elements, whose coefficients were discussed in \Sec\ref{sec:speccontgen}, and are provided explicitly at \ac{LO} in \Eqns\eqref{eq:explicitDim6Mes} and \eqref{eq:explicitDim7Mes}. As previously noted, these elements can be defined in terms of full QCD quark fields, or, more consistently for the power-counting in $1/m_c$, with heavy quark fields instead. For mesons, since \ac{HQET} estimates of the parameters are available \cite{KLR2021,LenzNote:2021}, we choose to work within \ac{HQET}.

At dimension-six, the relevant matrix elements can be parametrized, following \cite{LenzNote:2021}, as
\begin{align}
\langle D_q| \Opsixprime{i}{q}| D_q\rangle&= F_{D_q}(\mu)^2m_{D_q}B_i^q \,,  \label{eq:HQEdim6val} \\
\langle D_q|  \Opsixprime{i}{q'}| D_q\rangle&=F_{D_q}(\mu)^2m_{D_q}\delta_i^{q'q}
\,, \quad q \neq q' \,,
\label{eq:HQEdim6nonval}
\end{align}
where $F_{D_q}(\mu)$ is the static decay constant, which, in the $m_c \to \infty$ limit, is given by $F_{D_q} = f_{D_q} \sqrt{m^{\vspace{1ex}}_{D_q}}$. The first line describes valence quark contributions, and the second line describes non-valence contributions, with the parameters $\de_i^{q'q}$ denoting the ``eye contractions'' (schematically represented in \Fig\ref{fig:1-DiagramsEye}) \cite{Voloshin2001,LenzNote:2021,KLR2021}.\footnote{Non-valence four-quark operator contributions were extensively discussed in \cite{PU1998}.} The HQET bag parameters for the operators $\mathcal{O}_{1,2}^q$ are denoted by $B_{1,2}^q$,  while $B^q_{3,4}\equiv \eps_{1,2}^q$ denote the bag parameters of the colour octet operators $\mathcal{T}_{1,2}^q$. The isospin relations $B_i^u=B_i^d\equiv B_i^q$, $\delta_i^{ud}=\delta_i^{du}\equiv\delta_i^{qq}$, $\delta_i^{su}=\delta_i^{sd}\equiv\delta_i^{sq}$ and $\delta_i^{us}=\delta_i^{ds}\equiv\delta_i^{qs}$, are applied throughout. Recall that in the \ac{VIA}  $B_{1,2}=1$, while all other parameters vanish. Eye contractions that serve as corrections to the valence-quark matrix elements \eqref{eq:HQEdim6val}, $\de_i^{qq}$, are implicitly included in the bag parameters $B_i^q$. The $\de_i^{qq}$ and $\de_i^{q'q}$ parameters were calculated, using HQET sum rules, in \cite{KLR2021}, but have not been evaluated in the most recent lattice estimates \cite{Becirevic2000}. 
At tree level, eye contractions vanish for the octet operators, and the non-vanishing contributions come only from singlet operators. For charm quark decays, non-valence contributions with the $s$ quark into the loop can be non-negligible.
However, since these cannot be reliably calculated, and the
penguin contributions, of a similar size, are usually neglected (which is  justified for $c$-quark decays), these contributions have been neglected. For the non-valence contributions, we use the HQET evaluations provided in \cite{LenzNote:2021, KLR2021}, including the available eye contraction parameters.

The QCD (HQET) definitions of the decay constants of $D_q$ mesons, $f_{D_q}$ and $F_{D_q}$, respectively, are
\begin{align}
        \langle 0|\bar{q}\gamma^\mu\gamma_5 c| D_q(p)\rangle &= i f_{D_q} p^\mu \,, \nonumber\\
        \langle 0|\bar{q}\gamma^\mu\gamma_5 h_v| D_q(v)\rangle &= iF_{D_q}(\mu)\sqrt{m_{D_q}}v^\mu\,.
        \label{eq:fDandFDdefs}
    \end{align}
Employing the result, valid at the scale $\mu=m_c$, that the non-local dimension-seven matrix elements can be reabsorbed into the dimension-six matrix elements (see \cite{BlokShifman93Review,Neubert1993,KM1992,LenzNote:2021} and \App\ref{app:HQEtoQCD}), we further apply the replacement 
\begin{equation}
    F_{D_q}(\mu)^2 \to f_{D_q}^2 m_{D_q} \left( 1 + \frac{4}{3} \frac{\als(m_c)}{\pi} \right) 
    \label{eq:FDtofD}
\end{equation}
to express the dimension-six matrix elements \eqref{eq:HQEdim6val} and \eqref{eq:HQEdim6nonval} in terms of physical parameters. Values for all HQE parameters are taken from \cite{KLR2021,LenzNote:2021}.

The dimension-seven matrix elements are parametrized, again following \cite{LenzNote:2021}, by
\begin{align}
\langle D_q| \OpsevenPprime{1}{q}| D_q\rangle& = -m_q F_{D_q}^2 m_{D_q} B^P_{1}
\,, \nonumber \\
\langle D_q| \OpsevenPprime{2}{q}| D_q\rangle& =-\bar{\Lambda}_q F_{D_q}^2 m_{D_q}B^P_{2}
\,, \nonumber \\
\langle D_q| \OpsevenPprime{3}{q}| D_q\rangle& =-\bar{\Lambda}_q F_{D_q}^2 m_{D_q}B^P_{3}
\,, \nonumber \\
\langle D_q| \OpsevenRprime{1}{q}| D_q\rangle& = - F_{D_q}^2m_{D_q} (\bar{\La}_q - m_q)B^R_{1}\,, \nonumber \\
\langle D_q| \OpsevenRprime{2}{q}| D_q\rangle& =  F_{D_q}^2m_{D_q} (\bar{\La}_q - m_q)B^R_{2}\,,
\label{eq:dim7Mesmatelparam}
\end{align}
with the colour-octet operators having equivalent parametrizations on replacement of $B^{P,R}_i \to \eps^{P,R}_i$. Since there is no available computation of these parameters in HQET, we apply the \ac{VIA}, so that $B^{P,R}_i=1$ and $\eps^{P,R}_i=0$. Parametrizations of the non-local matrix elements, which formally contribute but here have been absorbed into the dimension-six matrix element via the replacement \eqref{eq:FDtofD}, are available in appendix C of \cite{LenzNote:2021}. We also apply the replacement $F_{D_q} \to f_{D_q} \sqrt{m^{\vspace{1ex}}_{D_q}}$. The parameter $\Lambda_q$ is of order the QCD scale; for the numerical evaluation we use
the numbers reported in \cite{LenzNote:2021}, namely
\begin{equation}
    \bar{\Lambda}=0.5\pm0.1\GeV\,,\qquad \qquad \bar{\Lambda}_s=0.6\pm0.1\GeV\,.
    \label{eq:labarvals}
\end{equation}

\subsection{Final numerical predictions for mesons}
\label{sec:results_mesons}

Our final numerical predictions for meson decay widths and lifetime ratios are presented in \Tab\ref{tab:fin_mesons}, while semileptonic branching fractions and ratios are presented in \Tab\ref{tab:fin_mesonsSL}. A detailed breakdown of various contributions, for two choices of the mass scheme for the charm quark, is shown in \Tabs\ref{Tab:MesonsCentral1} and \ref{Tab:MesonsCentral2}, provided in \App\ref{app:detailedtables}. Following \cite{LenzNote:2021}, we express lifetime ratios via the differences of the theoretical decay widths (denoted ``th''), scaled by the experimental lifetimes (denoted ``exp''):
\begin{equation}
    \frac{\tau(D_{(s)}^+)}{\tau(D^0)}=1+\bigg(\Gamma^{\text{th}}(D^0)-\Gamma^{\text{th}}(D_{(s)}^+)\bigg)\tau^{\text{exp}}(D_{(s)}^+)\,.\label{eq:MesonRatios}
\end{equation}
The benefit of such a definition lies in the cancellation between the, universal, non-spectator contributions. This is especially beneficial in the ratio $\tau{(D^+)}/\tau{(D^0)}$, where, due to isospin symmetry, there is even cancellation of the dimension-five and -six non-spectator contributions.

For the semileptonic contributions, we consider the inclusive decay channels involving electrons in the final states, eg $\Gamma^{(e)}(D^+)\equiv \Gamma(D^+\to X e\nu)$, with the definitions \cite{LenzNote:2021}
\begin{equation}
    BR^{(e)}(D)=\Gamma^{(e)}(D)\tau^{\text{exp}}(D)\,,\label{Eq:BRSL}
\end{equation}
and
\begin{equation}
    \frac{\Gamma^{(e)}(D_{(s)}^+)}{\Gamma^{(e)}(D^0)}=1+(\Gamma^{(e)\,\text{th}}(D_{(s)}^+)-\Gamma^{(e)\,\text{th}}(D^0))\bigg(\frac{\tau(D^0)}{BR^{(e)}(D^0)}\bigg)^{\text{exp}}\,.\label{Eq:Ratio_SL}
\end{equation}
The results in \Tabs\ref{tab:fin_mesons} and \ref{tab:fin_mesonsSL} are compatible with those in \cite{LenzNote:2021} and with experiment. We also agree with the criticism in \cite{LenzNote:2021} of the results presented in \cite{Cheng18c}. There is some minor difference in the uncertainty estimates, which can be attributed to a more conservative approach to uncertainties in our study. In our approach, the varied parameters have not been interpreted as following a probability distribution. Rather, the upper and lower errors are simply the maximal and minimal distances from the central values. For the estimate of the scale uncertainties we fixed hadronic parameters to their central values. 

 As with \cite{LenzNote:2021}, we also observe a slight tension with experiment in the theoretical value of the ratio $\tilde{\tau}(D_s^+)/\tau{(D^0)}$. This is a long-standing problem, to which several solutions have been proposed in the literature, such as large non-perturbative or non-valence WA contributions, albeit without a clear conclusion \cite{BU1993}. 

\begin{table}[ht]
\small
\centering
\begin{tabular}{|c|c|c|c|c||c|}
\hline
Observable & Pole & $\overline{\text{MS}}$ &  Kinetic & MSR & Experiment \\
\hhline{|=|=|=|=|=||=|}
$\Gamma(D^0)$  & $1.71^{+0.41+0.39}_{-0.47-0.36}$ & $1.43^{+0.36+0.48}_{-0.40-0.40}$  &  $1.77^{+0.40+0.53}_{-0.45-0.45}$  & $1.68^{+0.38+0.53}_{-0.43-0.44}$ & $2.44\pm 0.01$\\
\hline
$\Gamma(D^+)$ & $-0.07^{+0.76+0.31}_{-0.68-0.20}$ & $-0.27^{+0.66+0.03}_{-0.88-0.04}$ & $-0.07^{+0.73+0.20}_{-0.66-0.14}$ & $-0.13^{+0.71+0.13}_{-0.64-0.11}$ & $0.96\pm 0.01$\\
\hline
$\tilde{\Gamma}(D_s^+)$ & $1.71^{+0.49+0.44}_{-0.60-0.40}$ & $1.43^{+0.42+0.49}_{-0.52-0.41}$ & $1.77^{+0.47+0.55}_{-0.58-0.47}$ & $1.67^{+0.46+0.55}_{-0.56-0.46}$ & $1.88\pm 0.02$\\
\hhline{|=|=|=|=|=||=|}
$\tau(D^+)/\tau(D^0)$ & $2.85^{+0.68+0.10}_{-0.81-0.17}$ & $2.78^{+0.63+0.47}_{-0.73-0.37}$ &$2.91^{+0.68+0.35}_{-0.80-0.32}$ & $2.89^{+0.66+0.42}_{-0.78-0.35}$ & $2.54\pm 0.02$\\
\hline
$\tilde{\tau}(D_s^+)/\tau(D^0)$ & $1.00^{+0.24+0.02}_{-0.22-0.02}$ &
$1.00^{+0.21+0.01}_{-0.19-0.00}$&
$1.00^{+0.23+0.01}_{-0.21-0.01}$ & $1.00^{+0.23+0.01}_{-0.21-0.01}$&  $1.30\pm 0.01$\\
\hline
\end{tabular}
\caption{\small The total decay widths in units $\text{ps}^{-1}$, and their ratios for charmed mesons compared to the experimental values, using HQET parameters, see \Tab\ref{tab:lMesonLifetimesexpt} for the references to the experimental papers. The first uncertainties are coming from  independent variations of the hadronic matrix elements within the corresponding ranges, while the second uncertainties result from the variation of the renormalization scale $\mu$ in the range $[1,3]\GeV$. The references to the sources of the experimental data are given in \Sec\ref{subsec:charmed_mesons}.}
\label{tab:fin_mesons}
\end{table}

\begin{table}[ht]
\small
\centering
\begin{tabular}{|c|c|c|c|c||c|}
\hline
Observable & Pole & $\overline{\text{MS}}$ &  Kinetic & MSR & Experiment \\
\hhline{|=|=|=|=|=||=|}
$BR^{(e)}(D^0)\,[\%]$ & $4.07^{+2.21+0.84}_{-2.53-0.97}$ & $5.18^{+1.59+0.63}_{-1.82-0.55}$ & $5.87^{+1.94+0.22}_{-2.23-0.19}$ & $5.86^{+1.80+0.48}_{-2.07-0.41}$ & $6.49\pm 0.16$\\
\hline
$BR^{(e)}(D^+)\,[\%]$  &
$10.34^{+5.69+2.12}_{-6.52-2.44}$&
$13.15^{+4.10+1.61}_{-4.73-1.40}$ & $14.92^{+5.00+0.57}_{-5.75-0.49}$& $14.90^{+4.67+1.22}_{-5.37-1.06}$& $16.07\pm 0.30$\\
\hline
$BR^{(e)}(D_s^+)\,[\%]$ & $5.42^{+3.02+0.96}_{-3.44-1.10}$ &
$6.86^{+2.42+0.83}_{-2.83-0.72}$&
$7.67^{+2.80+0.34}_{-3.23-0.29}$ & $7.67^{+2.67+0.65}_{-3.10-0.56}$&  $6.30\pm 0.16$\\
\hhline{|=|=|=|=|=||=|}
$\Gamma^{(e)}(D^+)/\Gamma^{(e)}(D^0)$ & $1.00^{+0.02+0.00}_{-0.02-0.00}$ &
$1.00^{+0.01+0.00}_{-0.01-0.00}$&
$1.00^{+0.02+0.00}_{-0.02-0.00}$ & $1.00^{+0.02+0.00}_{-0.01-0.00}$&  $0.977\pm 0.031$\\
\hline
$\Gamma^{(e)}(D_s^+)/\Gamma^{(e)}(D^0)$ & $1.05^{+0.29+0.01}_{-0.31-0.01}$ &
$1.06^{+0.24+0.01}_{-0.27-0.01}$&
$1.07^{+0.28+0.01}_{-0.30-0.01}$ & $1.06^{+0.26+0.01}_{-0.29-0.01}$&  $0.790\pm 0.026$\\
\hline
\end{tabular}
\caption{\small Semileptonic decay widths in inclusive channel $X e\nu$ in units $\text{ps}^{-1}$, and their ratios for charmed mesons compared to the experimental values, using HQET parameters. The first uncertainties are coming from  independent variations of the hadronic matrix elements within the corresponding ranges, while the second uncertainties result from the variation of the renormalization scale $\mu$ in the range $[1,3]\GeV$. The references to the sources of the experimental data are given in \Sec\ref{subsec:charmed_mesons}.
}
\label{tab:fin_mesonsSL}
\end{table}

\begin{table}[ht]
	\centering
\begin{tabular}{*2c}
 \toprule
\multicolumn{2}{c}{$D^0$} \\
 \hline
WE & $-(0.01+0.22\,x)B_1^q+(0.01+0.21\,x) B_2^q-(2.27+0.93 \,x)\epsilon_1^q+(2.30+0.68\,x)\epsilon_2^q$\\
 \hline\hline
 \multicolumn{2}{c}{$D^+$} \\
 \hline
 PI & $-(1.25+1.02\,x)B_1^q-0.17\,x\,B_2^q+(7.46 + 5.24\,x )\epsilon_1^q-0.28\,x\,\epsilon_2^q $\\
  \hline
 WA & $-(0.13+0.05\,x)B_1^q+(0.13+0.05\,x)B_2^q-(0.02+0.06 \,x)\epsilon_1^q+(0.02+0.06\,x)\epsilon_2^q$\\
  \hline
 SL & $-(0.08 +0.01\,x)B_1^q+(0.08+0.01\,x)B_2^q-0.03\,x\, \epsilon_1^q+0.02\,x\,\epsilon_2^q$\\ 
 \hline\hline
 \multicolumn{2}{c}{$D_s^+$}\\
 \hline 
 PI & $-(0.09+0.08\,x)B_1^s-0.01\,x B_2^s+(0.55+0.39\, x)\epsilon_1^s-0.02\,x \epsilon_2^s$\\
 WA & $(-3.66-1.38\,x)B_1^s+(3.66+1.38\,x)B_2^s-(0.47+1.74 \,x)\epsilon_1^s+(0.47+1.60\,x)\epsilon_2^s$\\
 SL & $-(2.08+0.26\,x)B_1^s+(2.09+0.27\,x)B_2^s-0.72\,x \epsilon_1^s+0.67\,x \epsilon_2^s$\\
 \toprule
\end{tabular}
 \caption{\small Contributions of valence dimension-six spectator operators to the decay widths of charmed mesons, in units ps$^{-1}$, in the $\msbar$ scheme. The different contributions are separated by the topologies defined in \Fig\ref{fig:1-DiagramsMesons}. The HQET bag parameters have been left unevaluated, but are assumed to be renormalized at the scale $\mu_0 \sim 1.5 \GeV$. Their coefficients correspond to the scale $\mu=1.5\GeV$. The factor $x=1$ denotes the $\mathcal{O}(\als)$ contributions. Semileptonic contributions involve both $e$ and $\mu$ channels.}
    \label{tab:Dim6Mesons}
\end{table}
To discuss the spectator contributions in more detail, \Tab\ref{tab:Dim6Mesons} presents central values of \textit{valence} dimension-six spectator contributions to the decay widths of charmed mesons, evaluated in the $\msbar$ scheme at the scale $\mu=\mu_0=1.5\GeV$. The factor $x=1$ multiplies the contributions of order $\als$. 
 Evident from the expressions shown in \Tab\ref{tab:Dim6Mesons} is the well-known helicity suppression of the \acf{WE} and \acf{WA} contributions within the \ac{VIA}, while \acf{PI} drives a large suppression of the decay width of $D^+$ relative to that of $D^0$. The large value of the coefficient multiplying $\epsilon_1^q$ gives rise to a strong sensitivity to this hadronic parameter. While $\epsilon_1^q$ is close to zero \cite{KLR2021}, it comes with a large uncertainty. Sizeable perturbative $\als$ corrections boost \ac{PI} even further, driving the prediction of the $D^+$ lifetime towards the unphysical region for some choices of the hadronic parameters, see \Tab\ref{tab:fin_mesons}. This observation has been also made in \cite{Cheng18c,LenzNote:2021}. As a result, the prediction of the $D^+$ lifetime is particularly problematic. To resolve this problem, a lattice QCD determination of the dimension-six matrix elements will be necessary.
 
 The dimension-seven \ac{PI} contribution to the $D^+$ decay width also turns out to be sizeable, but positive, providing some cancellation of the dimension-six terms. The corresponding matrix elements are, however, currently estimated using the \ac{VIA} only. Hence, as with the dimension-six operators, a lattice determination of matrix elements of dimension-seven operators, and separately an \ac{NLO} computation at this order in the \ac{HQE}, would be welcome. It is possible that higher-order $\als$ spectator contributions would play a significant role in a more complete assessment of the \ac{PI} contribution. However, the issue of poor convergence of the $\als$ expansion, seen also in $\Gamma_3^{\text{\ac{SL}}}$ (cf. \Sec\ref{sec:alsseriescomments}), can also be expected to appear in the four-quark contribution. In this case, the theoretical precision would not necessarily be improved with further contributions, which would possibly motivate studies of some alternative approaches to the inclusive charm decays.

\section{Singly charmed baryons}
\label{sec:baryonsresultsmain}
\subsection{Matrix elements of spectator contributions and baryonic wavefunctions}
\label{sec:matrix_baryons}
The dimension-six spectator matrix elements for the operators $\Opsix{1,2}{u,d,s}$ between the baryon states, introduced in \Eqn\eqref{eq:Dim6BaryonBasis},
\begin{eqnarray}
M_i^q(\Bary_c) \equiv \frac{\langle \Bary_c | O_i^q | \Bary_c \rangle}{2 m_{\Bary_c}}\,, \qquad i=1,2 \quad \text{and} \quad q= u,d,s\,,
\label{eq:Barymatelshort}
\end{eqnarray} 
can be parametrized in constituent quark models as given in \Tab\ref{tab:matrixNRRel}. There, (N)RCQM denotes the expressions in nonrelativistic and relativistic constituent quark models. 
The remaining dimension-six matrix elements are related by \eqref{eq:BaryonMERelation1}, so that $ \tilde{M}_i^q(\Bary_c) = - \tilde{B} M_i^q(\Bary_c)$, where we will take $\tilde{B}=1$ throughout \cite{NS1996}.

\begin{table}[ht]\small
\renewcommand{\arraystretch}{3}
\centering
\begin{tabular}{|c|c|c|}
\hline
$M_i^q({\cal B}_c)\,, \; {\cal T}_c = \Lac, \Xi_c^+, \Xi_c^0$ & RCQM & \ac{NRCQM}  \\
\hhline{|=|=|=|}
$M_1^q({\cal T}_c) \equiv \cfrac{\langle {\cal T}_c| \Opsix{1}{q}|{\cal T}_c\rangle}{2m_{{\cal T}_c}}$  &  $ -(a_q+b_q)$ &  $-|\Psi_{cq}^{{\cal T}_c}(0)|^2$ 
\\
\hline
$M_2^q({\cal T}_c) \equiv \cfrac{\langle {\cal T}_c| \Opsix{2}{q}|{\cal T}_c\rangle}{2m_{{\cal T}_c}}$   &
$\frac{1}{2}(a_q+b_q)$ & 
$\frac{1}{2}\,|\Psi_{cq}^{{\cal T}_c}(0)|^2$
\\
\hhline{|=|=|=|}
$M_1^s(\Omc) \equiv \cfrac{\langle \Omc| \Opsix{1}{s}| \Omc\rangle}{2m_{\Omc}}$ & 
$-\frac{1}{3}(18 a_s+2b_s+32 c_s)$ & $-6\,|\Psi_{cs}^{\Omc}(0)|^2$ 
\\
\hline 
$M_2^s(\Omc) \equiv \cfrac{\langle \Omc| \Opsix{2}{s}| \Omc\rangle}{2m_{\Omc}}$ & 
$-(a_s-\frac{5}{3}b_s-\frac{16}{3}c_s)$
& $ -|\Psi_{cs}^{\Omc}(0)|^2$ 
\\
\hline
\end{tabular}
\caption{\small Generalized parametrizations of dimension-six matrix elements for baryons in relativistic (RCQM) and nonrelativistic (NRCQM) constituent quark models \cite{GRT1986,Cheng18c}. The definitions of $a_q$, $b_q$, and $c_q$ are in \eqref{eq:abcbag}. }
\label{tab:matrixNRRel}
\end{table}
In a relativistic constituent model, $a_q,\, b_q,\,$ and $c_q$ are the overlap integrals
\begin{align}
    a_q&=\int d^3 r[u^2_q(r)u^2_c(r)+v^2_q(r)v^2_c(r)]\,,\nonumber\\
    b_q&=\int d^3 r[u^2_q(r)v^2_c(r)+v^2_q(r)u^2_c(r)]\,,\nonumber\\
    c_q&=\int d^3 r[u_q(r)u_c(r)v_q(r)v_c(r)]\,,
    \label{eq:abcbag}
\end{align}
where $u_q(r)$ and $v_q(r)$ are the upper and lower components of the relativistic Dirac spinor. 
In the nonrelativistic limit, $b_q = c_q = 0$ and $a_q = |\psi_{cq}^{{\cal B}_c}(0)|^2$.
A popular relativistic model was the MIT bag model \cite{MITbag0,MITbag1,MITbag2,MITbag3,MITbag4,GNPR1979}, with some updated parameters for heavy baryons in \cite{BS2004,BS2008,BS2012}. 
The advantage of the MIT bag model is that with the only a few adjustable parameters the model can be easily applied for qualitative and quantitative predictions of mesonic and baryonic wavefunctions. The main source for the improvement of such QCD models is heavy hadron spectroscopy, but, despite much experimental progress, it is difficult to make a meaningful assessment of bag model parameters and so have a clear guidance for improvement to the model. Indeed, there are several versions of the MIT bag model, which are not compatible with each other \cite{BS2004}. Moreover, the standard MIT bag model problems, like the inclusion of center-of-mass motion or the value of the quark masses in the bag, are still not fully resolved and reliably treated in the models, while an estimation of the uncertainties in such models is questionable. However, we have checked that for the hydrogen-like MIT model, which could describe a singly charmed baryon configuration, the spectator matrix elements have too small values, leading to results that are not compatible with experimental values. We therefore turn to the \ac{NRCQM} approach.

The dynamics of a baryon state $(Qq_1q_2)$ is more complex than that of a meson $Q\bar{q}$. However, in the case of heavy baryons some simplification arises due to the heaviness of one quark, $Q$. The heavy quark is expected to have a very weak coupling to the light quarks, which themselves couple together as a light diquark system \cite{Copley1979,Jaffe2004,CMLW2016}. In such a picture, baryons can be treated as a quasi two-body system, and show similar dynamics to heavy-light mesons. 

The extraction of the wavefunction in the \ac{NRCQM} is based on the application of the seminal work by de Rujula, Georgi and Glashow \cite{GGR1975}, where the expression for heavy hadron masses is obtained by considering a two-body potential and the spin-spin interaction between the constituent quarks. For ground states we have  
\begin{eqnarray}
M_{H} = \sum_i m_i^{H} + \langle H_{\text{spin,H}}  \rangle   \,,
\label{eq:mHmaster}
\end{eqnarray}
where
\begin{eqnarray}
H_{\rm spin, \, mesons} = \frac{32 \pi\alpha_s}{9}  \frac{(\vec{s}_i\cdot\vec{s}_j)}{m_i^{\Mes} m_j^{\Mes}} \delta_{\Mes}^3(\vec{r}_{ij})  \,,
\nonumber \\
H_{\rm spin, \, baryons} = \sum_{i > j} \frac{16 \pi\alpha_s }{9} \frac{(\vec{s}_i\cdot\vec{s}_j)}{m_i^{\Bary} m_j^{\Bary}} \delta_{\Bary}^3(\vec{r}_{ij})  \,,
\label{eq:HFmass1}
\end{eqnarray}
are the spin-spin interactions for mesons and baryons respectively. By combining the expression \eqref{eq:mHmaster} for hadrons in different spin states and taking mass differences, the wavefunction $|\Psi(0)|_{ij}^2 \sim \delta^3(0)$ is extracted. The spectator contribution is then proportional to the squared modulus of the wavefunction for two quarks at the origin, as seen in the rightmost column of \Tab\ref{tab:matrixNRRel}.
The masses which appear in \eqref{eq:HFmass1} are constituent masses (sometimes also called effective masses), rather than bare masses, so for example $m_u^{\Mes,\Bary} \neq 0$ is non-negligible.
Constituent masses of quarks in baryons and mesons differ from one another, with $m_q^\Mes$ and $m_q^\Bary$ denoting, respectively, the constituent mass of the quark $q$ in a meson and baryon. The values of $m_q^{\Mes,\Bary}$ are obtained from the fits to experimentally determined hadron masses \cite{GR1981,KR2014}. We have for the constituent quark masses  \cite{KR2014} in mesons
\begin{equation}
m_{u,d}^\Mes = 310 \MeV\,, \quad m_s^\Mes = 483 \MeV \,, \quad m_c^\Mes = 1663.3 \MeV \,, \quad m_b^\Mes = 5003.8 \MeV \,,  
\label{eq:mqeffmes}
\end{equation}
and in the baryons
\begin{equation}
m_{u,d}^\Bary = 363 \MeV\,, \quad m_s^\Bary = 538 \MeV\,, \quad m_c^\Bary = 1710.5 \MeV \,, \quad m_b^\Bary = 5043.5 \MeV \,.
\label{eq:mqeffbary}
\end{equation}
As expected, the constituent quark masses are somewhat smaller in mesons.

There are several possibilities to build suitable heavy hadron mass differences in order to extract the wavefunction of a heavy baryon. 
The first method of extraction of the $\Lac$ wavefunction, which, to the best of our knowledge, was first proposed by Barger et al in \cite{BLS1980}, was driven by the knowledge of the experimentally measured mass of $\Sigma_c$, ie the mass difference $M_{\Sigma_c} - M_{\Lambda_c}$. This exploits the fact that $\Sigma_c$ has the same quark content as $\Lac$, although it is a member of the $SU(3)_{F}$ sextet rather than of the $SU(3)_F$ antitriplet, and that the light quarks in $\Lac$ are coupled to zero spin, so that their hyperfine interaction with the heavy quark is therefore zero. 
By accounting for different spins and spin interactions of sextet and antitriplet baryons, one arrives at
\begin{eqnarray}
M_{\Sigma_c} - M_{\Lac} = \frac{16 \pi\alpha_s }{9}  \frac{1}{m_c^\Bary m_u^\Bary} \left (
\frac{m_c^\Bary - m_u^\Bary}{m_u^\Bary}
\right ) \, |\Psi_{cq}^{{\Lac}}(0)|^2 \,,
\end{eqnarray}
where it has been assumed that the spatial wavefunctions of the baryons are equal. 

This relation, however, suffers from large uncertainty due to the value of the $\als$ coupling in the baryon, and is also dependent on the values of the constituent quark masses, which historically were not well-determined. As first recognized by Cortes and Sanches-Guillen \cite{Cortes1980}, this uncertainty can be reduced by exploiting a similar relation between charmed meson states from (\ref{eq:HFmass1}), and by relating the baryon and meson wavefunctions as 
\begin{eqnarray}
 |\Psi_{cq}^{{\Lac}}(0)|^2 &=& \frac{ 2 m_u^\Bary}{(M_D - m_u^\Bary)}  \, \frac{M_{\Sigma_c} - M_{\Lac}}{M_{D^{\ast}} - M_D} \; |\Psi_{cq}^{D_q}(0)|^2 \,,
\label{eq:ShWF}
\end{eqnarray}
where $m_c^\Bary$ has been replaced by $M_D$, and 
\begin{eqnarray}
|\Psi_{cq}^{D_q}(0)|^2 &=& \frac{f_D^2 M_D}{12} \,
\label{eq:DmesonWF}
\end{eqnarray}
is the mod-square wavefunction of the meson. 
This expression relies on the value of $\als$ being identical for meson and baryon states.

The formula \eqref{eq:ShWF}, which uses the mass differences of baryons with the same quark content
and having the same spin, was extensively used for the extraction of singly charmed baryon wavefunctions, until it was proposed by Rosner in \cite{Rosner1996} to exploit the hyperfine splittings in $\Sigma_c^{\ast}$ and $\Sigma_c$, with the same assumptions as above. Using
\begin{eqnarray}
M_{\Sigma_c^\ast} - M_{\Sigma_c} = \frac{16 \pi\alpha_s }{9} \frac{1}{m_c^\Bary m_u^\Bary}\,\frac{3}{2} \,    |\Psi_{cq}^{{\Lac}}(0)|^2 
\end{eqnarray}
this leads to
\begin{eqnarray}
 |\Psi_{cq}^{{\Lac}}(0)|^2 &=& \frac{4}{3} \frac{M_{\Sigma_c^\ast} - M_{\Sigma_c}}{M_{D^{\ast}} - M_D} \; |\Psi_{cq}^{D_q}(0)|^2 \,.
\label{eq:RosWF}
\end{eqnarray} 
This construction enables removal of the a priori unknown constituent mass $m_u^\Bary$ in \eqref{eq:ShWF}.
In addition, by taking the difference of $M_{\Sigma_c^\ast} - M_{\Sigma_c}$, one effectively performs the spin-weighted average of the hyperfine interactions in $\Sigma_c^\ast(3/2)$ and $\Sigma_c(1/2)$. 

Although the two formulas \eqref{eq:ShWF} and \eqref{eq:RosWF} do not look the same, they are derived from the same mass formula in \eqref{eq:HFmass1} and so should be numerically equivalent, which was not the case in the past, since the constituent quark masses
were not known precisely. By inserting explicitly the constituent quark masses given in \eqref{eq:mqeffbary}, we obtain for the difference of these wavefunctions
\begin{eqnarray}
\frac{ |\Psi_{cq}^{{\Lac}}(0)|^2}{|\Psi_{cq}^{D_q}(0)|^2}_{|\,(\ref{eq:RosWF})} -  \frac{ |\Psi_{cq}^{{\Lac}}(0)|^2}{|\Psi_{cq}^{D_q}(0)|^2}_{|\,( \ref{eq:ShWF})}  \sim 0.04 \,  \,,
\end{eqnarray}
which represents an approximately 6\% difference, and is therefore negligible at the present level of uncertainty. 

It is worth emphasizing that the derivation of the formulas \eqref{eq:ShWF} and \eqref{eq:RosWF} has relied on several assumptions, including that
\begin{enumerate}[(i)]
\item the wavefunctions of baryons with the same quark content are the same, even if they belong to different $SU(3)_F$ mutliplets or spin states;
\item strong couplings $\als$ for all interactions are approximately equal;
\item constituent quark masses in mesons and baryons are equal.
\end{enumerate}
Although all these assumptions seem to be plausible for such systems $(Qq_1q_2)$ with one heavy quark, they have to be critically examined. In particular, whereas the last assumption about the equality of the constituent masses in mesons and baryons 
has been used in previous studies \cite{Melic97c,Cheng18c}, it is not justified in view of the values in \eqref{eq:mqeffmes} and \eqref{eq:mqeffbary}. By using these values, the formulas \eqref{eq:ShWF} and \eqref{eq:RosWF} above should be multiplied by the correction factors
\begin{eqnarray}
y = \frac{m_c^\Bary m_u^\Bary}{m_c^\Mes m_u^\Mes} \simeq 1.20\,, \qquad \qquad
y_s = \frac{m_c^\Bary m_s^\Bary}{m_c^\Mes m_s^\Mes} \simeq 1.15 \,,
\label{eq:ConstituentMassRatio}
\end{eqnarray}
where the second factor is relevant for baryons containing an $s$ quark. We include the above correction factors in our numerical calculation. 

Bearing in mind all the considerations above, we will consider charmed baryon wavefunctions only in the \ac{NRCQM} approach, using the hyperfine mass-splitting relations \cite{GGR1975} and the method of \cite{Rosner1996} exemplified in \eqref{eq:RosWF}. The relevant baryon wavefunctions are then given by
\begin{equation}
\begin{split}
    |\Psi_{cq}^{\Lac}(0)|^2&=\frac{4}{3}\, \frac{m_{\Sigma_c^\ast}-m_{\Sigma_c}}{m_{D^\ast}-m_D} \,|\Psi_{cq}^{D_q}(0)|^2\,R_{cq}^{\Lac},\quad \text{for $q=u,d$},\\ |\Psi_{cq}^{\Xi_c}(0)|^2&=\frac{4}{3} \,\frac{m_{\Xi_c^\ast}-m_{\Xi'_c}}{m_{D_q^\ast}-m_{D_q}}\,|\Psi_{cq}^{D_q}(0)|^2R_{cq}^{\Xi_c}\,,\quad \text{for $q=u,d,s$}\\ |\Psi_{cs}^{\Omc}(0)|^2&=\frac{4}{3} \,\frac{m_{{\Omc}^\ast}-m_{\Omc}}{m_{D_s^\ast}-m_{D_s}}\,|\Psi_{cs}^{D_s}(0)|^2R_{cs}^{\Omega_c}\,,\label{eq:QuarkRel3}
\end{split}
\end{equation}
with the overall scaling coefficients $R_{cq}^{{\cal B}}$, such that $R^{{\cal B}}_{cq}=y$, for $q=u,d$ and $R^{{\cal B}}_{cs}=y_s$, with the values in \eqref{eq:ConstituentMassRatio}. 
Note that we also consider $SU(3)_F$-breaking in the $|\psi_{cq}^{\Xi_c}(0)|^2$ wavefunction. The relations above are taken to be valid at a low hadronic scale $\mu_H$, taken to be of order $1\GeV$.

The remaining question is how to treat the meson wavefunction in \eqref{eq:QuarkRel3}. The nonrelativistic $D$-meson wavefunction is given by\footnote{In general, four-quark operators are (re)normalized at the heavy quark scale. Their evolution from $m_c$ down to a hadronic scale  $\mu_{\rm had}\sim O(0.7-1 \,{\rm GeV})$ scale brings hybrid renormalization into account \cite{NS1996,SV1987,PW1988}, usually denoted by $\kappa(\mu)=(\alpha_s(\mu)/\alpha_s(m_c))^{3N_C/2\be_0}$, and the factor is sometimes in the literature explicitly associated with the wavefunctions as  $|\Psi_{cq}^{D_q}(0)|^2=\frac{1}{12}f_{D_q}^2m_{D_q} \kappa^{-4/9}$. Here, we perform explicit HQET and QCD matching using \Eqns\eqref{eq: C-matrix1} and \eqref{eq: C-matrix2}, so that the hybrid anomalous dimension is already included in the operators.}
\begin{equation}
    |\Psi_{cq}^{D_q}(0)|^2=\frac{1}{12}f_{D_q}^2m_{D_q} \,,
\end{equation}
in terms of the mesonic decay constant $f_{D_q}$. 
But, one has to keep in mind that the mesonic decay constant has its own $1/m_c$ expansion \cite{Neubert1992,KM1992,LenzNote:2021}.  
In \cite{BlokShifman93Review}, it was suggested to use the meson wavefunction defined in terms of the static decay constant $F_{D_q}$, for consistency within the \ac{HQE}, while the hadron mass differences in \eqref{eq:QuarkRel3} attain their static rather than physical values. The idea is that the renormalization of $F_{D_q}$ to $f_{D_q}$ in the dimension-six matrix elements by the non-local dimension-seven contributions, observed in mesons  and presented in \App\ref{app:HQEtoQCD}, would also occur for the $\Omc$, owing to the spin structure of its constituent $s$-quarks binded in a spin 1 diquark, but would not occur in the antitriplet of baryons $(\Lac,\,\Xicp,\,\Xico)$. This conjecture led also to the suggestion that $\Omc$ might be far longer-lived than was measured to be the case at the time of \cite{BlokShifman93Review}. In light of the new LHCb measurements increasing the $\Omc$ lifetime, this conjecture deserves further attention. 
A preliminary numerical test does suggest that predictions for the $\Lac$ and $\Xico$ lifetimes are brought closer to their experimental values, at the expense of a worse prediction for $\Lifetime{\Xicp}$, but any more concrete analysis at higher orders in the $1/m_c$ expansion will also require considering the non-local matrix elements, about which nothing is known for baryons.
For this reason, we prefer to restrict to considering only the \acs{QCD} matrix elements, for which all inputs attain their physical values.

We parametrize the dimension-seven matrix elements in the \ac{NRCQM} by relating them to those of dimension-six as follows:
\begin{alignat}{3}
 \frac{\langle {\cal T}_c| P_1^q| {\cal T}_c\rangle}{2m_{{\cal T}_c}} &\simeq m_q\frac{\langle {\cal T}_c| O_2^q| {\cal T}_c\rangle}{2m_{{\cal T}_c}}&=& \frac{1}{2}m_q|\Psi_{cq}^{{\cal T}_c}(0)|^2\rho_1\,,  \nonumber\\
  \frac{\langle {\cal T}_c| P_2^q| {\cal T}_c\rangle}{2m_{{\cal T}_c}} &\simeq\Lambda_{QCD}\frac{\langle T_c| O_1^q| {\cal T}_c\rangle}{2m_{{\cal T}_c}}&=&-\Lambda_{QCD}|\Psi_{cq}^{{\cal T}_c}(0)|^2\rho_2 \,, \nonumber \\
   \frac{\langle {\cal T}_c| P_3^q| {\cal T}_c\rangle}{2m_{{\cal T}_c}} &\simeq\Lambda_{QCD}\frac{\langle {\cal T}_c| O_2^q| {\cal T}_c\rangle}{2m_{{\cal T}_c}}&=&\frac{1}{2}\Lambda_{QCD}|\Psi_{cq}^{{\cal T}_c}(0)|^2\rho_3\,, \nonumber \\ 
  \nonumber \\
 \frac{\langle \Omc| P_1^q| \Omc\rangle}{2m_{\Omc}} &\simeq m_q\frac{\langle \Omc| O_2^q| \Omc\rangle}{2m_{\Omc}}&=&-m_q|\Psi_{cq}^{\Omc}(0)|^2\rho_1\,,  \nonumber\\
 \frac{\langle \Omc| P_2^q| \Omc\rangle}{2m_{\Omc}} &\simeq\Lambda_{QCD}\frac{\langle \Omc| O_1^q| \Omc\rangle}{2m_{\Omc}}&=&-6\Lambda_{QCD}|\Psi_{cq}^{\Omc}(0)|^2\rho_2 \,, \nonumber\\
 \frac{\langle \Omc| P_3^q| \Omc\rangle}{2m_{\Omc}} &\simeq\Lambda_{QCD}\frac{\langle \Omc| O_2^q| \Omc\rangle}{2m_{\Omc}}&=&-\Lambda_{QCD}|\Psi_{cq}^{\Omc}(0)|^2\rho_3\,,
 \label{eq:dim7Barmatel}
\end{alignat}
where ${\cal T}_c = \Lac, \Xi_c^+$ and $\Xi_c^0$ as before, and we expect the parameters $\rho_{1-3}$ to be of order 1.  The remaining dimension-seven matrix elements are, analogously to those of dimension-six, related by \Eqn\eqref{eq:BaryonMERelation1}, ie $ \tilde{P}_i^q(B) = - \tilde{B} P_i^q(B)$, where we will again take $\tilde{B}=1$ throughout \cite{NS1996}.  We note that our expectation for the matrix element of the operator $P_1$ differs from previous parametrizations in the literature \cite{GOP2003,GOP2004,Cheng18c}. We keep the explicit scaling with the light quark mass $m_q$, and use the quark model result for the matrix element $\langle P_1^q\rangle=m_q\langle O_2^q\rangle$, with $m_{u,d}$ set to $0 \GeV$ and $\overline{m}_s(\mu = 1.5 \GeV)=0.1 \GeV$. We also apply the relation $\langle P_{2,3}^q\rangle\simeq p_c\cdot p_q/m_c\langle O_{1,2}^q\rangle$, and estimate that $p_c\cdot p_q\sim m_c \Lambda_{QCD}$. 
For our central values we use $\Lambda_{QCD}=0.33\GeV$, evaluated for $n_f=3$ \cite{RunDec}.  This again differs from previous parametrizations \cite{GOP2003,GOP2004,Cheng18c}, but the resulting numerical difference between our parametrization of $P_{2,3}$ and that in previous literature is not much more than 20\%, and so falls within the range of uncertainties due to the wavefunctions in \eqref{eq:QuarkRel3}. 

Note, however, that there is currently no first-principles evaluation of dimension-seven four-quark matrix elements for baryons, and further scrutiny of this approach, and that in \cite{GOP2003,GOP2004,Cheng18c}, will be needed in the future, especially in view of the importance of such contributions in inclusive decay widths.

We finally evaluate the matrix elements of the Darwin operator, using \eqref{eq:Darwin4qbar} and \eqref{eq:Dim6BaryonBasis} with the matrix elements expressed in terms of baryon wavefunctions in the \ac{NRCQM}. Using the relations in \eqref{eq:BaryonMERelation1} with the fixed value $\tilde{B}=1$, we obtain
\begin{equation}
    2m_{\mathcal{B}_c}\rho_D^3(\mathcal{B}_c)=g_s^2\langle \mathcal{B}_c\vert  -\frac{1}{6} \Opsix{1}{q}+\frac{1}{3}\Opsix{2}{q}| {\cal B}_c\rangle+\mathcal{O}(1/m_c)\,,
    \label{eq:DarwinBary}
\end{equation}
where, with respect to \eqref{eq:Darwin4qbar}, we have expressed the result only in terms of $\Opsix{1,2}{q}$.
Using the relations in \Tab\ref{tab:matrixNRRel}, the wavefunctions from \eqref{eq:QuarkRel3}, and the value $g_s^2 \equiv 4\pi\als=4\pi$, leads to the values, already presented in \Tab\ref{tab:nonpertNScharm},
\begin{equation}
\begin{split}
    \rho_D(\Lac)&=0.04(1)\GeV^3\,,\qquad \rho_D(\Xicp)=0.05(2)\GeV^3\,,\qquad\\ 
    \rho_D(\Xico)&= 0.06(2)\GeV^3\,,\qquad\rho_D(\Omc)=0.06(2)\GeV^3\,,
    \label{eq:EstimateDarwinBary}
\end{split}
\end{equation}
including the uncertainties, which we conservatively set to $30\%$. 

\subsection{Final numerical predictions for baryons}
\label{sec:results_baryons}

\begin{figure}[th]
\centering
\includegraphics[scale=0.6]{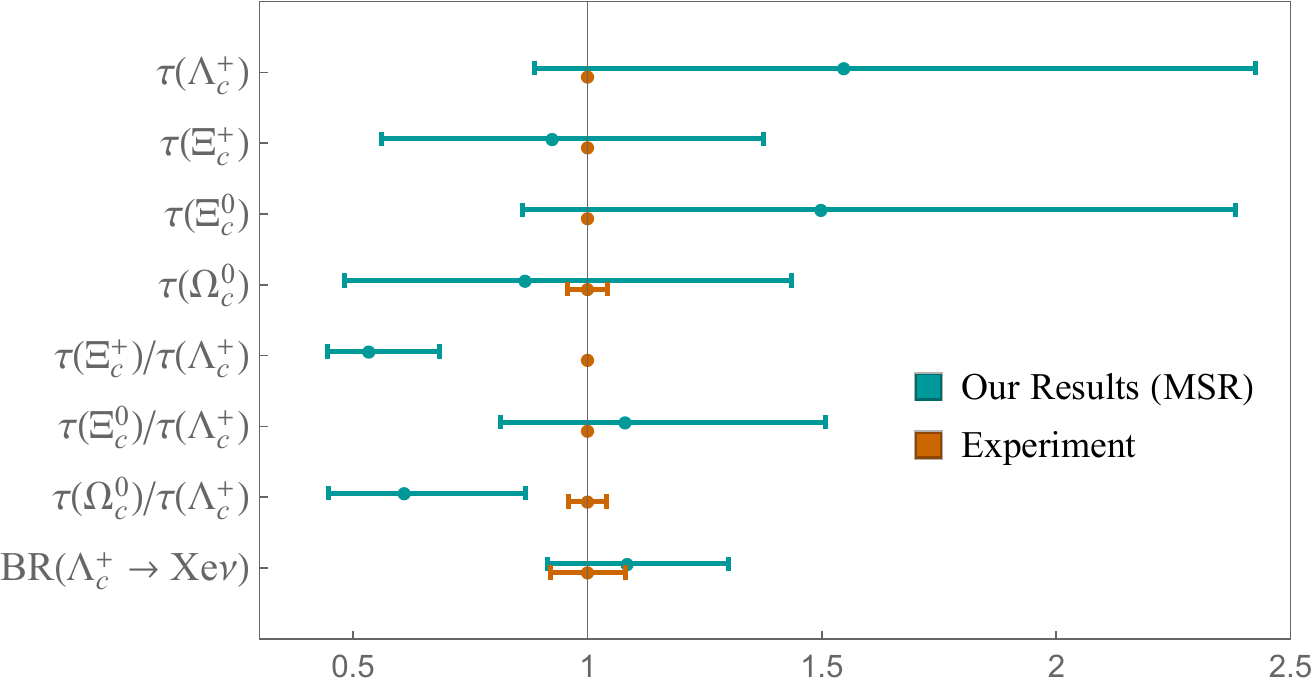}
\caption{\small Predictions for observables in the singly-charmed baryon sector, normalized to the corresponding experimental values, in the MSR scheme. Experimental values (lower of each pair) are in orange, with uncertainties provided when they are larger than $\sim1\%$. Our predictions are in bluegreen, with the uncertainties, from \Tab\ref{tab:BaryonnumericsMAIN}, added linearly. Tensions in the ratios $\Lifetime{\Xicp}/\Lifetime{\Lac}$ and $\Lifetime{\Omc}/\Lifetime{\Lac}$ are clearly visible, whereas predictions for the remaining observables are compatible with experimental values within uncertainties. A similar figure for meson observables is provided in \Fig7 of \cite{LenzNote:2021}.}
\label{fig:baryon_final}
\end{figure}

We present predictions for the following baryon observables: 
\begin{enumerate}[(i)]
    \item  the lifetimes of each baryon, $\Lifetime{\mathcal{B}_c}$;
    \item ratios compared with the experimental $\Lac$ lifetime, defined as
\begin{equation}
\frac{\tau(\mathcal{B}_c)}{\tau(\Lac)}\equiv \frac{1}{1+(\Gamma^{\text{th}}(\mathcal{B}_c)-\Gamma^{\text{th}}(\Lac))\tau^{\text{exp}}(\Lac)}\,;
\label{eq:RatioDefBar}
\end{equation}
\item inclusive semileptonic branching fractions involving the electrons in the final states, defined as
\begin{equation}
    BR(\mathcal{B}_c\to X e \nu)\equiv \Gamma(\mathcal{B}_c\to X e \nu)\,\tau^{\text{exp}}(\mathcal{B}_c)\,.
    \label{eq:BRDefBar}
\end{equation}
\end{enumerate}
As in the case of mesons, ratios are defined via the differences of the theoretical widths, which results in cancellations of the universal non-spectator terms, leading to reduction of theoretical uncertainties.  Our final predictions are presented in \Tab\ref{tab:BaryonnumericsMAIN}, while central values of individual contributions are given in \Tabs\ref{Tab:BaryonsCentral1} and \ref{Tab:BaryonsCentral2}, serving as an illustration of their relative sizes. In \Fig\ref{fig:baryon_final} we show a comparison of all our predictions in singly charm baryon sector, normalized to the corresponding experimental central values, similar to the one for charmed mesons in \cite{LenzNote:2021}. 

Central values are obtained using the \ac{NRCQM} expressions for the baryon wavefunctions given in \Eqn\eqref{eq:QuarkRel3}, with the remaining hadronic parameters given in \Tab\ref{tab:nonpertNScharm}. The corresponding uncertainties are estimated by allowing for $30\%$ variations around these values. The matrix elements of the dimension-seven operators involve the overall scaling coefficients $\rho_i$, which we set to 1, so that all hadronic uncertainties from the dimension-seven contribution follow from the corresponding uncertainties of the wavefunctions. 

\begin{table}[th]
\small
\centering
\begin{tabular}{|c|c|c|c|c||c|}
\hline
Observable & Pole & $\overline{\text{MS}}$ &  Kinetic & MSR & Experiment \\
\hline
$\Lifetime{\Lac}$/$10^{-13}$s  & $3.04^{+0.72+0.78}_{-0.51-0.62}$ & $3.50^{+0.87+1.26}_{-0.61-0.95}$  & $3.00^{+0.68+0.95}_{-0.49-0.74}$  & $3.12^{+0.73+1.05}_{-0.52-0.81}$ & $2.02\pm 0.03$\\
\hline
$\Lifetime{\Xicp}$/$10^{-13}$s & $4.25^{+0.79+0.93}_{-0.63-0.78}$ & $4.82^{+0.97+1.38}_{-0.76-1.22}$ & $4.03^{+0.73+1.16}_{-0.59-0.93}$ & $4.21^{+0.78+1.28}_{-0.63-1.02}$ & $4.56\pm 0.05$\\
\hline
$\Lifetime{\Xico}$/$10^{-13}$s & $2.31^{+0.66+0.52}_{-0.43-0.41}$ & $2.50^{+0.75+0.82}_{-0.49-0.63}$ &  $2.21^{+0.62+0.63}_{-0.41-0.50}$ & $2.28^{+0.65+0.69}_{-0.43-0.54}$ & $1.52\pm 0.02$\\
\hline
$\Lifetime{\Omc}$/$10^{-13}$s & $2.59^{+0.72+0.73}_{-0.49-0.50}$ & $2.62^{+0.79+1.05}_{-0.52-0.70}$ & $2.33^{+0.66+0.81}_{-0.44-0.56}$ & $2.37^{+0.68+0.88}_{-0.45-0.60}$ & $2.74\pm 0.12$\\
\hhline{|=|=|=|=|=||=|}
$\Lifetime{\Xicp}/\Lifetime{\Lac}$ & $1.23^{+0.24+0.10}_{-0.14-0.07}$ & $1.19^{+0.20+0.11}_{-0.12-0.07}$ & $1.21^{+0.23+0.11}_{-0.14-0.07}$ & $1.20^{+0.22+0.12}_{-0.13-0.07}$ & $2.25\pm 0.04$\\
\hline
$\Lifetime{\Xico}/\Lifetime{\Lac}$ & $0.83^{+0.16+0.02}_{-0.17-0.01}$ &
$0.81^{+0.15+0.03}_{-0.16-0.03}$&
$0.81^{+0.16+0.03}_{-0.17-0.03}$ & $0.81^{+0.29+0.03}_{-0.17-0.03}$&  $0.75\pm 0.02$\\
\hline
$\Lifetime{\Omc}/\Lifetime{\Lac}$ & $0.90^{+0.34+0.03}_{-0.19-0.03}$&
$0.84^{+0.28+0.05}_{-0.17-0.04}$&
$0.84^{+0.31+0.05}_{-0.18-0.04}$ & $0.83^{+0.30+0.05}_{-0.18-0.04}$&  $1.36\pm 0.06$\\
\hhline{|=|=|=|=|=||=|}
$BR(\Lac \to X e \nu)$/\% & $3.80^{+0.49+0.31}_{-0.39-0.41}$ &
$3.71^{+0.45+0.47}_{-0.35-0.36}$&
$4.42^{+0.48+0.25}_{-0.38-0.20}$ & $4.28^{+0.47+0.39}_{-0.37-0.30}$&  $3.95\pm 0.35$\\
\hline
$BR(\Xicp \to X e \nu)$/\% & $12.74^{+2.51+0.38}_{-2.24-1.00}$ &
$13.46^{+2.70+1.89}_{-2.42-1.69}$&
$15.20^{+2.80+1.12}_{-2.52-1.17}$ & $14.95^{+2.66+1.59}_{-2.45-1.50}$&  not measured \\
\hline
$BR(\Xico \to X e \nu)$/\% & $4.31^{+0.86+0.12}_{-0.77-0.33}$ &
$4.56^{+0.92+0.64}_{-0.83-0.58}$&
$5.13^{+0.96+0.38}_{-0.86-0.40}$ & $5.06^{+0.91+0.54}_{-0.84-0.51}$& not measured\\
\hline
$BR(\Omc \to X e \nu)$/\% & $7.59^{+2.49+0.00}_{-2.23-0.21}$ &
$10.40^{+2.98+2.28}_{-2.71-2.30}$&
$10.93^{+3.07+1.53}_{-2.81-1.75}$ & $11.19^{+3.01+1.94}_{-2.89-2.09}$&  not measured\\
\hline
\end{tabular}
\caption{\small Results for baryons in different mass schemes, including the lifetimes, lifetime ratios compared to the $\Lac$, and semileptonic branching fractions. The lifetime ratios are determined using \Eqn\eqref{eq:RatioDefBar}, and the semileptonic branching fractions using \Eqn\eqref{eq:BRDefBar}. The first and second errors correspond to hadronic and renormalization scale uncertainties, respectively. As for mesons, central values correspond to the scale choice $\mu=\mu_0=1.5\GeV$, while the scale uncertainties are estimated for fixed values of the hadronic parameters by varying the scale $\mu$ in the range $[1,3]\GeV$.}
\label{tab:BaryonnumericsMAIN}
\end{table}
Our values of baryon lifetimes turn out consistent, in all mass schemes and within sizeable theoretical uncertainties, with experimental measurements. The preferred value for $\Lifetime{\Omc}$ is larger than previous theoretical estimates \cite{Melic97c,Cheng1997c}, and favours the most recent LHCb results \cite{LHCbOmegac2018,LHCb2021Omega0}. However, our central value for the lifetime of $\Lac$ is $50\%$ larger than the measured value, and we similarly overestimate the lifetime of $\Xico$, although in both cases the measured lifetimes fall within our estimate of theoretical uncertainties. On the other hand, we observe a tension in the lifetime ratios $\tau(\Xicp)/\tau(\Lac)$ and $\tau(\Omc)/\tau(\Lac)$, both of which are smaller than the corresponding experimental values, which can be attributed to our larger-than-measured lifetime prediction of $\Lac$. We nevertheless can accommodate the newly-established hierarchy of experimental lifetimes
\begin{eqnarray}
\Lifetime{\Xico} < \Lifetime{\Lac}< \Lifetime{\Omc} < \Lifetime{\Xicp}\,,
\end{eqnarray}
although our results do not rule out $\tau(\Xico)/\tau(\Lac)>1$ or $\tau(\Omc)/\tau(\Lac)<1$ with certainty.

Our value for the semileptonic branching fraction $BR(\Lac\to Xev)$ is consistent with experiment. We also give predictions for the semileptonic branching fractions of the remaining baryons, which are yet to be measured experimentally.

As shown in \Tabs\ref{Tab:BaryonsCentral1} and \ref{Tab:BaryonsCentral2}, NLO contributions improve agreement of the $\Lac$ and $\Xi_c^+$ lifetimes with  experiment,\footnote{There was an earlier concern by experimentalists that the $\Omc \to \Xicp \pi^-$ decay might generate a substantial systematic error in measurements of $\Lifetime{\Xicp}$ and enhance the lifetime, which was theoretically not confirmed in \cite{BGMS2004}.} and moreover help to stabilize the dependence of our results on the mass scheme, which would otherwise be significant if only the \ac{LO} results were used in predicting lifetimes. It can be expected that higher order $\alpha_s$ contributions, with the caveat that the question of convergence of the $\als$ series beyond NLO must be addressed in charm decays, could improve the agreement still further. Likewise, the missing NLO corrections of dimension-seven spectator contributions might be important for improving this picture. 

Tables~\ref{Tab:BaryonsCentral1} and \ref{Tab:BaryonsCentral2} also show the relative sizes of different contributions to the decay widths. It can be seen, for example, that the $\Lac$ and $\Xico$ widths receive large contributions due to weak exchange, $\tilde{\Ga}_{6,\exc}$. If this contribution were to be enhanced by some unknown mechanism, then it is possible that the $\Lac$ and $\Xico$ lifetimes would be more consistent with their measured values. It is noteworthy that this contribution is related, by comparing \Figs\ref{fig:1-DiagramsMesons} and \Figs\ref{fig:1-DiagramsBaryons}, to the \ac{PI} contribution to the $D^+$ decay width, and appears at the same level in the \ac{CKM} hierarchy. In that case, as noted in \Sec\ref{sec:results_mesons}, the large \ac{PI} contribution drove the width towards unphysical negative values. It could be speculated that better theoretical control of the contributions arising from these topologies would alleviate tensions in the charmed hadron lifetimes, but such a simple resolution, without affecting other contributions, seems implausible. 

We now compare our results to the most recent previous study of singly charmed baryon lifetimes, which was performed in \cite{Cheng18c} and subsequently reviewed in \cite{Cheng2021RevI,Cheng2021RevII}. Firstly, we should stress that in these calculations, several contributions, that we have found to be numerically significant, were not included. The Darwin term was, at the time, unavailable, being first computed for charm decays in \cite{LenzNote:2021}, but represents a $\sim20\%$ enhancement of the non-spectator contributions at \ac{LO}. The author of \cite{Cheng18c} also chose to neglect NLO-QCD contributions, to both the dimension-three and -six contributions, both of which we find to be significant: the available \ac{NLO} contributions improve the fit in all cases, as well as reducing dependence of results on the mass scheme. 

In addition, the author of \cite{Cheng18c} enhanced wavefunctions of charmed baryons by an arbitrary coefficient $y=1.75$, presumably with the intention of bringing lifetimes of antitriplet baryons into agreement with experiment. After then obtaining a large lifetime for $\Omc$, and a negative semileptonic decay rate $\Gamma^{\text{SL}}(\Omc)$, a second arbitrary factor, designed to suppress large and negative $\Gamma_{7,\intp}$ contributions, was introduced. Our results show, however, that the predicted lifetime of $\Omc$ is compatible with the most recent experimental value without any need for such arbitrary factors. 
This is true whether or not the Darwin and $\als$ contributions are accounted for in the decay width, and may therefore be traced to our different parametrization of the $P_1$ matrix element (cf. discussion after \Eqn\eqref{eq:dim7Barmatel}) as compared with \cite{Cheng18c,GOP2003,GOP2004}.\footnote{In \cite{Cheng18c} it was argued that the factor $(m_{\Lac}^2-m_{[ud]}^2)/m_c^2-1$, where $m_{[ud]}$ is the effective mass of the light diquark pair in an antisymmetric spin state, is of order $m_q/m_c$, which was used to justify the choice of parametrization of the $P_1$ matrix element made in \cite{Cheng18c}. However, for the given value of $m_{[ud]}$ \cite{EFG2011} we find $((m_{\Lac}^2-m_{[ud]}^2)/m_c^2-1) \sim 1.1$, which is no longer of order $m_q/m_c$. It is perhaps relevant that the first analysis of the dimension-seven corrections \cite{GOP2003,GOP2004}, on which the approach in \cite{Cheng18c} was based, focused on the $b$ sector. In inclusive $b$ hadron decays, a potential overestimate of the size of the matrix elements $P_{1,2,3}$ is more tolerable in view of the $1/m_b^3$ suppression of all four-quark contributions.} That we have been able to accommodate the new $\Omc$ lifetime without needing to introduce any such arbitrary factors lends support to our approach.  We therefore do not find evidence for the claim advanced by \cite{Cheng2021RevI,Cheng2021RevII} that the \ac{HQE} fails to apply for $\Omc$. Further consideration of the dimension-seven matrix elements, and beyond to higher-order terms in the $1/m_c$ expansion, will be necessary in order to settle the question. 

\begin{figure}[t!]
    \centering
\includegraphics[width=\textwidth]{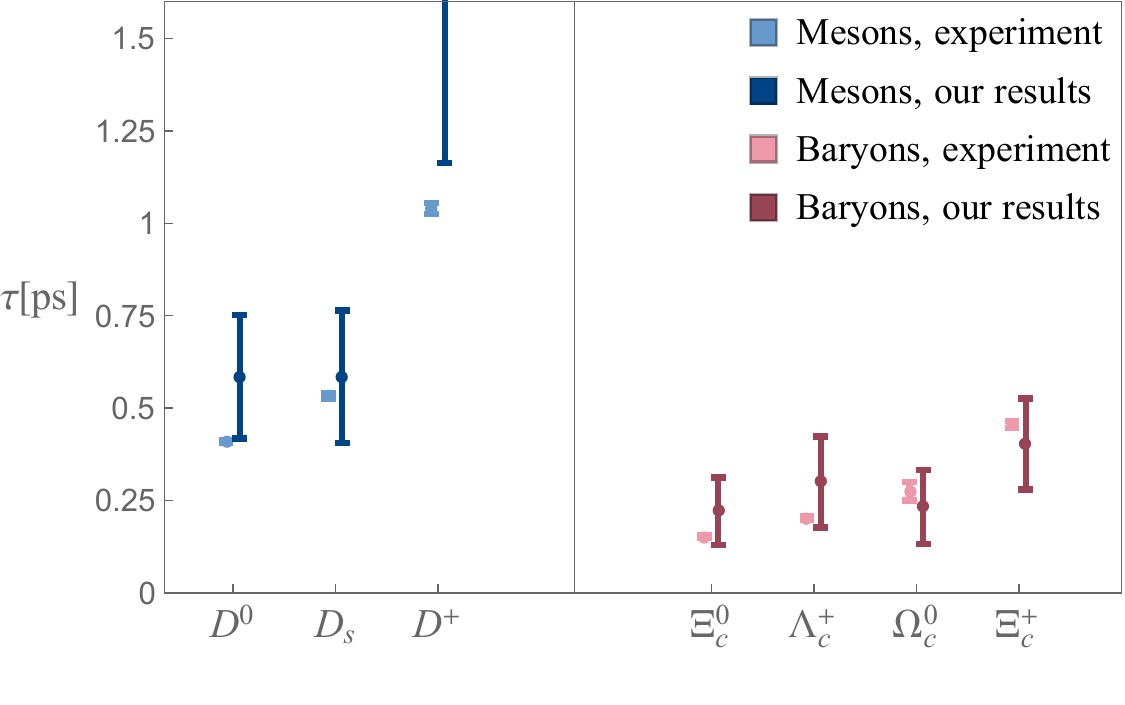}
\caption{\small Hierarchy of lifetimes, in$\pico$, of charmed mesons (left, in blue) and singly charmed baryons (right, in red). Experimental values, including the latest LHCb results for $\Omc$ and $\Xico$ \cite{LHCb2021Omega0,PDG2020}, are on the left of each pair of values; our predictions, using the kinetic scheme, are on the right.}
\label{fig:hierarchy_summary}
\end{figure}

\section{Conclusions and outlook}
\label{sec:conclusion}
In this work, we have predicted the lifetimes of singly charmed hadrons, with the main results presented in \Tabs\ref{tab:fin_mesons} and \ref{tab:fin_mesonsSL} for mesons, and \Tab\ref{tab:BaryonnumericsMAIN} for baryons. In particular, the results for baryons include the most complete set of contributions to date, and in this sense supersede previous theoretical predictions in \cite{Melic97c,Cheng18c}. A summary of our predictions, illustrating the lifetime hierarchies among charmed mesons and singly charmed baryons, is exhibited in \Fig\ref{fig:hierarchy_summary}.

While we agree with the recent results for charm meson observables in \cite{LenzNote:2021}, we generally disagree with the results and conclusions of the analysis of inclusive singly charmed hadron decays from \cite{Cheng18c,Cheng2021RevI,Cheng2021RevII}. This arises from several considerations. Firstly, the analysis therein focused on the \ac{LO} contributions (with the exception of the, at the time unavailable, Darwin contribution, which is itself sizeable), but it is apparent from our results (cf. the tables in \App\ref{app:detailedtables}) that the \ac{NLO} contributions to inclusive charm decays are large, and therefore cannot be neglected. Secondly, the analysis in \cite{Cheng18c} did not present  uncertainty estimates. In our analysis, however, supported by \cite{LenzNote:2021}, the uncertainties of the results  are large, so that no conclusions should be drawn merely by focusing on the central values obtained for a given choice of input parameters. Thirdly, the study in \cite{Cheng18c} made use of arbitrary parameters, in baryon lifetime predictions particularly, in order to compensate for missing contributions and with an eye on reproducing the experimental results. In light of the previous two points, this cannot be justified. Furthermore, we disagree with the claim made in \cite{Cheng18c,Cheng2021RevI,Cheng2021RevII} that the \ac{HQE} is not applicable to $\Omc$. Although the applicability of the \ac{HQE} in charm decays remains unclear, there is no indication in our results for such a strong statement only in inclusive $\Omc$ decays. 

A further reason that neglecting \ac{NLO} contributions, as was done in \cite{Cheng18c}, is insufficient to predict inclusive charm decays reliably is that these contributions significantly stabilize results between different mass schemes. By contrast, there is a large dependence on the choice of charm mass scheme at \ac{LO}, as the tables in \App\ref{app:detailedtables} show. However, the pole mass scheme exhibits huge $\alpha_s$ corrections, with signs that the picture may be even worse at higher orders in the $\als$ expansion. It is reasonable to expect a similar pattern in the full \ac{NLO} results, in particular the dimension-seven spectator contributions, supporting the long-standing argument that the pole mass should not be used for reliable predictions in inclusive charm decays. It will be important to examine up to at least NNLO contributions, consistently across both nonleptonic and semileptonic channels, in order to gain a more complete picture. This includes the missing NLO contributions to the dimension-five and -six two-quark contributions. Taken together, these could help to understand how best to address the issues surrounding the convergence of the $\als$ expansion, most prominently visible in the leading dimension-three term to semileptonic decays \cite{FSS2020}.

Comparing our results to experimental data, we see some signs of slight tension. For mesons, the decay width of $D^+$ can be driven to negative values by a sizeable Pauli interference contribution, and we find the ratio $\Lifetime{D_s}/\Lifetime{D^0}$ to be closer to unity than in experiment; both observations were also made in \cite{LenzNote:2021}. For baryons, while our predictions for the lifetimes are \emph{individually} compatible with experimental measurements, some of the ratios, particularly that of $\Lifetime{\Xicp}/\Lifetime{\Lac}$, are not. Such tension can be largely attributed to our central value of the $\Lac$ lifetime being an overestimate compared with experiment. Nevertheless, the picture elsewhere, with notably the $\Omc$ lifetime prediction being compatible with the new LHCb results \cite{LHCbOmegac2018,LHCb2021Omega0}, suggests that the application of the \ac{HQE} to inclusive charm decays remains plausible. In view of the large hadronic uncertainties, chiefly arising from spectator contributions, it is difficult to reach any firm conclusions on this point. Alternative approaches to arranging the \ac{HQE} for charm, such as those presented in \cite{FMV2019,MMP2021}, are also worth considering for the future, with in particular \cite{MMP2021} likely to be relevant in addressing the issue of a slowly-converging series of four-quark operators. 

Since spectator contributions are present in semileptonic decays, one could be tempted to try to extract the main ingredient of these contributions, the hadron wavefunctions, from precisely-measured semileptonic decay rates. Unfortunately, due to large uncertainties present in other matrix elements, this seems not to be possible at the moment, but the situation can improve once the non-spectator matrix elements and their higher-order contributions are known with larger precision. Lattice computations of all relevant matrix elements, both for baryons and mesons, with the latter complementing the HQET estimates in \cite{KLR2021}, could serve to address this. On the other hand, without any control over the uncertainties of such matrix elements, merely going to higher orders in the $1/m_c$ expansion is unlikely to improve the theoretical predictions meaningfully.

Finally, one should not forget that in charmed hadron decays there is a possibility of quark-hadron duality violation, one of the crucial elements in the calculation of the inclusive widths within the \ac{HQE}, which goes beyond the operator product expansion. This could bring non-negligible contributions $\sim e^{-m_c/\mu_{\text{hadr}}}$ that could be in particular notable in semileptonic decays, as discussed in \cite{Shifman1994,BDS1994,BBBF2003} for example, and recently analysed in the context of heavy meson decays in \cite{Umeeda2021}. More generally, it could be that the charm quark mass is not heavy enough for duality to set in. In any case, the concept of duality is worthy of further investigation. 

We urge for a new, independent measurement of the $\Omc$ lifetime, to be compared with the surprisingly large LHCb result,\footnote{There is a dedicated group at Belle II working on an independent determination of $\Lifetime{\Omc}$, and it will be interesting to see their results soon \cite{SchwartzComms}.} which radically changed the long-standing lifetime hierarchy of singly charmed baryons. Measurements of semileptonic branching fractions of other baryons, apart from $\Lac$, would likewise be more than welcome.

\subsection*{Acknowledgments}
We wish to thank Goran Duplan\v ci\' c, Matteo Fael, Daniel King, Alexander Lenz, Maria Laura Piscopo, Aleksey Rusov, Alan Schwartz, and Keri Vos for useful discussions, and Kenji Nishiwaki for collaborating at early stages of this work.
BM and JG have been supported by the Croatian Science Foundation (HRZZ) project “Heavy hadron decays and lifetimes” IP-2019-04-7094. Sponsorship has also been provided by the Alexander von Humboldt Foundation in the framework of the Research Group Linkage Programme, funded by the German Federal Ministry of Education and Research. BM would like to thank the organizers of the MIAPP workshop ``Charming Clues for Existence'' for very pleasant atmosphere they created, and to acknowledge support and hospitality of the Munich Institute for Astro- and Particle Physics (MIAPP), which is funded by the Deutsche Forschungsgemeinschaft (DFG, German Research Foundation) under Germany´s Excellence Strategy – EXC-2094 – 390783311, where the paper was finalized.

\appendix

\section{Numerical inputs}
\label{app:parameterinputs}

In this section we collect the numerical inputs used in determining the lifetime values. Values of the inputs are taken from PDG \cite{PDG2020} except where stated.

Table~\ref{tab:inputs} contains the input values of quark (in the $\msbar$ scheme) and lepton masses, alongside input parameters relevant for $\als$ running. Tables~\ref{tab:Dmes_masses} and \ref{tab:cBary_masses} contain the masses of mesons and baryons respectively, alongside the meson decay constants. Tables~\ref{tab:Mesres} and \ref{tab:Baryres} contain resonance masses useful for computing $\muG$ (see \Sec\ref{sec:muG}) and $\mukin$ (see \Sec\ref{sec:mukin}).

\begin{table}[ht]
	\centering
	\begin{tabular}{ |c|c| }
		\hline
		Parameter & Numerical value \\
			\hhline{|=|=|}
		$G_F$ & $1.1663787(6)\cdot 10^{-5}\,\GeV^{-2}$ \\
		\hline
		$\overline{m}_s(2 \GeV)$ & $0.093\GeV$ \\
		\hline
		$\overline{m}_c(\overline{m}_c)$ \cite{FLAG2019,FLAG2021} & $1.280(13)\GeV$  \\
		\hline
		$\overline{m}_b(\overline{m}_b)$ \cite{FLAG2019,FLAG2021} & $4.198(12)\GeV$  \\
		\hline
		$\alpha_s{(m_Z)}$ & $0.1180(7)$ \\
			\hline
		$m_Z$ & $91.1876\GeV$ \\
			\hline
		$m_{\mu}$ & $0.105658\GeV$ \\
		\hline
	\end{tabular}
	\caption{\small Values of input parameters used in the numerical analysis. Uncertainties in the final digit(s), which are neglected in this study, are given in brackets. 
	The value of $\overline{m}_c(\overline{m}_c)$ is the average by the Flavour Lattice Averaging Group (FLAG) \cite{FLAG2019,FLAG2021} of the lattice QCD results in \cite{ETM2014I,ETM2014II,HPQCD2014I,MILC2018,HPQCD2018}. The value of $\overline{m}_b(\overline{m_b})$ is the FLAG average of the lattice QCD results in \cite{HPQCD2014II,ETM2016,GMS2017}.}
	\label{tab:inputs}
\end{table}

The magnitudes of the CKM parameters are \cite{PDG2020}
\begin{align}
|V_{\textrm{CKM}}| = \begin{pmatrix}
|V_{ud}| & |V_{us}| & |V_{ub}| \\ 
|V_{cd}| & |V_{cs}| & |V_{cb}| \\
 |V_{td}| & |V_{ts}| & |V_{tb}|  \end{pmatrix}&= \begin{pmatrix}
0.97401 & 0.22650 & 0.00361 \\ 
0.22636 & 0.97320 & 0.04053 \\
 0.00854 & 0.03978 & 0.999172  \end{pmatrix}\,,
\end{align}
where we have neglected the corresponding uncertainties, which are negligible relative to other uncertainties in this paper.

\begin{table}[ht]
    \centering
    \begin{tabular}{|c|c|c|c|} \hline
         & $D^\pm$ & $D^0$ & $D_s$ \\ \hline
    $m_M$     & $1.86966(5)$ & $ 1.86484(5) $ & $ 1.96835(7)$ \\
    $f_M$     & $0.2120(7)$ & $0.2120(7)$ & $0.2499(5) $ \\\hline
    \end{tabular}
    \caption{\small Masses and decay constants of $D$ mesons in GeV, from the latest PDG \cite{PDG2020} and FLAG \cite{FLAG2019,ETM2014III,MILC2017} values. Uncertainties in the final digit(s), which are neglected in this study as they are dominated by other effects, are given in brackets.}
    \label{tab:Dmes_masses}
\end{table}

\begin{table}[ht]
    \centering
    \begin{tabular}{|c|c|c|c|c|} \hline
        & $\Lac$ & $\Xicp$ & $\Xico$ & $\Omc$   \\ \hline
    $m_H$    & $2.28646(14)$ & $2.46771(23)$ & $2.47044(28)$ & $2.6952(17)$   \\
    \hline
    \end{tabular}
    \caption{\small Masses of singly charmed baryons, in GeV, taken from PDG \cite{PDG2020}. Uncertainties in the final digit(s) are neglected in this study, but are given in brackets.}
    \label{tab:cBary_masses}
\end{table}

\begin{table}[ht]
    \centering
    \begin{tabular}{|c|c|c|c||c|c|} \hline
    & $D_0^*$ & $D_{\pm}^*$ & $D_s^*$ &$B^*$ & $B_s^*$ \\ \hline
   $m_{M^*}$ & $2.00685(5)$ & $2.01026(5)$ & $2.1122(4)$ & $5.32470(21)$ & $5.4154^{+0.0018}_{-0.0015}$ \\ \hline
    \end{tabular}
    \caption{\small Masses of the excited mesons in GeV, relevant for spectroscopic determinations, taken from PDG values \cite{PDG2020}. Uncertainties in the final digit(s) are neglected in this study, but are given in brackets, except for $B_s^*$ which currently has asymmetric uncertainty.}
    \label{tab:Mesres}
\end{table}

\begin{table}[ht]
\centering
\begin{tabular}{|c|c|c|c|c|c|} \hline
     & $\Om_c^{*} $ & $\Sigma_c $ & $\Sigma_c^{*} $ & $ \Xi_c^{'} $ & $\Xi_c^{*} $ \\ \hline
  $m_H$   & 2.7659(20) & 2.4529(4) & 2.5175(23) & 2.5782(5) & 2.64616(25) \\  \hline
\end{tabular}
\caption{\small Masses, in GeV, of (excited) baryon states relevant for determining the baryon wavefunctions in \Sec\ref{sec:matrix_baryons} using the representation in \Eqn\eqref{eq:QuarkRel3} \cite{Rosner1996}, taken from PDG \cite{PDG2020}. Uncertainties in the final digit(s) are neglected in this study, but are given in brackets.}
\label{tab:Baryres}
\end{table}

\subsection{$\als$ and Wilson coefficients}
For the evaluation of the two-loop running of the strong coupling constant $\als$ we use the function \verb+AsRunDec+ from the version $3.1$ of the software package \verb+RunDec+ \cite{RunDec, RunDecv3}. This function automatically performs the flavour decoupling across the $b$- and $c$ quark thresholds at the default values $\mu_c=1.5\GeV$ and $\mu_b=4.8\GeV$. We used $\alpha_s(M_Z)=0.1180$. For easier future comparisons, we list the values of $\als(\mu)$ for few reference values of the scale in \Tab\ref{tab:AlphasNumeric}. For our evaluations we used the five-loop values.
\begin{table}[ht]
	\centering
	\begin{tabular}{ |c|c|c| }
		  \hline 
		Scale & $\alpha_s(\mu)$ (two-loop) & $\alpha_s(\mu) $ (five-loop) \\
		\hline\hline
		$\mu=1.28\GeV$ & $0.371$ & $0.385$ \\
		\hline
		$\mu=1.50\GeV$ & $0.340$ & $0.349$ \\
		\hline
		$\mu=3.0\GeV$ & $0.251$ & $0.253$ \\
		\hline
	\end{tabular}
	\caption{\small Numerical values of $\alpha_s(\mu)$ evaluated with two-loop and five-loop running. For our numerical evaluations we use the values obtained with five-loop running. We used the initial value of the strong coupling constant $\alpha_s^{(5)}(m_Z=91.1876\GeV)=0.1180$. See the text for further details.}
	\label{tab:AlphasNumeric}
\end{table}
The coefficients of the $\als$ expansion of \Eqn\eqref{eq:WilsonCoeffsSchematic} depend on the Wilson coefficients $C_{1,2}$. In the case of a specific contribution for which the expansion in $\als$ is known beyond the leading order, we use the $C^{\NLO}_{1,2}(\mu)$-values for the evaluation of the leading coefficient $\mathcal{C}_n^{(0)}$, and the $C^{\LO}_{1,2}(\mu)$-values for evaluation of the next-to-leading coefficient $\mathcal{C}_n^{(1)}$. However, in the case that only the leading order coefficient $\mathcal{C}_n^{(0)}$ is known, we employ the LO results $C^{\LO}_{1,2}(\mu)$. We illustrate the running of $C_{1,2}$ for few reference renormalization scale points in \Tab\ref{tab:WilsonCoeffsNumeric}.
 \begin{table}[t]
	\centering
	\begin{tabular}{ |c|c|c||c|c| }
		\hline
		  Scale & $C_1^\text{\,LO}(\mu)$ & $C_2^\text{\,LO}(\mu)$ & $C_1^\text{\,NLO}(\mu)$& $C_2^\text{\,NLO}(\mu)$\\
		\hline\hline
		$\mu=1.28\GeV$ & $-0.52$ & $1.27$ & $-0.40$ & $1.20$ \\
		\hline
		$\mu=1.50\GeV$ & $-0.47$ & $1.24$ & $-0.36$ & $1.18$ \\
		\hline
		$\mu=3.0\GeV$ & $-0.32$ & $1.15$ & $-0.24$ & $1.10$ \\
		\hline
	\end{tabular}
	\caption{\small Numerical values of the Wilson coefficients $C_{1,2}$ at the leading order (LO) and next-to-leading-order (NLO) evaluated with the five-loop running of $\als(\mu)$.}
	\label{tab:WilsonCoeffsNumeric}
\end{table}

\section{Analytic forms for leading order coefficients in the HQE}
\label{app:c356}
Here we compile analytic expressions for the coefficients of the contributions to the inclusive decay width \eqref{eq:HQEsystematic}, up to leading order. 
\subsection{Non-spectator contributions}
Recall from \eqref{eq:decompcns} that the \ac{LO} coefficients $c_i$ have the general form
\begin{equation}
c_n^{(0)} = N_C C_1^2 \mathcal{K}^{(0)}_{n,11}  + 2 C_1 C_2 \mathcal{K}^{(0)}_{n,12}  + N_C C_2^2 \mathcal{K}^{(0)}_{n,22}  +\mathcal{K}^{(0)}_{n,\textrm{SL}} \, ,
\end{equation}
where the $\mathcal{K}_{n,ij}$ can then be written in terms of phase space functions. For $c_{3,\pi,G}$, the relevant functions appearing in \eqref{eq:c3atLO} and \eqref{eq:c5atLO} are $I_{0,1,2}(x,y,z)$, where, explicitly \cite{CPT1982,Koyrakh1993,Cheng18c,BUV1992HQE,BS1992I,BS1992II,BBSUV92HQE},
\begin{align}
I_0(0,0,0) &= 1 \,, \nonumber \\
I_0(x,0,0) &= I_0(0,x,0) = I_0(0,0,x) = 1 - 8x + 8x^3 - x^4 - 12x^2 \ln x  \,, \nonumber \\
I_0(x,x,0) &= \sqrt{1 - 4x} \left(1 - 14 x - 2x^2 - 12x^3 \right) + 24 x^2 (1-x^2) \ln \frac{1 + \sqrt{1 - 4x}}{1 - \sqrt{1 - 4x}}  \,, \nonumber \\
I_0(x,y,0) &= \sqrt{\la} \left(1 - 7(x+y) - 7(x^2+y^2) + x^3 + y^3 +xy\left(12 - 7(x + y) \right) \right)\nonumber \\ 
& \quad {} + 12 x^2(1 - y^2) \ln \frac{1 + v_x}{1 - v_x} +  12 y^2(1 - x^2) \ln \frac{1 + v_y}{1 - v_y} \,, 
\label{eq:phaseI0}
\end{align}
where $\la \equiv \la(1,x,y)$ is the K\"all\'en function, and $v_x$, $v_y$ can be interpreted as the maximal velocities of particles $x$ and $y$. They are given by
\begin{equation}
\la(a,b,c)  = a^2 + b^2 + c^2  - 2 (ab + bc + ca) \, , \quad v_x  = \frac{\sqrt{\la}}{1 + x - y} \, , \quad v_y  = \frac{\sqrt{\la}}{1 +y - x} \, .
\label{eq:Kallen}
\end{equation}
In the limit $ y = x$, then $v_x = v_y = \sqrt{1 - 4x}$. The  function $I_1$ is related to $I_0$ by 
\begin{align}
I_1(x,y,0) &= \frac{1}{2}\left(2 -  x \frac{\partial}{\partial x} -  y \frac{\partial}{\partial y} \right) I_0(x,y,0) \, , \nonumber \\
I_1(x,x,0) &= \frac{1}{2}\left(2 -  x \frac{d}{d x} \right) I_0(x,x,0)  \, , \nonumber \\
I_1(x,0,0) &= \frac{1}{2}\left(2 -  x \frac{d}{d x} \right) I_0(x,0,0) \, , 
\label{eq:phaseI1}
\end{align}
whereas $I_2$ is new:
\begin{align}
I_2(x,0,0) &= (1 - x)^3 \,, \nonumber \\
I_2(x,x,0) &= \sqrt{1 - 4x} \left(1 + \frac{1}{2}x + 3x^2 \right) - 3x (1 - 2x^2) \ln \frac{1+ \sqrt{1 - 4x}}{1- \sqrt{1 - 4x}} \, . 
\label{eq:phaseI2}
\end{align}
To the best of our knowledge, no explicit form for $I_2(x,y,0)$ is available, but fortunately for charmed hadrons the contribution never arises, so the explicit forms above suffice in computing the decay width.

For the Darwin term, we give the results of \cite{LenzNote:2021}, translated to our notation and conventions. Starting again from \eqref{eq:decompcns}, we have 
\begin{align}
\mathcal{K}^{(0)}_{\rho,ij} &= \sum\limits_{q_1,q_2=d,s} |V_{cq_1} V_{uq_2}|^2 \mathcal{K}^{(q_1\bar{q}_2)}_{\rho,ij} \,, \nonumber \\ 
\mathcal{K}^{(0)}_{\rho,\text{SL}} &=  |V_{cs}|^2 ( \cdarwinlepc{s}{e} + \cdarwinlepc{s}{\mu}) +  |V_{cd}|^2 ( \cdarwinlepc{d}{e} + \cdarwinlepc{d}{\mu}) \,.
\end{align}
Note that the Darwin operator mixes with the four-quark operators under operator renormalization \cite{FMV2019,LPR2020}. In the results below, this gives rise to terms proportional to $\log \mu_0^2/m_c^2$, where $\mu_0$ is the renormalization scale for the operators. In our results, we will take $\mu_0 = 1.5 \GeV$ throughout.

The LO coefficients for nonleptonic decays are \cite{LenzNote:2021}
\begin{align}
\cdarwinc{1}{1}{d}{d} &= \cdarwinc{2}{2}{d}{d} = 6 + 8 \ln \frac{\mu_0^2}{m_c^2} \,, \qquad \cdarwinc{1}{2}{d}{d} = - \frac{34}{3} \,,  \\
\cdarwinc{1}{1}{d}{s} &= \cdarwinc{2}{2}{d}{s} = \frac{2}{3}(1- x_s) \left[9 + 11 x_s - 25 x_s^2 +5 x_s^3 - 12 x_s^2 \ln x_s  - 24 (1-x_s^2) \ln (1 - x_s) \right] \nonumber \\
& \qquad \qquad {} + 8 (1- x_s)(1-x_s^2) \ln \frac{\mu_0^2}{m_c^2} \,,  \displaybreak[0]\nonumber\\
\cdarwinc{1}{2}{d}{s} &= -\frac{2}{3}\left[17 -26 x_s  + 18 x_s^2 -38 x_s^3 +5 x_s^4 + 12 x_s(5 + 2x_s - 2x_s^2) \ln x_s \right. \nonumber \\ & \quad \left. {} + 48 (1- x_s)(1-x_s^2) \ln (1 - x_s) \right]  - 16 x_s(1+ x_s -x_s^2) \ln \frac{\mu_0^2}{m_c^2} \,,  \displaybreak[0]\\
\cdarwinc{1}{1}{s}{d} &= \frac{2}{3}(1-x_s)\left[9 + 11 x_s - 25 x_s^2 +5 x_s^3 - 12 x_s^2 \ln x_s  - 24 (1-x_s^2) \ln (1 - x_s) \right] \nonumber \\
& \quad  {} + 8 (1- x_s)(1-x_s^2) \ln \frac{\mu_0^2}{m_c^2} \,, \nonumber \displaybreak[0]\\
\cdarwinc{1}{2}{s}{d} &= -\frac{2}{3} \left[ 17 - 50 x_s + 90x_s^2 - 54x_s^3 +5x_s^4 +12 x_s^2 (3-x_s) \ln x_s - 24 (1-x_s)^3 \ln (1-x_s) \right] \nonumber \\
& \quad {} + 8 x_s (3-3x_s+x_s^2) \ln \frac{\mu_0^2}{m_c^2} \,,  \displaybreak[0]\nonumber \\
\cdarwinc{2}{2}{s}{d} &= \frac{2}{3} \left[9 - 16 x_s - 12x_s^2 + 16x_s^3 - 5x_s^4 \right]+ 8 \ln \frac{\mu_0^2}{m_c^2} \,,  \displaybreak[0]\\
\cdarwinc{1}{1}{s}{s} &= \frac{2}{3} \left[ \vphantom{\frac{1 + \sqrt{1- 4x_s}}{1- \sqrt{1- 4x_s}}}\sqrt{1-4x_s} \left(9 + 22 x_s - 34 x_s^2 -60x_s^3 + 24 \ln x_s - 24 \ln (1-4x_s) \right) \right. \nonumber \\
& \left.\quad {} + 24 \left(1 - 2 x_s - x_s^2 - 2 x_s^3 - 5x_s^4 \right) \ln \frac{1 + \sqrt{1- 4x_s}}{1- \sqrt{1- 4x_s}}\right] + 8 \sqrt{1-4x_s} \ln \frac{\mu_0^2}{m_c^2} \,,  \displaybreak[0]\nonumber \\
\cdarwinc{1}{2}{s}{s} &= \frac{2}{3} \left[ \vphantom{\frac{1 + \sqrt{1- 4x_s}}{1- \sqrt{1- 4x_s}}} \sqrt{1- 4 x_s} \left(- 33 + 46 x_s - 106 x_s^2 -60x_s^3 + 24 \ln x_s - 24 \ln (1-x_s) \right) + 4 (1-x_s)^2 (4 - x_s) \right. \nonumber \\
& \quad\left. {} +  12(1-x_s)^3 \ln x_s   + 12 \left(3 - 2 x_s + 4x_s^2 - 16 x_s^3 - 10x_s^4 \right) \ln \frac{1 + \sqrt{1- 4x_s}}{1- \sqrt{1- 4x_s}} \right] \nonumber \\
& \quad {} - 8 \left(1 - \sqrt{1 - 4 x_s} - 3 x_s + x_s^2 - x_s^3 \right) \ln \frac{\mu_0^2}{m_c^2} \,, \nonumber  \displaybreak[0] \\
\cdarwinc{2}{2}{s}{s} & = \frac{2}{3} \left[  \vphantom{\frac{1 + \sqrt{1- 4x_s}}{1- \sqrt{1- 4x_s}}}\sqrt{1-4x_s}\left(17 + 8 x_s - 22 x_s^2 - 60 x_s^3\right) -4 (2 - 3 x_s + x_s^3) \right. \nonumber \\
& \quad \left. {} - 12 \left(1 -  x_s  -2 x_s^2 + 2 x_s^3 + 10x_s^4 \right) \ln \frac{1 + \sqrt{1- 4x_s}}{1- \sqrt{1- 4x_s}} - 12 (1-x_s)(1-x_s^2) \ln x_s\right] \nonumber \\
& \quad  {} + 8 (1- x_s)(1-x_s^2) \ln \frac{\mu_0^2}{m_c^2} \,. 
\end{align}
For semileptonic decays, the LO coefficients are
\begin{align}
\cdarwinlepc{d}{e} &=  6 + 8 \ln \frac{\mu_0^2}{m_c^2} \,,  \nonumber \\
\cdarwinlepc{d}{\mu} &= \frac{2}{3}(1- x_\mu) \left[9 + 11 x_\mu - 25 x_\mu^2 +5 x_\mu^3 - 12 x_\mu^2 \ln x_\mu  - 24 (1-x_\mu^2) \ln (1 - x_\mu) \right] \nonumber \\
& \qquad \qquad {} + 8 (1- x_\mu)(1-x_\mu^2) \ln \frac{\mu_0^2}{m_c^2} \,,  \displaybreak[0]\nonumber\\
\cdarwinlepc{s}{e} &= \frac{2}{3} \left[9 - 16 x_s - 12x_s^2 + 16x_s^3 - 5x_s^4 \right]+ 8 \ln \frac{\mu_0^2}{m_c^2} \,, \nonumber \displaybreak[2]\\
\cdarwinlepc{s}{\mu} & = \frac{2}{3} \left[  \vphantom{\frac{1 + \sqrt{1- 4x_s}}{1- \sqrt{1- 4x_s}}}\sqrt{1-4x_s}\left(17 + 8 x_s - 22 x_s^2 - 60 x_s^3\right) -4 (2 - 3 x_s + x_s^3) \right. \nonumber \\
& \quad \left. {} - 12 \left(1 -  x_s  -2 x_s^2 + 2 x_s^3 + 10x_s^4 \right) \ln \frac{1 + \sqrt{1- 4x_s}}{1- \sqrt{1- 4x_s}} - 12 (1-x_s)(1-x_s^2) \ln x_s\right] \nonumber \\
& \quad  {} + 8 (1- x_s)(1-x_s^2) \ln \frac{\mu_0^2}{m_c^2} \,, 
\end{align}
where the approximation $m_\mu \approx m_s$ has been made in the last expression, as the result for the semileptonic decay $c \to s \mu^+ \nu_\mu$ is not yet available. The \ac{SL} contributions can also be obtained from the \ac{NL} contributions by the replacement rules $\NC \to 1, \,C_1 \to 0, \, C_2 \to 1$, and $x_s \to x_\mu$ as appropriate.

\subsection{Spectator contributions}
 The leading-order expressions for the four-quark operators, the spectator contributions, are provided below. In many cases these have been known for a long time, but we bring the results into a symmetric form and present results for arbitrary masses in the loop, thus unifying previous presentations.
 
 The leading-order dimension-six results in the meson basis are (eg \cite{LenzRauh2013,Cheng18c,LenzNote:2021})
\begin{align}
\widehat{\Ga}^q_{6,\text{WA}}(x_1, x_2)&=\frac{\Ga_0}{2m_M}\frac{16\pi^2\sqrt{\la}}{m_Q^3}\Bigg\{\Big[\frac{1}{\NC}\big((x_1-x_2)^2+x_1+x_2-2\big)(C_1+\NC C_2)^2\Big]\Opsixprime{1}{q} \nonumber\\
&-\Big[\frac{2}{\NC} \big(2(x_1-x_2)^2-x_1-x_2-1\big)(C_1+\NC C_2)^2\Big]\Opsixprime{2}{q} \nonumber\\
&+2\Big[\big((x_1-x_2)^2+x_1+x_2-2\big)C_1^2\Big] \OpsixTprime{1}{q}-4\Big[\big(2(x_1-x_2)^2-x_1-x_2-1\big)C_1^2\Big]\OpsixTprime{2}{q}\Bigg\}\,, \nonumber \\
    \widehat{\Ga}^q_{6,\text{PI}}(x_1,x_2)&=\frac{\Gamma_0}{2m_M}\frac{16\pi^2\sqrt{\la}}{m_Q^3}\Big\{\Big[\frac{6(1-x_1-x_2)}{\NC}(C_1^2+2\NC C_1 C_2+C_2^2)\Big]\Opsixprime{1}{q}\nonumber \\
    &+\Big[4\NC(1-x_1-x_2)(C_1^2+C_2^2)\Big]\OpsixTprime{1}{q}\Big\}\,, \nonumber\\
   \nonumber  \\
    \widehat{\Ga}^q_{6, \text{WE}}(x_1,x_2)&=\widehat{\Ga}^{q\prime}_{6, \text{WA}}(x_1, x_2)\big|_{C_1\longleftrightarrow C_2}\,, \nonumber \\
    \nonumber \\
    \widehat{\Ga}^{q,\text{SL}}_{6, \text{WA}}(x_\ell,0)&=\widehat{\Ga}^{q}_{6, \text{WA}}(x_\ell,0)\big|_{C_1 \to 0, C_2 \to 1, N_C \to 1}\,,
\label{eq:explicitDim6Mes}
\end{align}
where $\la \equiv\lambda(1,x_1,x_2)$ is the K\"all\'en function, defined in \Eqn\eqref{eq:Kallen}, and $x_a=m_{a}^2/m_c^2$, where $a = 1,2,\ell$, and  $m_{q_{1,2}}$ and $m_\ell$ denote the masses of the light quarks and leptons in the loop, see \Fig\ref{fig:1-DiagramsMesons}.
Note that the right-hand side of the \ac{WE}-\ac{WA} relation also involves appropriate replacement of the light quark flavour. The above expressions are supplemented with the $\als$ corrections \cite{CFLM2001,BBGLN2002,FLMT2002,LenzRauh2013}, which are here not shown explicitly, but are taken into account in our evaluations of the lifetimes. The results below are given in the HQET basis of operators; the equivalent expressions in the QCD basis can be recovered by replacing $\Opsixprime{i}{q} \to \Opsix{i}{q}$ and $\OpsixTprime{i}{q} \to \OpsixT{i}{q}$.

The equivalent expressions for the dimension-seven contribution are \cite{LenzRauh2013,Cheng18c,LenzNote:2021}
\begin{align}
    \widehat{\Ga}^q_{7,\text{WA}}&=\frac{\Gamma_0}{2m_M}\frac{16\pi^2}{m_c^4}\sqrt{\lambda}\frac{\big(C_1+N_C C_2)^2}{N_C}\bigg\{\bigg[(x_1-x_2)^2+x_1+x_2-2\bigg]\OpsevenRprime{1}{q}\nonumber\\
    &-2\bigg[1-2(x_1-x_2)^2+x_1+x_2\bigg](\OpsevenPprime{1}{q}+{\OpsevenPprime{1}{q}}^\dagger-\mathcal{R}_2^q)\nonumber\\
    &{}+\frac{2}{\lambda}\bigg[(x_1+x_2-1)\Big((x_1-x_2)^2+x_1+x_2-2\Big)+\lambda\Big(2(x_1-x_2)^2+x_1+x_2\Big)\bigg]\OpsevenPprime{2}{q}\nonumber\\
    &{}+\frac{4}{\lambda}\bigg[(1-x_1-x_2)\Big(\lambda+(x_1-x_2)^2+x_1+x_2-2\Big)+\lambda\Big(1+2x_1+2x_2-6(x_1-x_2)^2\Big)\bigg]\OpsevenPprime{3}{q}\bigg\}\nonumber\\
    &{} +
    \bigg\{P_i^q\rightarrow S_i^q, \mathcal{R}^q_i\rightarrow \mathcal{U}^q_i,\frac{(C_1+N_C C_2)^2}{N_C}\rightarrow 2C_1^2\bigg\}\,,\nonumber \\
    \widehat{\Ga}^q_{7,\text{PI}}&=\frac{\Gamma_0}{2m_M}\frac{16\pi^2\sqrt{\la}}{m_Q^4}\frac{C_1^2+2 \NC C_1 C_2+C_2^2}{\NC}\bigg\{\Big[\frac{12\left((1-x_1-x_2)^2+(x_1+x_2)\la\right)}{\la}\Big]P^q_2\nonumber\\
    &+6(1-x_1-x_2)\OpsevenRprime{1}{q}\Big \}\nonumber\\
    &+\bigg\{\OpsevenPprime{2}{q}\rightarrow
    \mathcal{S}^q_2\,, \OpsevenRprime{1}{q}\rightarrow \mathcal{U}^q_1\,,\frac{C_1^2+2\NC C_1 C_2+C_2^2}{\NC}\rightarrow 2(C_1^2+C_2^2)\bigg\}\,, \nonumber \\ 
    \nonumber\\
     \widehat{\Ga}^q_{7,\text{WE}}&=  \widehat{\Ga}^{q'}_{7,\text{WA}}\big|_{C_1\longleftrightarrow C_2}\,, \nonumber \\
     \nonumber\\
    \widehat{\Ga}^{q,\text{SL}}_{7, \text{WA}}(x_\ell,0)&=\widehat{\Ga}^{q}_{7, \text{WA}}(x_\ell,0)\big|_{C_1 \to 0, C_2 \to 1, N_C \to 1}\,, \label{eq:explicitDim7Mes}
\end{align}
with the same notation as in \Eqn\eqref{eq:explicitDim6Mes} and the $(x_1,x_2)$ dependence suppressed. To switch between the HQET and QCD bases, one replaces the operators $\OpsevenPprime{i}{q} \to \OpsevenP{i}{q}$ and $\OpsevenRprime{i}{q} \to 0$. Note also that, in the HQET basis, additional non-local matrix elements are generated, but since these are reabsorbed by replacing the static decay constant $F_M$ with the physical decay constant $f_M$, cf. \App\ref{app:HQEtoQCD} and \cite{Neubert1992,Neubert1993,LenzNote:2021,KM1992}, we do not explicitly present them here.

The equivalent expressions to \Eqns\eqref{eq:explicitDim6Mes} and \eqref{eq:explicitDim7Mes} for baryons can be derived by applying the change of basis in \eqref{eq:colourid}, but are provided explicitly for convenience:
\begin{align}
\widehat{\Ga}^{q}_{6,\intp}(x_1, x_2)&=\frac{\Ga_0}{2m_M}\frac{16\pi^2\sqrt{\la}}{m_Q^3}\Bigg\{\Big[\big((x_1-x_2)^2+x_1+x_2-2\big)(2C_1 C_2+\NC C_2^2)\Big]O^q_1\nonumber \\
&-\Big[ 2\big(2(x_1-x_2)^2-x_1-x_2-1\big)(2C_1 C_2+\NC C_2^2)\Big]O^q_2\nonumber \\
&+\Big[\big((x_1-x_2)^2+x_1+x_2-2\big)C_1^2\Big] \tilde{O}^q_1-2\Big[\big(2(x_1-x_2)^2-x_1-x_2-1\big)C_1^2\Big]\,\tilde{O}^q_2\Bigg\}\,, \nonumber \\
\widehat{\Ga}^q_{6,\exc}(x_1,x_2)&=\frac{\Gamma_0}{2m_B}\frac{16\pi^2}{m_Q^3} \left(2 \NC \sqrt{\la} (1-x_1-x_2) \right)\Big\{\Big[2 C_1 C_2\Big]O^q_1+\Big[C_1^2+C_2^2\Big]\tilde{O}^q_1\Big\}\,,\nonumber \\
\nonumber\\
\widehat{\Ga}^{q}_{6,\intm}(x_1,x_2)&=\widehat{\Ga}^{q}_{6,\intp}(x_1, x_2)\big|_{C_1\longleftrightarrow C_2}\,,\nonumber \\
\nonumber\\
\widehat{\Ga}^{q,\text{SL}}_{6,\intp}(x_\ell,0)&=\widehat{\Ga}^{q}_{6,\intp}(x_\ell,0)\big|_{C_1 \to 0, C_2 \to 1, N_C \to 1}\,,
\label{eq:explicitDim6Bar}
\end{align}
and for dimension-seven,
\begin{align}
    \widehat{\Ga}^{q}_{7,\intp}&=\frac{\Ga_0}{2m_M}\frac{16\pi^2\sqrt{\la}}{m_Q^4}(2 C_1 C_2 + N_C C_2^2) \bigg\{2 \left[2(x_1- x_2)^2  - x_1 -x_2 -1\right] \left(P_1^q +{P_1^{q}}^{\dagger}\right) \nonumber \\
    &{}+\frac{2}{\lambda}\bigg[(x_1+x_2-1)\Big((x_1-x_2)^2+x_1+x_2-2\Big)+\lambda\Big(2(x_1-x_2)^2+x_1+x_2\Big)\bigg]\OpsevenP{2}{q}\nonumber \\
    &{}+\frac{4}{\lambda}\bigg[(1-x_1-x_2)\Big(\lambda+(x_1-x_2)^2+x_1+x_2-2\Big)+\lambda\Big(1+2x_1+2x_2-6(x_1-x_2)^2\Big)\bigg]\OpsevenP{3}{q}\bigg\}\nonumber \\
    &{} + \bigg\{\OpsevenP{i}{q} \to \OpsevenPt{i}{q}, (2 C_1 C_2 + N_C C_2^2) \to C_1^2 \bigg\} \,, \nonumber \\
    \widehat{\Ga}^q_{7,\exc}&=\frac{\Gamma_0}{2m_B}\frac{16\pi^2}{m_Q^4} \Big[\frac{12\left((1-x_1-x_2)^2+(x_1+x_2)\la\right)}{\sqrt{\la}}\Big]\Big\{\Big[2 C_1 C_2\Big]P^q_2+\Big[C_1^2+C_2^2\Big]\tilde{P}^q_2\Big\}\,, \nonumber \\
    \nonumber\\
    \widehat{\Ga}^{q}_{7,\intm}&=\widehat{\Ga}^{q}_{7,\intp}\big|_{C_1\longleftrightarrow C_2}\,, \nonumber \\
    \nonumber\\
    \widehat{\Ga}^{q,\text{SL}}_{7,\intp}(x_\ell,0)&=\widehat{\Ga}^{q}_{7,\intp}(x_\ell,0)\big|_{C_1 \to 0, C_2 \to 1, N_C \to 1}\,,
\label{eq:explicitDim7Bar}
\end{align}
with again the same notation as in \Eqn\eqref{eq:explicitDim6Mes} and the $(x_1,x_2)$ dependence suppressed. Note that results throughout for baryons are presented in the QCD basis, since we do not consider HQET matrix elements for baryons.

\section{A note on the relation between the HQET and QCD four-quark matrix elements at $O(1/m_Q^3)$}
\label{app:HQEtoQCD}

Given that the basis of \ac{HQET} operators and \acs{QCD} operators are related, but are structurally different, it might well be asked how free is the choice to use one basis or the other, and what is the difference between each choice. In one sense, the answer is clear, because the two ought to be related up to higher-order corrections in the $1/m_c$ and $\als$ expansions. However, this is far from trivial: at the very least, it is important to be able to quantify this difference.

To illustrate the point, we consider spectator contributions up to dimension-seven in semileptonic decay width for mesons. Helicity suppression implies that, in the limit $m_{s,\mu} \to 0 $, $ \hat{\Ga}_{6+7,\text{SL}} \to 0$, with equality in the \ac{VIA}. At the operator level, this can be written in the QCD basis as
\begin{equation}
    \hat{\Ga}_{6+7,\text{SL}} = -\frac{G_F^2 |V_{cs}|^2 m_c^2}{6\pi}\frac{1}{2m_M}   \left( O_1 - O_2 + \frac{2}{m_c}(P_1 - P_2) + O(m_s^2,m_\ell^2) \right) \, ,
    \label{eq:helicityQCD}
\end{equation}
where we neglect corrections from $m_s$ and $m_\mu$. Using the QCD parametrizations within the \ac{VIA} \cite{NS1996}
\begin{alignat}{2}
    \langle D | O_1 | D \rangle  &= f_D^2 m_D^2  \,, \quad &
    \langle D | O_2 | D \rangle  &= f_D^2 m_D^2 \left(\frac{m_D}{m_c+m_q}\right)^2 \,, \nonumber \\
    \langle D | P_1 | D \rangle  &= -m_q f_D^2 m_D^2  \,, \quad  &
    \langle D | P_2 | D \rangle  &= - f_D^2 m_D^2 \frac{m_D^2- m_c^2-m_q^2}{2m_c}  \,, 
\end{alignat} 
we can see that this is indeed the case: the combination \eqref{eq:helicityQCD} vanishes in the limit $m_q \to 0$. Note that the parametrization of $O_2$ includes the factor $(m_D/(m_c+m_q))^2$, which should be set to one if dimension-seven contributions were excluded. The equivalent expression in the \ac{HQE} basis is 
\begin{align}
      \hat{\Ga}_{6+7,\text{SL}} =-\frac{G_F^2 |V_{cs}|^2 m_c^2}{6\pi}\frac{1}{2m_M}& \left( \mathcal{O}_1 - \mathcal{O}_2 + \frac{1}{m_c}(2\mathcal{P}_1 - 2\mathcal{P}_2 + \mathcal{R}_1 - \mathcal{R}_2 \right. \nonumber \\ & {} \left.+ \mathcal{M}_{\pi,1} - \mathcal{M}_{\pi,2} + \mathcal{M}_{G,1} - \mathcal{M}_{G,2}  ) + O(m_q^2) \right) \, ,
     \label{eq:helicityHQE}
\end{align}
where the $\mathcal{M}$ are non-local operators whose forms are given explicitly in \cite{Neubert1992,Neubert1993,LenzNote:2021} (and see also \cite{KM1992}). Inserting the parametrizations in \Eqns\eqref{eq:HQEdim6val} and \eqref{eq:dim7Mesmatelparam}, supplemented by the non-local parametrizations from appendix C in \cite{LenzNote:2021}, again shows that this cancels, but this time order-by-order in the $1/m_c$ expansion rather than mixing across dimensions, as was the case in the QCD basis. 

Moreover, applying the relation  \cite{Neubert1992,KM1992,LenzNote:2021} 
\begin{equation}
    F_D^2 \left(1 - \frac{\bar{\La}}{m_c} + \frac{2 G_1}{m_c} + \frac{12 G_2}{m_c}\right) = f_D^2 m_D + O(1/m_c^2) \, ,
\end{equation}
(where $G_{1,2}$ parametrize nonlocal, $1/m_c$ suppressed, matrix elements defined in appendix C in \cite{LenzNote:2021}), which is valid at a fixed scale $\mu = m_c$ and within the \ac{VIA}, then the two expressions \eqref{eq:helicityQCD} and \eqref{eq:helicityHQE} can be directly related by making the identification
\begin{align}
    \langle\mathcal{O}_1\rangle + \frac{1}{m_c}(\langle\mathcal{R}_1\rangle + \langle\mathcal{M}_{\pi,1}\rangle + \langle\mathcal{M}_{G,1}\rangle) \to \langle\mathcal{O}_1\rangle|_{F_D^2 \to f_D^2 m_D} \, , \nonumber \\
    \langle\mathcal{O}_2\rangle + \frac{1}{m_c}(\langle\mathcal{R}_2\rangle + \langle\mathcal{M}_{\pi,2}\rangle + \langle\mathcal{M}_{G,2}\rangle) \to \langle\mathcal{O}_2\rangle|_{F_D^2 \to f_D^2 m_D} + \frac{2}{m_c} \langle\mathcal{R}_2\rangle \,.
\end{align}
Finally, we may relate them to the full QCD basis by noting that 
\begin{equation}
    \langle\mathcal{O}_2\rangle + \frac{2}{m_c} \langle\mathcal{R}_2\rangle = f_D^2 m_D^2 \left(1+ 2 \frac{\bar{\La}-m_q}{m_c} \right) \approx f_D^2 m_D^2 \left(\frac{m_D}{m_c+m_q}\right)^2 = \langle O_2\rangle \,,
\end{equation}
where the second step is valid as usual up to higher-order corrections in the \ac{HQE}. This justifies the parametrizations employed in \ref{sec:mesonmateldefs}.

This re-absorption effect can be extended, in principle, to include higher-order terms arising from the HQE or the $\als$ expansion, although in the latter case the discussion is more complicated owing to operator mixing \cite{Neubert1993}. 

The discussion does not, however, naturally extend to baryons, not least because there is no factorization of 4-quark matrix elements in baryons and no information about the baryon matrix elements is available beyond QCD. The conjecture of \cite{BlokShifman93Review}, that such a reabsorption of dimension-seven non-local contributions in the $f_D$ decay constant does \emph{not} occur for the triplet ($\Lac$, $\Xicp$, $\Xico$), but \emph{would} occur in the $\Omc$, therefore remains to be tested. 
%
\section{Supplementary tables of results}
\label{app:detailedtables}

The following tables provide a detailed breakdown of contributions to meson and baryon decay widths, arranged so that it is possible to see the relative sizes of contributions entering at different orders in the $1/m_Q$ and $\als$ expansion. We present results in the pole and kinetic schemes, for illustrative purposes; the remaining two mass schemes employing in this paper, the MSR and $\msbar$ schemes, are not presented explicitly. Specifically:
\begin{itemize}
\item \Tab\ref{Tab:MesonsCentral1} presents contributions to decay widths of mesons in the pole mass scheme;
\item \Tab\ref{Tab:MesonsCentral2} presents contributions to decay widths of mesons in the kinetic mass scheme;
\item \Tab\ref{Tab:BaryonsCentral1} presents contributions to decay widths of baryons in the pole mass scheme;
\item \Tab\ref{Tab:BaryonsCentral2} presents contributions to decay widths of baryons in the kinetic mass scheme.
\end{itemize}
In all cases, only the central values are given, so it should be borne in mind that each contribution therefore comes with, potentially sizeable, uncertainties due to variations of the hadronic parameters or the scale $\mu$. 
The size of the uncertainties of the relevant contributions can be seen in tables~\ref{tab:fin_mesons}, \ref{tab:fin_mesonsSL}, and \ref{tab:BaryonnumericsMAIN} in the main text. 
\\
\clearpage
\begin{sidewaystable}[ht]
\centering
\begin{tabular}{|cc||c|c|c|c|c|c|c|c|c|c|c|c|c|c||c|c|}
\hline
  \multicolumn{2}{|c||}{} & ${\Ga_{3+5}}^{\text{NL}}$ & ${\Ga_{3+5}}^{\textrm{SL}}$ & $\Ga_\rho^{\text{NL}}$ &$\Ga_\rho^{\text{SL}}$ & $\tilde{\Ga}_{6,\textrm{WE}}$ & $\tilde{\Ga}_{6,\text{PI}}$ & $\tilde{\Ga}_{6,\text{WA}}$ & $\tilde{\Ga}_{6,\text{WA}}^{\text{SL}}$ & $\tilde{\Ga}_{7,\text{WE}}$ & $\tilde{\Ga}_{7,\text{PI}}$ & $\tilde{\Ga}_{7,\text{WA}}$ & $\tilde{\Ga}_{7,\text{WA}}^{\text{SL}}$ &$\Ga_{\text{tot}}^{\text{SL}}$& $\Ga_{\text{tot}}$ & 
  $\tau$ & $\tau_{\textrm{exp.}}$ \\\hline
  \multirow{2}{*}{$D^0$}   & LO & 1.05 &0.30& 0.19&0.05 &  0.05& 0.0& 0.0& 0.0& 0.0& 0.0& 0.0& 0.0&0.35& 1.64& 0.61 & \multirow{2}{*}{0.41} \\ \cline{2-17}  & NLO &1.31 &0.15 &0.19 &0.05 & 0.04 &-0.01 &-0.01 &-0.01 &0.0 &0.0 &0.0 &0.0 &0.19 &1.71 &0.59 & \\ 
  \hhline{|==|=|=|=|=|=|=|=|=|=|=|=|=|=|=||=|=|}
  \multirow{2}{*}{$D^+$}  & LO &1.05 &0.30 &0.19 &0.05&  0.0 &-1.87 &0.0 &0.0 &0.0 &1.14 &0.0 &0.0 &0.35 & 0.87 &1.15 &\multirow{2}{*}{1.04}\\ \cline{2-17}  & NLO &1.31 &0.15 &0.19 &0.05  & 0.0 &-2.90 &-0.01 &-0.01 &0.0 &1.14 &0.0 &0.0 &0.19 &-0.07 &- & \\
  \hhline{|==|=|=|=|=|=|=|=|=|=|=|=|=|=|=||=|=|}
  \multirow{2}{*}{$D_s^+$}   & LO & 1.06& 0.28&0.29 &0.08 &0.0 &-0.13 &-0.01 &0.0 &0.0 &0.11 &0.0 &0.0 &0.36 & 1.66 &0.60 &\multirow{2}{*}{0.53}\\ \cline{2-17}  & NLO &1.31 &0.13 &0.29 &0.08 &0.0 &-0.22 &0.0 &0.01 &0.0 &0.11 &0.0 &0.0 &0.22 &1.71 &0.58 & \\\hline
\end{tabular}
\caption{\small Central values of various contributions to decay widths of charmed mesons, in the \emph{pole} scheme for the charm quark mass, indicating also the \acf{LO} and \acf{NLO} contributions where available. Decay widths are given in units $\pico^{-1}$, and lifetimes in units of $\pico$. The NLO contributions to $\Gamma_\rho$ and $\tilde{\Gamma}_7$ are not known; for the corresponding NLO entries we repeat the LO results, for the reader's convenience. The \acf{SL} contributions involve both electron and muon channels. Note that, for the $D_s^+$ meson, the given value of $\tau^{exp}$ results after subtracting the decay width of the channel $D_s\to\tau\nu$, see \Eqn\eqref{eq:ModifiedDs0}. All hadronic matrix elements and other numerical inputs used to obtain these results correspond to those in \Sec\ref{sec:mesonmateldefs}.}
\label{Tab:MesonsCentral1}
\end{sidewaystable}
\clearpage
\begin{sidewaystable}[ht]
\centering
\begin{tabular}{|cc||c|c|c|c|c|c|c|c|c|c|c|c|c|c||c|c|}
\hline
  \multicolumn{2}{|c||}{} & ${\Ga_{3+5}}^{\text{NL}}$ & ${\Ga_{3+5}}^{\textrm{SL}}$ & $\Ga_\rho^{\text{NL}}$ &$\Ga_\rho^{\text{SL}}$ & $\tilde{\Ga}_{6,\textrm{WE}}$ & $\tilde{\Ga}_{6,\text{PI}}$ & $\tilde{\Ga}_{6,\text{WA}}$ & $\tilde{\Ga}_{6,\text{WA}}^{\text{SL}}$ & $\tilde{\Ga}_{7,\text{WE}}$ & $\tilde{\Ga}_{7,\text{PI}}$ & $\tilde{\Ga}_{7,\text{WA}}$ & $\tilde{\Ga}_{7,\text{WA}}^{\text{SL}}$  & $\Ga_{\text{tot}}^{\text{SL}}$ &  $\Ga_{\text{tot}}$ &
  $\tau$ & $\tau_{exp.}$ \\\hline
  \multirow{2}{*}{$D^0$}   & LO & 0.79 &0.20& 0.19& 0.05& 0.04& 0.0& 0.0& 0.0& 0.0& 0.0& 0.0& 0.0&0.25 & 1.28  & 0.78&\multirow{2}{*}{0.41} \\ \cline{2-17}  & NLO &1.28 &0.23 &0.19 &0.05 &0.04 &-0.01 &-0.01 &-0.01 &0.0 &0.0 &0.0 &0.0 &0.28 &1.77 &  0.56 &\\
  \hhline{|==|=|=|=|=|=|=|=|=|=|=|=|=|=|=||=|=|}
  \multirow{2}{*}{$D^+$}  & LO &0.79 &0.20 &0.19 &0.05 &0.0 &-1.66 &0.0 &0.0 &0.0 &1.08 &0.0 &0.0 & 0.25&0.65 &1.54 &\multirow{2}{*}{1.04}\\ \cline{2-17}  & NLO &1.28 &0.23 &0.19 &0.05 &0.0 &-2.89 &-0.01 &-0.01 &0.0 &1.08 &0.0 &0.0 & 0.28 &-0.07  &- & \\
   \hhline{|==|=|=|=|=|=|=|=|=|=|=|=|=|=|=||=|=|}
  \multirow{2}{*}{$D_s^+$} & LO &0.79 &0.19 &0.28 &0.08 &0.0 &-0.12 &-0.01 &0.0 &0.0 &0.10 &0.0 &0.0 &0.27 &1.31  & 0.76&\multirow{2}{*}{0.53}\\ \cline{2-17}  & NLO &1.29 &0.22 &0.28 &0.08 &0.0 &-0.22 &0.0&0.01 &0.0 &0.10 &0.0 &0.0 &0.30 &1.77  &0.56 & \\\hline
\end{tabular}
\caption{\small Central values of various contributions to decay widths of charmed mesons, in the \emph{kinetic} scheme for the charm quark mass, indicating also the \acf{LO} and \acf{NLO} contributions where available. Decay widths are given in units $\pico^{-1}$, and lifetimes in units $\pico$. See the caption of \Tab\ref{Tab:MesonsCentral1} for further details.}
\label{Tab:MesonsCentral2}
\end{sidewaystable}
\clearpage

\begin{sidewaystable}[ht]
\centering
\begin{tabular}{|cc||c|c|c|c|c|c|c|c|c|c|c|c|c|c||c|c|}
\hline
  \multicolumn{2}{|c||}{} & ${\Ga_{3+5}}^{\text{NL}}$ & ${\Ga_{3+5}}^{\textrm{SL}}$ &  $\Ga_\rho^{\text{NL}}$ &$\Ga_\rho^{\text{SL}}$ & $\tilde{\Ga}_{6,\exc}$ & $\tilde{\Ga}_{6,\intm}$ & $\tilde{\Ga}_{6,\intp}$ & $\tilde{\Ga}^{\text{SL}}_{6,\intp}$ & $\tilde{\Ga}_{7,\exc}$ & $\tilde{\Ga}_{7,\intm}$ & $\tilde{\Ga}_{7,\intp}$ & $\tilde{\Ga}^{\text{SL}}_{7,\intp}$  & $\Ga_{\text{tot}}^{\text{SL}}$& $\Ga_{\text{tot}}$& 
  $\tau$ & $\tau_{\textrm{exp.}}$ \\\hline
  \multirow{2}{*}{$\Lac$}   & LO & $0.90$ &$0.43$ &$0.15$ &$0.04$ & $1.04$& $-0.38$&$0.03$ &$0.02$ &$0.46$ &$0.11$ &$-0.01$ & $-0.01$& $0.48$& $2.78$&$3.59$ & \\ \cline{2-17}  & NLO & $1.15$& $0.32$& $0.15$&$0.04$ & $1.28$&$-0.25$ &$0.03$ &$0.01$ &$0.46$ &$0.11$ &$-0.01$ & $-0.01$&  $0.37$ & $3.29$&$3.04$ &2.02 \\
  \hhline{|==|=|=|=|=|=|=|=|=|=|=|=|=|=|=||=|=|}
  \multirow{2}{*}{$\Xicp$}   & LO & $0.89$& $0.42$&$0.18$&$0.05$  &$0.08$ &$-0.39$ &$0.81$ &$0.50$ &$0.04$ &$0.12$ &$-0.29$ &$-0.18$ & $0.80$ & $2.22$& $4.51$& \\ \cline{2-17}  & NLO &$1.14$ &$0.32$ &$0.18$ &$0.05$ &$0.09$ &$-0.26$ &$0.79$ &$0.37$ &$0.04$ &$0.12$ &$-0.29$ &$-0.18$ & $0.55$ &$2.36$ &$4.25$ &4.56\\
  \hhline{|==|=|=|=|=|=|=|=|=|=|=|=|=|=|=||=|=|}
  \multirow{2}{*}{$\Xico$}  & LO &$0.89$ &$0.42$ &$0.18$& $0.05$ &$1.15$ &$0$ &$0.84$ &$0.52$ &$0.52$ &$0$ &$-0.30$ &$-0.19$ & $0.81$ & $4.10$&$2.45$ & \\ \cline{2-17}  & NLO &$1.14$ &$0.32$ &$0.18$&$0.05$  &$1.42$ &$0$ &$0.82$ &$0.38$ &$0.52$ &$0$ & $-0.30$& $-0.19$& $0.56$ & $4.35$ &$2.31$ &1.52\\
  \hhline{|==|=|=|=|=|=|=|=|=|=|=|=|=|=|=||=|=|}
  \multirow{2}{*}{$\Omc$}  & LO & $0.98$& $0.34$&$0.23$&$0.06$  &$0.48$ &$0$ &$2.85$ &$1.77$ &$0.22$ &$0$ &$-1.50$ &$-0.94$ & $1.23$ &  $4.49$& $2.23$ & \\ \cline{2-17}  & NLO &$1.23$ &$0.20$ &$0.23$&$0.06$  &$0.59$ &$0$ &$2.54$ &$1.21$ &$0.22$ &$0$ &$-1.50$ &$-0.94$ & $0.54$ &$3.86$ &$2.59$ &  2.74\\ \hline
\end{tabular}
\caption{\small Central values of various contributions to decay widths of baryons, in the {\it pole} scheme for the charm quark mass, indicating also the \acf{LO} and \acf{NLO} contributions where available. Decay widths are given in units $\pico^{-1}$, and lifetimes in units $10^{-13}$s.  See the caption of \Tab\ref{Tab:MesonsCentral1} for further details.}
\label{Tab:BaryonsCentral1}
\end{sidewaystable}

\begin{sidewaystable}[ht]
\centering
\begin{tabular}{|cc||c|c|c|c|c|c|c|c|c|c|c|c|c|c||c|c|}
\hline
  \multicolumn{2}{|c||}{} & ${\Ga_{3+5}}^{\text{NL}}$ & ${\Ga_{3+5}}^{\textrm{SL}}$ &  $\Ga_\rho^{\text{NL}}$ &$\Ga_\rho^{\text{SL}}$ & $\tilde{\Ga}_{6,\exc}$ & $\tilde{\Ga}_{6,\intm}$ & $\tilde{\Ga}_{6,\intp}$ & $\tilde{\Ga}_{6,\intp}^{\text{SL}}$ & $\tilde{\Ga}_{7,\exc}$ & $\tilde{\Ga}_{7,\intm}$ & $\tilde{\Ga}_{7,\intp}$ & $\tilde{\Ga}^{\text{SL}}_{7,\intp}$ & $\Ga_{\text{tot}}^{\text{SL}}$ & $\Ga_{\text{tot}}$& 
  $\tau$ & $\tau_{\textrm{exp.}}$ \\\hline
  \multirow{2}{*}{$\Lac$}   & LO & $0.65$ &$0.31$ &$0.15$ &0.04 & $0.92$& $-0.34$&$0.03$ &$0.02$ &$0.44$ &$0.11$ &$-0.01$ & $-0.01$& 0.37 & $2.31$&$4.32$ & \\ \cline{2-17}  & NLO & $1.15$& $0.38$&$0.15$& 0.04 & $1.32$& $-0.29$&$0.03$ &$0.02$ &$0.44$ &$0.11$ &$-0.01$ & $-0.01$& 0.43 & $3.33$ &$3.00$ &2.02 \\
   \hhline{|==|=|=|=|=|=|=|=|=|=|=|=|=|=|=||=|=|}
  \multirow{2}{*}{$\Xicp$}   & LO &$0.65$ &$0.31$ &$0.18$& 0.05 &$0.07$ & $-0.35$&$0.72$ &$0.45$ &$0.04$ &$0.11$ &$-0.28$ &$-0.17$ & 0.64 & $1.77$& $5.65$& \\ \cline{2-17}  & NLO &$1.15$ &$0.37$ & $0.18$&0.05 &$0.10$ &$-0.30$ &$0.83$ &$0.40$ &$0.04$ &$0.11$ &$-0.28$ &$-0.17$ & 0.65 &$2.48$ &$4.03$ &4.56\\
   \hhline{|==|=|=|=|=|=|=|=|=|=|=|=|=|=|=||=|=|}
  \multirow{2}{*}{$\Xico$}  & LO & $0.65$&$0.31$ &$0.18$& 0.05 &$1.02$ &$0$ &$0.75$ &$0.47$ &$0.49$ &$0$ &$-0.29$ &$-0.18$ & 0.65 & $3.45$&$2.90$ & \\ \cline{2-17}  & NLO &$1.15$ &$0.37$ &$0.18$&0.05  &$1.47$ &$0$ &$0.86$ &$0.42$ &$0.49$ &$0$ &$-0.29$ &$-0.18$ & 0.66 & $4.53$&$2.21$ &1.52\\
   \hhline{|==|=|=|=|=|=|=|=|=|=|=|=|=|=|=||=|=|}
  \multirow{2}{*}{$\Omc$}  & LO &$0.72$ &$0.24$ &$0.22$& 0.06 &$0.43$ &$0$ &$2.53$ &$1.57$ &$0.21$ &$0$ &$-1.41$ &$-0.88$ & 0.99 & $3.69$& $2.71$& \\ \cline{2-16}  & NLO &$1.22$ &$0.29$ &$0.22$&0.06  &$0.61$ &$0$ &$2.66$ &$1.32$ &$0.21$ &$0$ &$-1.41$ &$-0.88$ & 0.78 &$4.30$ &$2.33$ & 2.74\\ \hline
\end{tabular}
\caption{\small Central values of various contributions to decay widths of baryons, in the {\it kinetic} scheme for the charm quark mass, indicating also the \acf{LO} and \acf{NLO} contributions where available. Decay widths are given in units $\pico^{-1}$, and lifetimes in units $10^{-13}$s.  See the caption of \Tab\ref{Tab:MesonsCentral1} for further details.}
\label{Tab:BaryonsCentral2}
\end{sidewaystable}

\clearpage

\bibliographystyle{JHEP}
\bibliography{ReferencesLifetimes_updated.bib}

\providecommand{\href}[2]{#2}\begingroup\raggedright\begin{thebibliography}{100}

\bibitem{LHCbOmegac2018}
{\scshape LHCb} collaboration, \emph{{Measurement of the $\Omega_c^0$ baryon
  lifetime}}, \href{https://doi.org/10.1103/PhysRevLett.121.092003}{\emph{Phys.
  Rev. Lett.} {\bfseries 121} (2018) 092003}
  [\href{https://arxiv.org/abs/1807.02024}{{\ttfamily 1807.02024}}].

\bibitem{LHCbcharmedLifetimes2019}
{\scshape LHCb} collaboration, \emph{{Precision measurement of the
  $\Lambda_c^+$, $\Xi_c^+$, and $\Xi_c^0$ baryon lifetimes}},
  \href{https://doi.org/10.1103/PhysRevD.100.032001}{\emph{Phys. Rev. D}
  {\bfseries 100} (2019) 032001}
  [\href{https://arxiv.org/abs/1906.08350}{{\ttfamily 1906.08350}}].

\bibitem{LHCb2021Omega0}
{\scshape LHCb} collaboration, \emph{{Measurement of the lifetimes of promptly
  produced $\Omega^{0}_{c}$ and $\Xi^{0}_{c}$ baryons}},
  \href{https://doi.org/10.1016/j.scib.2021.11.022}{\emph{Sci. Bull.}
  {\bfseries 67} (2022) 479}
  [\href{https://arxiv.org/abs/2109.01334}{{\ttfamily 2109.01334}}].

\bibitem{SELEXLambdac}
{\scshape SELEX} collaboration, \emph{{Precision measurements of the
  $\Lambda^+_c$ and $D^0$ lifetimes}},
  \href{https://doi.org/10.1103/PhysRevLett.86.5243}{\emph{Phys. Rev. Lett.}
  {\bfseries 86} (2001) 5243}
  [\href{https://arxiv.org/abs/hep-ex/0010014}{{\ttfamily hep-ex/0010014}}].

\bibitem{CLEOLambdac}
{\scshape CLEO} collaboration, \emph{{Measurement of the $\Lambda^+_c$
  lifetime}}, \href{https://doi.org/10.1103/PhysRevLett.86.2232}{\emph{Phys.
  Rev. Lett.} {\bfseries 86} (2001) 2232}
  [\href{https://arxiv.org/abs/hep-ex/0011049}{{\ttfamily hep-ex/0011049}}].

\bibitem{FOCUSLambdac2002}
{\scshape FOCUS} collaboration, \emph{{A high statistics measurement of the
  $\Lambda^+_c$ lifetime}},
  \href{https://doi.org/10.1103/PhysRevLett.88.161801}{\emph{Phys. Rev. Lett.}
  {\bfseries 88} (2002) 161801}
  [\href{https://arxiv.org/abs/hep-ex/0202001}{{\ttfamily hep-ex/0202001}}].

\bibitem{CLEOXic}
{\scshape CLEO} collaboration, \emph{{Measurement of the $\Xi^+_c$ lifetime}},
  \href{https://doi.org/10.1103/PhysRevD.65.031102}{\emph{Phys. Rev. D}
  {\bfseries 65} (2002) 031102}
  [\href{https://arxiv.org/abs/hep-ex/0110058}{{\ttfamily hep-ex/0110058}}].

\bibitem{FOCUSXicp2001}
{\scshape FOCUS} collaboration, \emph{{A new measurement of the $\Xi^+_c$
  lifetime}}, \href{https://doi.org/10.1016/S0370-2693(01)01322-3}{\emph{Phys.
  Lett. B} {\bfseries 523} (2001) 53}
  [\href{https://arxiv.org/abs/hep-ex/0110002}{{\ttfamily hep-ex/0110002}}].

\bibitem{FOCUSXic02002}
{\scshape FOCUS} collaboration, \emph{{A new measurement of the $\Xi^0_c$
  lifetime}}, \href{https://doi.org/10.1016/S0370-2693(02)02239-6}{\emph{Phys.
  Lett. B} {\bfseries 541} (2002) 211}
  [\href{https://arxiv.org/abs/hep-ex/0206069}{{\ttfamily hep-ex/0206069}}].

\bibitem{PDG2018}
{\scshape Particle Data Group} collaboration, \emph{{Review of Particle
  Physics}}, \href{https://doi.org/10.1103/PhysRevD.98.030001}{\emph{Phys. Rev.
  D} {\bfseries 98} (2018) 030001}.

\bibitem{FOCUSOmegac2003}
{\scshape FOCUS} collaboration, \emph{{Measurement of the $\Omega^0_c$
  lifetime}}, \href{https://doi.org/10.1016/S0370-2693(03)00388-5}{\emph{Phys.
  Lett. B} {\bfseries 561} (2003) 41}
  [\href{https://arxiv.org/abs/hep-ex/0302033}{{\ttfamily hep-ex/0302033}}].

\bibitem{SELEXOmegac}
{\scshape SELEX} collaboration, \emph{{Measurement of the $\Omega^0_c$
  Lifetime}},  \href{https://arxiv.org/abs/hep-ex/0701021}{{\ttfamily
  hep-ex/0701021}}.

\bibitem{Melic97c}
B.~Guberina and B.~Meli\'c, \emph{{Inclusive charmed baryon decays and
  lifetimes}}, \href{https://doi.org/10.1007/s100520050173}{\emph{Eur. Phys. J.
  C} {\bfseries 2} (1998) 697}
  [\href{https://arxiv.org/abs/hep-ph/9704445}{{\ttfamily hep-ph/9704445}}].

\bibitem{Cheng1997c}
H.-Y.~Cheng, \emph{{A Phenomenological analysis of heavy hadron lifetimes}},
  \href{https://doi.org/10.1103/PhysRevD.56.2783}{\emph{Phys. Rev. D}
  {\bfseries 56} (1997) 2783}
  [\href{https://arxiv.org/abs/hep-ph/9704260}{{\ttfamily hep-ph/9704260}}].

\bibitem{SV1985}
M.A.~Shifman and M.B.~Voloshin, \emph{{Preasymptotic effects in inclusive weak
  decays of charmed particles}}, {\emph{Sov. J. Nucl. Phys.} {\bfseries 41}
  (1985) 120}.

\bibitem{Chay:1990da}
J.~Chay, H.~Georgi and B.~Grinstein, \emph{{Lepton energy distributions in
  heavy meson decays from QCD}},
  \href{https://doi.org/10.1016/0370-2693(90)90916-T}{\emph{Phys. Lett. B}
  {\bfseries 247} (1990) 399}.

\bibitem{BUV1992HQE}
I.I.Y.~Bigi, N.G.~Uraltsev and A.I.~Vainshtein, \emph{{Nonperturbative
  corrections to inclusive beauty and charm decays: QCD versus phenomenological
  models}}, \href{https://doi.org/10.1016/0370-2693(92)90908-M}{\emph{Phys.
  Lett. B} {\bfseries 293} (1992) 430}
  [\href{https://arxiv.org/abs/hep-ph/9207214}{{\ttfamily hep-ph/9207214}}].

\bibitem{BBSUV92HQE}
I.I.Y.~Bigi, B.~Blok, M.A.~Shifman, N.G.~Uraltsev and A.I.~Vainshtein, \emph{{A
  QCD `manifesto' on inclusive decays of beauty and charm}},  in \emph{{7th
  Meeting of the APS Division of Particles Fields}}, 11, 1992
  [\href{https://arxiv.org/abs/hep-ph/9212227}{{\ttfamily hep-ph/9212227}}].

\bibitem{BSUV1993I}
I.I.Y.~Bigi, M.A.~Shifman, N.G.~Uraltsev and A.I.~Vainshtein, \emph{{QCD
  predictions for lepton spectra in inclusive heavy flavor decays}},
  \href{https://doi.org/10.1103/PhysRevLett.71.496}{\emph{Phys. Rev. Lett.}
  {\bfseries 71} (1993) 496}
  [\href{https://arxiv.org/abs/hep-ph/9304225}{{\ttfamily hep-ph/9304225}}].

\bibitem{BSUV1993II}
I.I.Y.~Bigi, M.A.~Shifman, N.G.~Uraltsev and A.I.~Vainshtein, \emph{{On the
  motion of heavy quarks inside hadrons: Universal distributions and inclusive
  decays}}, \href{https://doi.org/10.1142/S0217751X94000996}{\emph{Int. J. Mod.
  Phys. A} {\bfseries 9} (1994) 2467}
  [\href{https://arxiv.org/abs/hep-ph/9312359}{{\ttfamily hep-ph/9312359}}].

\bibitem{PDG2020}
{\scshape Particle Data Group} collaboration, \emph{{Review of Particle
  Physics}}, \href{https://doi.org/10.1093/ptep/ptaa104}{\emph{PTEP} {\bfseries
  2020} (2020) 083C01}.

\bibitem{GNPR1979}
B.~Guberina, S.~Nussinov, R.D.~Peccei and R.~R{\"{u}}ckl, \emph{{$D$ meson
  lifetimes and decays}},
  \href{https://doi.org/10.1016/0370-2693(79)90086-8}{\emph{Phys. Lett. B}
  {\bfseries 89} (1979) 111}.

\bibitem{Cortes1980}
J.L.~Cortes and J.~Sanchez-Guillen, \emph{{The lifetimes of charmed baryons}},
  \href{https://doi.org/10.1103/PhysRevD.24.2982}{\emph{Phys. Rev. D}
  {\bfseries 24} (1981) 2982}.

\bibitem{KS1983}
V.A.~Khoze and M.A.~Shifman, \emph{{Heavy quarks}},
  \href{https://doi.org/10.1070/PU1983v026n05ABEH004398}{\emph{Sov. Phys. Usp.}
  {\bfseries 26} (1983) 387}.

\bibitem{BGT1984}
N.~Bili{\'c}, B.~Guberina and J.~Trampeti{\'c}, \emph{{Pauli interference
  effect in $D^+$ lifetime}},
  \href{https://doi.org/10.1016/0550-3213(84)90596-0}{\emph{Nucl. Phys. B}
  {\bfseries 248} (1984) 261}.

\bibitem{GRT1986}
B.~Guberina, R.~R{\"{u}}ckl and J.~Trampeti{\'c}, \emph{{Charmed baryon
  lifetime differences}}, \href{https://doi.org/10.1007/BF01411150}{\emph{Z.
  Phys. C} {\bfseries 33} (1986) 297}.

\bibitem{SV1986}
M.A.~Shifman and M.B.~Voloshin, \emph{{Hierarchy of lifetimes of charmed and
  beautiful hadrons}}, {\emph{Sov. Phys. JETP} {\bfseries 64} (1986) 698}.

\bibitem{LenzRauh2013}
A.~Lenz and T.~Rauh, \emph{{$D$-meson lifetimes within the heavy quark
  expansion}}, \href{https://doi.org/10.1103/PhysRevD.88.034004}{\emph{Phys.
  Rev. D} {\bfseries 88} (2013) 034004}
  [\href{https://arxiv.org/abs/1305.3588}{{\ttfamily 1305.3588}}].

\bibitem{Lenz2014}
A.~Lenz, \emph{{Lifetimes and heavy quark expansion}},
  \href{https://doi.org/10.1142/S0217751X15430058}{\emph{Int. J. Mod. Phys. A}
  {\bfseries 30} (2015) 1543005}
  [\href{https://arxiv.org/abs/1405.3601}{{\ttfamily 1405.3601}}].

\bibitem{FLMT2002}
E.~Franco, V.~Lubicz, F.~Mescia and C.~Tarantino, \emph{{Lifetime ratios of
  beauty hadrons at the next-to-leading order in QCD}},
  \href{https://doi.org/10.1016/S0550-3213(02)00262-6}{\emph{Nucl. Phys. B}
  {\bfseries 633} (2002) 212}
  [\href{https://arxiv.org/abs/hep-ph/0203089}{{\ttfamily hep-ph/0203089}}].

\bibitem{BBGLN2002}
M.~Beneke, G.~Buchalla, C.~Greub, A.~Lenz and U.~Nierste, \emph{{The $B^+
  -B^0_d$ lifetime difference beyond leading logarithms}},
  \href{https://doi.org/10.1016/S0550-3213(02)00561-8}{\emph{Nucl. Phys. B}
  {\bfseries 639} (2002) 389}
  [\href{https://arxiv.org/abs/hep-ph/0202106}{{\ttfamily hep-ph/0202106}}].

\bibitem{GOP2004}
F.~Gabbiani, A.I.~Onishchenko and A.A.~Petrov, \emph{{Spectator effects and
  lifetimes of heavy hadrons}},
  \href{https://doi.org/10.1103/PhysRevD.70.094031}{\emph{Phys. Rev. D}
  {\bfseries 70} (2004) 094031}
  [\href{https://arxiv.org/abs/hep-ph/0407004}{{\ttfamily hep-ph/0407004}}].

\bibitem{BUZ2005}
I.I.~Bigi, N.~Uraltsev and R.~Zwicky, \emph{{On the nonperturbative charm 
effects in inclusive $B \to X_c \ell \nu$ decays}},
\href{https://doi.org/10.1140/epjc/s10052-007-0216-8}{\emph{Eur.\ Phys.\ J.\ C}
{\bfseries 50} (2007) 539}
[\href{https://arxiv.org/abs/hep-ph/0511158}{{\ttfamily hep-ph/0511158}}].

\bibitem{DMT2006}
B.M.~Dassinger, T.~Mannel and S.~Turczyk, \emph{{Inclusive semi-leptonic $B$
  decays to order $1 / m_b^4$}},
  \href{https://doi.org/10.1088/1126-6708/2007/03/087}{\emph{JHEP} {\bfseries
  03} (2007) 087} [\href{https://arxiv.org/abs/hep-ph/0611168}{{\ttfamily
  hep-ph/0611168}}].

\bibitem{MTU2010}
T.~Mannel, S.~Turczyk and N.~Uraltsev, \emph{{Higher order power corrections in
  inclusive $B$ decays}},
  \href{https://doi.org/10.1007/JHEP11(2010)109}{\emph{JHEP} {\bfseries 11}
  (2010) 109} [\href{https://arxiv.org/abs/1009.4622}{{\ttfamily 1009.4622}}].

\bibitem{GHT2016}
P.~Gambino, K.J.~Healey and S.~Turczyk, \emph{{Taming the higher power
  corrections in semileptonic $B$ decays}},
  \href{https://doi.org/10.1016/j.physletb.2016.10.023}{\emph{Phys. Lett. B}
  {\bfseries 763} (2016) 60}
  [\href{https://arxiv.org/abs/1606.06174}{{\ttfamily 1606.06174}}].

\bibitem{FMV2019}
M.~Fael, T.~Mannel and K.K.~Vos, \emph{{The heavy quark expansion for inclusive
  semileptonic charm decays revisited}},
  \href{https://doi.org/10.1007/JHEP12(2019)067}{\emph{JHEP} {\bfseries 12}
  (2019) 067} [\href{https://arxiv.org/abs/1910.05234}{{\ttfamily
  1910.05234}}].

\bibitem{MMP2021}
T.~Mannel, D.~Moreno and A.A.~Pivovarov, \emph{{The heavy quark expansion for
  the charm quark}},  \href{https://arxiv.org/abs/2103.02058}{{\ttfamily
  2103.02058}}.

\bibitem{LenzNote:2021}
D.~King, A.~Lenz, M.L.~Piscopo, T.~Rauh, A.~Rusov and C.~Vlahos,
  \emph{{Revisiting inclusive decay widths of charmed mesons}},
  \href{https://arxiv.org/abs/2109.13219}{{\ttfamily 2109.13219}}.

\bibitem{Cheng18c}
H.-Y.~Cheng, \emph{{Phenomenological study of heavy hadron lifetimes}},
  \href{https://doi.org/10.1007/JHEP11(2018)014}{\emph{JHEP} {\bfseries 11}
  (2018) 014} [\href{https://arxiv.org/abs/1807.00916}{{\ttfamily
  1807.00916}}].

\bibitem{MMP2020}
T.~Mannel, D.~Moreno and A.~Pivovarov, \emph{{Heavy quark expansion for heavy
  hadron lifetimes: completing the $ 1/{m}_b^3 $ corrections}},
  \href{https://doi.org/10.1007/JHEP08(2020)089}{\emph{JHEP} {\bfseries 08}
  (2020) 089} [\href{https://arxiv.org/abs/2004.09485}{{\ttfamily
  2004.09485}}].

\bibitem{Moreno2020}
D.~Moreno, \emph{{Completing $1/m_b^3$ corrections to non-leptonic
  bottom-to-up-quark decays}},
  \href{https://doi.org/10.1007/JHEP01(2021)051}{\emph{JHEP} {\bfseries 01}
  (2021) 051} [\href{https://arxiv.org/abs/2009.08756}{{\ttfamily
  2009.08756}}].

\bibitem{LPR2020}
A.~Lenz, M.L.~Piscopo and A.V.~Rusov, \emph{{Contribution of the Darwin
  operator to non-leptonic decays of heavy quarks}},
  \href{https://doi.org/10.1007/JHEP12(2020)199}{\emph{JHEP} {\bfseries 12}
  (2020) 199} [\href{https://arxiv.org/abs/2004.09527}{{\ttfamily
  2004.09527}}].

\bibitem{GOP2003}
F.~Gabbiani, A.I.~Onishchenko and A.A.~Petrov, \emph{{$\Lambda_b$ lifetime
  puzzle in heavy quark expansion}},
  \href{https://doi.org/10.1103/PhysRevD.68.114006}{\emph{Phys. Rev. D}
  {\bfseries 68} (2003) 114006}
  [\href{https://arxiv.org/abs/hep-ph/0303235}{{\ttfamily hep-ph/0303235}}].

\bibitem{CFLM2001}
M.~Ciuchini, E.~Franco, V.~Lubicz and F.~Mescia, \emph{{Next-to-leading order
  QCD corrections to spectator effects in lifetimes of beauty hadrons}},
  \href{https://doi.org/10.1016/S0550-3213(02)00006-8}{\emph{Nucl. Phys. B}
  {\bfseries 625} (2002) 211}
  [\href{https://arxiv.org/abs/hep-ph/0110375}{{\ttfamily hep-ph/0110375}}].

\bibitem{BESIII2021}
{\scshape BESIII} collaboration, \emph{{Measurement of the absolute branching
  fraction of inclusive semielectronic $D_s^+$ decays}},
  \href{https://doi.org/10.1103/PhysRevD.104.012003}{\emph{Phys. Rev. D}
  {\bfseries 104} (2021) 012003}
  [\href{https://arxiv.org/abs/2104.07311}{{\ttfamily 2104.07311}}].

\bibitem{BelleII2021}
{\scshape Belle-II} collaboration, \emph{{Precise measurement of the $D^0$ and
  $D^+$ lifetimes at Belle II}},
  \href{https://doi.org/10.1103/PhysRevLett.127.211801}{\emph{Phys. Rev. Lett.}
  {\bfseries 127} (2021) 211801}
  [\href{https://arxiv.org/abs/2108.03216}{{\ttfamily 2108.03216}}].

\bibitem{BESIII:2018mug}
{\scshape BESIII} collaboration, \emph{{Measurement of the absolute branching
  fraction of the inclusive semileptonic $\Lambda_c^+$ decay}},
  \href{https://doi.org/10.1103/PhysRevLett.121.251801}{\emph{Phys. Rev. Lett.}
  {\bfseries 121} (2018) 251801}
  [\href{https://arxiv.org/abs/1805.09060}{{\ttfamily 1805.09060}}].

\bibitem{BBL1995}
G.~Buchalla, A.J.~Buras and M.E.~Lautenbacher, \emph{{Weak decays beyond
  leading logarithms}},
  \href{https://doi.org/10.1103/RevModPhys.68.1125}{\emph{Rev. Mod. Phys.}
  {\bfseries 68} (1996) 1125}
  [\href{https://arxiv.org/abs/hep-ph/9512380}{{\ttfamily hep-ph/9512380}}].

\bibitem{LukeThm}
M.E.~Luke, \emph{{Effects of subleading operators in the heavy quark effective
  theory}}, \href{https://doi.org/10.1016/0370-2693(90)90568-Q}{\emph{Phys.
  Lett. B} {\bfseries 252} (1990) 447}.

\bibitem{Mannel1994}
T.~Mannel, \emph{{Higher order $1/m$ corrections at zero recoil}},
  \href{https://doi.org/10.1103/PhysRevD.50.428}{\emph{Phys. Rev. D} {\bfseries
  50} (1994) 428} [\href{https://arxiv.org/abs/hep-ph/9403249}{{\ttfamily
  hep-ph/9403249}}].

\bibitem{BSUV1994}
I.I.Y.~Bigi, M.A.~Shifman, N.G.~Uraltsev and A.I.~Vainshtein, \emph{{Sum rules
  for heavy flavor transitions in the SV limit}},
  \href{https://doi.org/10.1103/PhysRevD.52.196}{\emph{Phys. Rev. D} {\bfseries
  52} (1995) 196} [\href{https://arxiv.org/abs/hep-ph/9405410}{{\ttfamily
  hep-ph/9405410}}].

\bibitem{Neubert1993}
M.~Neubert, \emph{{Heavy quark symmetry}},
  \href{https://doi.org/10.1016/0370-1573(94)90091-4}{\emph{Phys. Rept.}
  {\bfseries 245} (1994) 259}
  [\href{https://arxiv.org/abs/hep-ph/9306320}{{\ttfamily hep-ph/9306320}}].

\bibitem{Manohar:2000dt}
A.V.~Manohar and M.B.~Wise, \emph{{Heavy quark physics}}, vol.~10 (2000).

\bibitem{BSUV1994II}
I.I.Y.~Bigi, M.A.~Shifman, N.G.~Uraltsev and A.I.~Vainshtein, \emph{{The Pole
  mass of the heavy quark. Perturbation theory and beyond}},
  \href{https://doi.org/10.1103/PhysRevD.50.2234}{\emph{Phys. Rev. D}
  {\bfseries 50} (1994) 2234}
  [\href{https://arxiv.org/abs/hep-ph/9402360}{{\ttfamily hep-ph/9402360}}].

\bibitem{CPT1982}
J.L.~Cortes, X.-Y.~Pham and A.~Tounsi, \emph{{Mass effects in weak decays of
  heavy particles}}, \href{https://doi.org/10.1103/PhysRevD.25.188}{\emph{Phys.
  Rev. D} {\bfseries 25} (1982) 188}.

\bibitem{Koyrakh1993}
L.~Koyrakh, \emph{{Nonperturbative corrections to the heavy lepton energy
  distribution in the inclusive decays $H_b \to \tau \bar{\nu} X$}},
  \href{https://doi.org/10.1103/PhysRevD.49.3379}{\emph{Phys. Rev. D}
  {\bfseries 49} (1994) 3379}
  [\href{https://arxiv.org/abs/hep-ph/9311215}{{\ttfamily hep-ph/9311215}}].

\bibitem{BS1992I}
B.~Blok and M.A.~Shifman, \emph{{The rule of discarding $1/N_c$ in inclusive
  weak decays. 1.}},
  \href{https://doi.org/10.1016/0550-3213(93)90504-I}{\emph{Nucl. Phys. B}
  {\bfseries 399} (1993) 441}
  [\href{https://arxiv.org/abs/hep-ph/9207236}{{\ttfamily hep-ph/9207236}}].

\bibitem{BS1992II}
B.~Blok and M.A.~Shifman, \emph{{The rule of discarding $1/N_c$ in inclusive
  weak decays. 2.}},
  \href{https://doi.org/10.1016/0550-3213(93)90505-J}{\emph{Nucl. Phys. B}
  {\bfseries 399} (1993) 459}
  [\href{https://arxiv.org/abs/hep-ph/9209289}{{\ttfamily hep-ph/9209289}}].

\bibitem{GK1996}
M.~Gremm and A.~Kapustin, \emph{{Order $1/m_b^3$ corrections to $B \to X_c l
  \bar{\nu}_l$ decay and their implication for the measurement of
  $\bar{\Lambda}$ and $\lambda_1$}},
  \href{https://doi.org/10.1103/PhysRevD.55.6924}{\emph{Phys. Rev. D}
  {\bfseries 55} (1997) 6924}
  [\href{https://arxiv.org/abs/hep-ph/9603448}{{\ttfamily hep-ph/9603448}}].

\bibitem{MRS2017}
T.~Mannel, A.V.~Rusov and F.~Shahriaran, \emph{{Inclusive semitauonic $B$
  decays to order ${\cal O} (\Lambda_{QCD}^3/m_b^3)$}},
  \href{https://doi.org/10.1016/j.nuclphysb.2017.05.016}{\emph{Nucl. Phys. B}
  {\bfseries 921} (2017) 211}
  [\href{https://arxiv.org/abs/1702.01089}{{\ttfamily 1702.01089}}].

\bibitem{HokimPham84}
Q.~Ho-kim and X.-Y.~Pham, \emph{{Exact one gluon corrections for inclusive weak
  processes}}, \href{https://doi.org/10.1016/0003-4916(84)90258-6}{\emph{Annals
  Phys.} {\bfseries 155} (1984) 202}.

\bibitem{BBBG1994}
E.~Bagan, P.~Ball, V.M.~Braun and P.~Gosdzinsky, \emph{{Charm quark mass
  dependence of QCD corrections to nonleptonic inclusive $B$ decays}},
  \href{https://doi.org/10.1016/0550-3213(94)90591-6}{\emph{Nucl. Phys. B}
  {\bfseries 432} (1994) 3}
  [\href{https://arxiv.org/abs/hep-ph/9408306}{{\ttfamily hep-ph/9408306}}].

\bibitem{BBFG1995}
E.~Bagan, P.~Ball, B.~Fiol and P.~Gosdzinsky, \emph{{Next-to-leading order
  radiative corrections to the decay $b \to c\bar{c}s$}},
  \href{https://doi.org/10.1016/0370-2693(95)00437-P}{\emph{Phys. Lett. B}
  {\bfseries 351} (1995) 546}
  [\href{https://arxiv.org/abs/hep-ph/9502338}{{\ttfamily hep-ph/9502338}}].

\bibitem{KLR2013}
F.~Krinner, A.~Lenz and T.~Rauh, \emph{{The inclusive decay $b \to c\bar{c}s$
  revisited}},
  \href{https://doi.org/10.1016/j.nuclphysb.2013.07.028}{\emph{Nucl. Phys. B}
  {\bfseries 876} (2013) 31} [\href{https://arxiv.org/abs/1305.5390}{{\ttfamily
  1305.5390}}].

\bibitem{CST2005}
A.~Czarnecki, M.~{\'S}lusarczyk and F.V.~Tkachov, \emph{{Enhancement of the
  hadronic $b$ quark decays}},
  \href{https://doi.org/10.1103/PhysRevLett.96.171803}{\emph{Phys. Rev. Lett.}
  {\bfseries 96} (2006) 171803}
  [\href{https://arxiv.org/abs/hep-ph/0511004}{{\ttfamily hep-ph/0511004}}].

\bibitem{CJK1994}
A.~Czarnecki, M.~Je\.zabek and J.H.~K{\"u}hn, \emph{{Radiative corrections to
  $b \to c \tau \bar{\nu}_\tau$}},
  \href{https://doi.org/10.1016/0370-2693(94)01681-2}{\emph{Phys. Lett. B}
  {\bfseries 346} (1995) 335}
  [\href{https://arxiv.org/abs/hep-ph/9411282}{{\ttfamily hep-ph/9411282}}].

\bibitem{LSW1994}
M.E.~Luke, M.J.~Savage and M.B.~Wise, \emph{{Charm mass dependence of the $O
  (\alpha_s^2 n_f)$ correction to inclusive $B \to X_c e \bar{\nu}_e$ decay}},
  \href{https://doi.org/10.1016/0370-2693(94)01573-U}{\emph{Phys. Lett. B}
  {\bfseries 345} (1995) 301}
  [\href{https://arxiv.org/abs/hep-ph/9410387}{{\ttfamily hep-ph/9410387}}].

\bibitem{Ritbergen1999}
T.~van Ritbergen, \emph{{The second order QCD contribution to the semileptonic
  $b \to u$ decay rate}},
  \href{https://doi.org/10.1016/S0370-2693(99)00407-4}{\emph{Phys. Lett. B}
  {\bfseries 454} (1999) 353}
  [\href{https://arxiv.org/abs/hep-ph/9903226}{{\ttfamily hep-ph/9903226}}].

\bibitem{PC2008I}
A.~Pak and A.~Czarnecki, \emph{{Mass effects in muon and semileptonic $ b \to
  c$ decays}},
  \href{https://doi.org/10.1103/PhysRevLett.100.241807}{\emph{Phys. Rev. Lett.}
  {\bfseries 100} (2008) 241807}
  [\href{https://arxiv.org/abs/0803.0960}{{\ttfamily 0803.0960}}].

\bibitem{PC2008II}
A.~Pak and A.~Czarnecki, \emph{{Heavy-to-heavy quark decays at NNLO}},
  \href{https://doi.org/10.1103/PhysRevD.78.114015}{\emph{Phys. Rev. D}
  {\bfseries 78} (2008) 114015}
  [\href{https://arxiv.org/abs/0808.3509}{{\ttfamily 0808.3509}}].

\bibitem{BM2009}
S.~Biswas and K.~Melnikov, \emph{{Second order QCD corrections to inclusive
  semileptonic $b \to X_c l \bar{\nu}_l$ decays with massless and massive
  lepton}}, \href{https://doi.org/10.1007/JHEP02(2010)089}{\emph{JHEP}
  {\bfseries 02} (2010) 089} [\href{https://arxiv.org/abs/0911.4142}{{\ttfamily
  0911.4142}}].

\bibitem{FSS2020}
M.~Fael, K.~Sch\"onwald and M.~Steinhauser, \emph{{Third order corrections to
  the semi-leptonic $b\to c$ and the muon decays}},
  \href{https://doi.org/10.1103/PhysRevD.104.016003}{\emph{Phys. Rev. D}
  {\bfseries 104} (2021) 016003}
  [\href{https://arxiv.org/abs/2011.13654}{{\ttfamily 2011.13654}}].

\bibitem{CCD2021}
M.~Czakon, A.~Czarnecki and M.~Dowling, \emph{{Three-loop corrections to the
  muon and heavy quark decay rates}},
  \href{https://doi.org/10.1103/PhysRevD.103.L111301}{\emph{Phys. Rev. D}
  {\bfseries 103} (2021) L111301}
  [\href{https://arxiv.org/abs/2104.05804}{{\ttfamily 2104.05804}}].

\bibitem{AGN2013}
A.~Alberti, P.~Gambino and S.~Nandi, \emph{{Perturbative corrections to power
  suppressed effects in semileptonic $B$ decays}},
  \href{https://doi.org/10.1007/JHEP01(2014)147}{\emph{JHEP} {\bfseries 01}
  (2014) 147} [\href{https://arxiv.org/abs/1311.7381}{{\ttfamily 1311.7381}}].

\bibitem{MPR2014}
T.~Mannel, A.A.~Pivovarov and D.~Rosenthal, \emph{{Inclusive semileptonic $B$
  decays from QCD with NLO accuracy for power suppressed terms}},
  \href{https://doi.org/10.1016/j.physletb.2014.12.058}{\emph{Phys. Lett. B}
  {\bfseries 741} (2015) 290}
  [\href{https://arxiv.org/abs/1405.5072}{{\ttfamily 1405.5072}}].

\bibitem{MPR2015}
T.~Mannel, A.A.~Pivovarov and D.~Rosenthal, \emph{{Inclusive weak decays of
  heavy hadrons with power suppressed terms at NLO}},
  \href{https://doi.org/10.1103/PhysRevD.92.054025}{\emph{Phys. Rev. D}
  {\bfseries 92} (2015) 054025}
  [\href{https://arxiv.org/abs/1506.08167}{{\ttfamily 1506.08167}}].

\bibitem{MP2019}
T.~Mannel and A.A.~Pivovarov, \emph{{QCD corrections to inclusive heavy hadron
  weak decays at $\Lambda_{\rm QCD}^3 /m_Q^3$}},
  \href{https://doi.org/10.1103/PhysRevD.100.093001}{\emph{Phys. Rev. D}
  {\bfseries 100} (2019) 093001}
  [\href{https://arxiv.org/abs/1907.09187}{{\ttfamily 1907.09187}}].

\bibitem{MMP2021II}
T.~Mannel, D.~Moreno and A.A.~Pivovarov,
\emph{{NLO QCD corrections to inclusive $b \rightarrow c \ell \bar{\nu}$decay 
spectra up to~$1/m_Q^3$}},
\href{https://doi.org/10.1103/PhysRevD.105.054033}{\emph{Phys.\ Rev.\ D} 
{\bfseries 105} (2022) 054033}
[\href{https://arxiv.org/abs/2112.03875}{{\ttfamily 2112.03875}}].

\bibitem{BlokShifman93Review}
B.~Blok and M.A.~Shifman, \emph{{Lifetimes of charmed hadrons revisited. Facts
  and fancy}},  in \emph{{3rd Workshop on the Tau-Charm Factory}}, 11, 1991
  [\href{https://arxiv.org/abs/hep-ph/9311331}{{\ttfamily hep-ph/9311331}}].

\bibitem{BSU97ReviewHQE}
I.I.Y.~Bigi, M.A.~Shifman and N.~Uraltsev, \emph{{Aspects of heavy quark
  theory}}, \href{https://doi.org/10.1146/annurev.nucl.47.1.591}{\emph{Ann.
  Rev. Nucl. Part. Sci.} {\bfseries 47} (1997) 591}
  [\href{https://arxiv.org/abs/hep-ph/9703290}{{\ttfamily hep-ph/9703290}}].

\bibitem{Voloshin2001}
M.B.~Voloshin, \emph{{Nonfactorization effects in heavy mesons and
  determination of $|V_{ub}$ from inclusive semileptonic $B$ decays}},
  \href{https://doi.org/10.1016/S0370-2693(01)00812-7}{\emph{Phys. Lett. B}
  {\bfseries 515} (2001) 74}
  [\href{https://arxiv.org/abs/hep-ph/0106040}{{\ttfamily hep-ph/0106040}}].

\bibitem{KLR2021}
D.~King, A.~Lenz and T.~Rauh, \emph{{$SU(3)$ breaking effects in $B$ and $D$
  meson lifetimes}},  \href{https://arxiv.org/abs/2112.03691}{{\ttfamily
  2112.03691}}.

\bibitem{Voloshin96}
M.B.~Voloshin, \emph{{Spectator effects in semileptonic decay of charmed
  baryons}}, \href{https://doi.org/10.1016/0370-2693(96)00837-4}{\emph{Phys.
  Lett. B} {\bfseries 385} (1996) 369}
  [\href{https://arxiv.org/abs/hep-ph/9604335}{{\ttfamily hep-ph/9604335}}].

\bibitem{NS1996}
M.~Neubert and C.T.~Sachrajda, \emph{{Spectator effects in inclusive decays of
  beauty hadrons}},
  \href{https://doi.org/10.1016/S0550-3213(96)00559-7}{\emph{Nucl. Phys. B}
  {\bfseries 483} (1997) 339}
  [\href{https://arxiv.org/abs/hep-ph/9603202}{{\ttfamily hep-ph/9603202}}].

\bibitem{PU1998}
D.~Pirjol and N.~Uraltsev, \emph{{Four fermion heavy quark operators and light
  current amplitudes in heavy flavor hadrons}},
  \href{https://doi.org/10.1103/PhysRevD.59.034012}{\emph{Phys. Rev. D}
  {\bfseries 59} (1999) 034012}
  [\href{https://arxiv.org/abs/hep-ph/9805488}{{\ttfamily hep-ph/9805488}}].

\bibitem{KM1992}
W.~Kilian and T.~Mannel, \emph{{QCD corrected $1/m_b$ contributions to $B
  \bar{B}$ mixing}},
  \href{https://doi.org/10.1016/0370-2693(93)91167-L}{\emph{Phys. Lett. B}
  {\bfseries 301} (1993) 382}
  [\href{https://arxiv.org/abs/hep-ph/9211333}{{\ttfamily hep-ph/9211333}}].

\bibitem{FMV2018}
M.~Fael, T.~Mannel and K.K.~Vos, \emph{{$V_{cb}$ determination from inclusive
  $b \to c$ decays: an alternative method}},
  \href{https://doi.org/10.1007/JHEP02(2019)177}{\emph{JHEP} {\bfseries 02}
  (2019) 177} [\href{https://arxiv.org/abs/1812.07472}{{\ttfamily
  1812.07472}}].

\bibitem{Beneke1998Renormalon}
M.~Beneke, \emph{{Renormalons}},
  \href{https://doi.org/10.1016/S0370-1573(98)00130-6}{\emph{Phys. Rept.}
  {\bfseries 317} (1999) 1}
  [\href{https://arxiv.org/abs/hep-ph/9807443}{{\ttfamily hep-ph/9807443}}].

\bibitem{Beneke2021}
M.~Beneke, \emph{{Pole mass renormalon and its ramifications}},
  \href{https://doi.org/10.1140/epjs/s11734-021-00268-w}{\emph{Eur. Phys. J.
  ST} {\bfseries 230} (2021) 2565}
  [\href{https://arxiv.org/abs/2108.04861}{{\ttfamily 2108.04861}}].

\bibitem{CS2000}
K.G.~Chetyrkin and M.~Steinhauser, \emph{{The relation between the
  $\overline{\textrm{MS}}$ and the on-shell quark mass at order $\alpha_s^3$}},
  \href{https://doi.org/10.1016/S0550-3213(99)00784-1}{\emph{Nucl. Phys. B}
  {\bfseries 573} (2000) 617}
  [\href{https://arxiv.org/abs/hep-ph/9911434}{{\ttfamily hep-ph/9911434}}].

\bibitem{MR2000}
K.~Melnikov and T.v.~Ritbergen, \emph{{The three loop relation between the
  $\overline{\textrm{MS}}$ and the pole quark masses}},
  \href{https://doi.org/10.1016/S0370-2693(00)00507-4}{\emph{Phys. Lett. B}
  {\bfseries 482} (2000) 99}
  [\href{https://arxiv.org/abs/hep-ph/9912391}{{\ttfamily hep-ph/9912391}}].

\bibitem{ETM2014I}
{\scshape European Twisted Mass} collaboration, \emph{{Up, down, strange and
  charm quark masses with $N_f = 2+1+1$ twisted mass lattice QCD}},
  \href{https://doi.org/10.1016/j.nuclphysb.2014.07.025}{\emph{Nucl. Phys. B}
  {\bfseries 887} (2014) 19} [\href{https://arxiv.org/abs/1403.4504}{{\ttfamily
  1403.4504}}].

\bibitem{ETM2014II}
{\scshape European Twisted Mass} collaboration, \emph{{Baryon spectrum with
  $N_f=2+1+1$ twisted mass fermions}},
  \href{https://doi.org/10.1103/PhysRevD.90.074501}{\emph{Phys. Rev. D}
  {\bfseries 90} (2014) 074501}
  [\href{https://arxiv.org/abs/1406.4310}{{\ttfamily 1406.4310}}].

\bibitem{HPQCD2014I}
{\scshape HPQCD} collaboration, \emph{{High-precision quark masses and QCD
  coupling from $n_f=4$ lattice QCD}},
  \href{https://doi.org/10.1103/PhysRevD.91.054508}{\emph{Phys. Rev. D}
  {\bfseries 91} (2015) 054508}
  [\href{https://arxiv.org/abs/1408.4169}{{\ttfamily 1408.4169}}].

\bibitem{MILC2018}
{\scshape Fermilab Lattice, MILC, TUMQCD} collaboration, \emph{{Up-, down-,
  strange-, charm-, and bottom-quark masses from four-flavor lattice QCD}},
  \href{https://doi.org/10.1103/PhysRevD.98.054517}{\emph{Phys. Rev. D}
  {\bfseries 98} (2018) 054517}
  [\href{https://arxiv.org/abs/1802.04248}{{\ttfamily 1802.04248}}].

\bibitem{HPQCD2018}
{\scshape HPQCD} collaboration, \emph{{Determination of quark masses from
  $n_f=4$ lattice QCD and the RI-SMOM intermediate scheme}},
  \href{https://doi.org/10.1103/PhysRevD.98.014513}{\emph{Phys. Rev. D}
  {\bfseries 98} (2018) 014513}
  [\href{https://arxiv.org/abs/1805.06225}{{\ttfamily 1805.06225}}].

\bibitem{BSUV1996}
I.I.Y.~Bigi, M.A.~Shifman, N.G.~Uraltsev and A.I.~Vainshtein, \emph{{High power
  $n$ of $m_b$ in beauty widths and $n=5 \to \infty$ limit}},
  \href{https://doi.org/10.1103/PhysRevD.56.4017}{\emph{Phys. Rev. D}
  {\bfseries 56} (1997) 4017}
  [\href{https://arxiv.org/abs/hep-ph/9704245}{{\ttfamily hep-ph/9704245}}].

\bibitem{FSS2020I}
M.~Fael, K.~Sch{\"{o}}nwald and M.~Steinhauser, \emph{{Kinetic heavy quark mass
  to three loops}},
  \href{https://doi.org/10.1103/PhysRevLett.125.052003}{\emph{Phys. Rev. Lett.}
  {\bfseries 125} (2020) 052003}
  [\href{https://arxiv.org/abs/2005.06487}{{\ttfamily 2005.06487}}].

\bibitem{FSS2020II}
M.~Fael, K.~Sch{\"{o}}nwald and M.~Steinhauser, \emph{{Relation between the
  $\overline{\mathrm{MS}}$ and the kinetic mass of heavy quarks}},
  \href{https://doi.org/10.1103/PhysRevD.103.014005}{\emph{Phys. Rev. D}
  {\bfseries 103} (2021) 014005}
  [\href{https://arxiv.org/abs/2011.11655}{{\ttfamily 2011.11655}}].

\bibitem{HJS2008}
A.H.~Hoang, A.~Jain, I.~Scimemi and I.W.~Stewart, \emph{{Infrared
  renormalization group flow for heavy quark masses}},
  \href{https://doi.org/10.1103/PhysRevLett.101.151602}{\emph{Phys. Rev. Lett.}
  {\bfseries 101} (2008) 151602}
  [\href{https://arxiv.org/abs/0803.4214}{{\ttfamily 0803.4214}}].

\bibitem{HJLMPSS2017}
A.H.~Hoang, A.~Jain, C.~Lepenik, V.~Mateu, M.~Preisser, I.~Scimemi et~al.,
  \emph{{The MSR mass and the $
  \mathcal{O}\left({\Lambda}_{\mathrm{QCD}}\right) $ renormalon sum rule}},
  \href{https://doi.org/10.1007/JHEP04(2018)003}{\emph{JHEP} {\bfseries 04}
  (2018) 003} [\href{https://arxiv.org/abs/1704.01580}{{\ttfamily
  1704.01580}}].

\bibitem{Pineda2001}
A.~Pineda, \emph{{Determination of the bottom quark mass from the
  $\Upsilon(1S)$ system}},
  \href{https://doi.org/10.1088/1126-6708/2001/06/022}{\emph{JHEP} {\bfseries
  06} (2001) 022} [\href{https://arxiv.org/abs/hep-ph/0105008}{{\ttfamily
  hep-ph/0105008}}].

\bibitem{FLAG2019}
{\scshape Flavour Lattice Averaging Group} collaboration, \emph{{FLAG Review
  2019: Flavour Lattice Averaging Group (FLAG)}},
  \href{https://doi.org/10.1140/epjc/s10052-019-7354-7}{\emph{Eur. Phys. J. C}
  {\bfseries 80} (2020) 113}
  [\href{https://arxiv.org/abs/1902.08191}{{\ttfamily 1902.08191}}].

\bibitem{FLAG2021}
Y.~Aoki et~al., \emph{{FLAG Review 2021}},
  \href{https://arxiv.org/abs/2111.09849}{{\ttfamily 2111.09849}}.

\bibitem{RunDec}
K.G.~Chetyrkin, J.H.~K{\"u}hn and M.~Steinhauser, \emph{{RunDec: A Mathematica
  package for running and decoupling of the strong coupling and quark masses}},
  \href{https://doi.org/10.1016/S0010-4655(00)00155-7}{\emph{Comput. Phys.
  Commun.} {\bfseries 133} (2000) 43}
  [\href{https://arxiv.org/abs/hep-ph/0004189}{{\ttfamily hep-ph/0004189}}].

\bibitem{RunDecv3}
F.~Herren and M.~Steinhauser, \emph{{Version 3 of RunDec and CRunDec}},
  \href{https://doi.org/10.1016/j.cpc.2017.11.014}{\emph{Comput. Phys. Commun.}
  {\bfseries 224} (2018) 333}
  [\href{https://arxiv.org/abs/1703.03751}{{\ttfamily 1703.03751}}].

\bibitem{HLM1999I}
A.H.~Hoang, Z.~Ligeti and A.V.~Manohar, \emph{{$B$ decay and the Upsilon
  mass}}, \href{https://doi.org/10.1103/PhysRevLett.82.277}{\emph{Phys. Rev.
  Lett.} {\bfseries 82} (1999) 277}
  [\href{https://arxiv.org/abs/hep-ph/9809423}{{\ttfamily hep-ph/9809423}}].

\bibitem{HLM1999II}
A.H.~Hoang, Z.~Ligeti and A.V.~Manohar, \emph{{$B$ decays in the upsilon
  expansion}}, \href{https://doi.org/10.1103/PhysRevD.59.074017}{\emph{Phys.
  Rev. D} {\bfseries 59} (1999) 074017}
  [\href{https://arxiv.org/abs/hep-ph/9811239}{{\ttfamily hep-ph/9811239}}].

\bibitem{BS1999}
M.~Beneke and A.~Signer, \emph{{The Bottom $\overline{\textrm{MS}}$ quark mass
  from sum rules at next-to-next-to-leading order}},
  \href{https://doi.org/10.1016/S0370-2693(99)01348-9}{\emph{Phys. Lett. B}
  {\bfseries 471} (1999) 233}
  [\href{https://arxiv.org/abs/hep-ph/9906475}{{\ttfamily hep-ph/9906475}}].

\bibitem{Uraltsev2004}
N.~Uraltsev, \emph{{Heavy quark expansion in beauty: Recent successes and
  problems}},  in \emph{{Workshop on Continuous Advances in QCD 2004}},
  pp.~100--114, 9, 2004, \href{https://doi.org/10.1142/9789812702326_0009}{DOI}
  [\href{https://arxiv.org/abs/hep-ph/0409125}{{\ttfamily hep-ph/0409125}}].

\bibitem{Bigi2006}
I.I.~Bigi, \emph{{The physics of beauty (\& charm [\& $\tau$]) at the LHC and
  in the era of the LHC}}, {\emph{Acta Phys. Polon. B} {\bfseries 38} (2007)
  867} [\href{https://arxiv.org/abs/hep-ph/0608225}{{\ttfamily
  hep-ph/0608225}}].

\bibitem{AG2021}
J.~Aebischer and B.~Grinstein, \emph{{Standard Model prediction of the $B_c$
  lifetime}}, \href{https://doi.org/10.1007/JHEP07(2021)130}{\emph{JHEP}
  {\bfseries 07} (2021) 130}
  [\href{https://arxiv.org/abs/2105.02988}{{\ttfamily 2105.02988}}].

\bibitem{FaelComms}
M. Fael, private communication.

\bibitem{FN92I}
A.F.~Falk and M.~Neubert, \emph{{Second order power corrections in the heavy
  quark effective theory. 1. Formalism and meson form-factors}},
  \href{https://doi.org/10.1103/PhysRevD.47.2965}{\emph{Phys. Rev. D}
  {\bfseries 47} (1993) 2965}
  [\href{https://arxiv.org/abs/hep-ph/9209268}{{\ttfamily hep-ph/9209268}}].

\bibitem{FN92II}
A.F.~Falk and M.~Neubert, \emph{{Second order power corrections in the heavy
  quark effective theory. 2. Baryon form-factors}},
  \href{https://doi.org/10.1103/PhysRevD.47.2982}{\emph{Phys. Rev. D}
  {\bfseries 47} (1993) 2982}
  [\href{https://arxiv.org/abs/hep-ph/9209269}{{\ttfamily hep-ph/9209269}}].

\bibitem{Neubert1996}
M.~Neubert, \emph{{$B$ decays and $CP$ violation}},
  \href{https://doi.org/10.1142/S0217751X96001966}{\emph{Int. J. Mod. Phys. A}
  {\bfseries 11} (1996) 4173}
  [\href{https://arxiv.org/abs/hep-ph/9604412}{{\ttfamily hep-ph/9604412}}].

\bibitem{Grozin:2007fh}
A.G.~Grozin, P.~Marquard, J.H.~Piclum and M.~Steinhauser, \emph{{Three-loop
  chromomagnetic interaction in HQET}},
  \href{https://doi.org/10.1016/j.nuclphysb.2007.08.012}{\emph{Nucl. Phys. B}
  {\bfseries 789} (2008) 277}
  [\href{https://arxiv.org/abs/0707.1388}{{\ttfamily 0707.1388}}].

\bibitem{Neubert1997Btheory}
M.~Neubert, \emph{{Theory of inclusive $B$ decays}},
  \href{https://doi.org/10.1016/S0920-5632(97)00432-5}{\emph{Nucl. Phys. B
  Proc. Suppl.} {\bfseries 59} (1997) 101}
  [\href{https://arxiv.org/abs/hep-ph/9702310}{{\ttfamily hep-ph/9702310}}].

\bibitem{AGHN2014}
A.~Alberti, P.~Gambino, K.J.~Healey and S.~Nandi, \emph{{Precision
  determination of the Cabibbo-Kobayashi-Maskawa element $V_{cb}$}},
  \href{https://doi.org/10.1103/PhysRevLett.114.061802}{\emph{Phys. Rev. Lett.}
  {\bfseries 114} (2015) 061802}
  [\href{https://arxiv.org/abs/1411.6560}{{\ttfamily 1411.6560}}].

\bibitem{BMU2011}
I.I.~Bigi, T.~Mannel and N.~Uraltsev, \emph{{Semileptonic width ratios among
  beauty hadrons}}, \href{https://doi.org/10.1007/JHEP09(2011)012}{\emph{JHEP}
  {\bfseries 09} (2011) 012} [\href{https://arxiv.org/abs/1105.4574}{{\ttfamily
  1105.4574}}].

\bibitem{Voloshinmukin}
M.B.~Voloshin, \emph{{Optical sum rule for form-factors of heavy mesons}},
  \href{https://doi.org/10.1103/PhysRevD.46.3062}{\emph{Phys. Rev. D}
  {\bfseries 46} (1992) 3062}.

\bibitem{BU1993}
I.I.Y.~Bigi and N.G.~Uraltsev, \emph{{$D_s$ lifetime, $m_b$, $m_c$ and
  $|V_{cb}|$ in the heavy quark expansion}},
  \href{https://doi.org/10.1007/BF01574165}{\emph{Z. Phys. C} {\bfseries 62}
  (1994) 623} [\href{https://arxiv.org/abs/hep-ph/9311243}{{\ttfamily
  hep-ph/9311243}}].

\bibitem{GS2013}
P.~Gambino and C.~Schwanda, \emph{{Inclusive semileptonic fits, heavy quark
  masses, and $V_{cb}$}},
  \href{https://doi.org/10.1103/PhysRevD.89.014022}{\emph{Phys. Rev. D}
  {\bfseries 89} (2014) 014022}
  [\href{https://arxiv.org/abs/1307.4551}{{\ttfamily 1307.4551}}].

\bibitem{Becirevic2000}
D.~Be\v{c}irevi\'c, \emph{{Heavy quark phenomenology from lattice QCD}},
  \href{https://doi.org/10.1016/S0920-5632(01)00959-8}{\emph{Nucl. Phys. B
  Proc. Suppl.} {\bfseries 94} (2001) 337}
  [\href{https://arxiv.org/abs/hep-lat/0011075}{{\ttfamily hep-lat/0011075}}].

\bibitem{MITbag0}
A.~Chodos, R.L.~Jaffe, K.~Johnson, C.B.~Thorn and V.F.~Weisskopf, \emph{{A new
  extended model of hadrons}},
  \href{https://doi.org/10.1103/PhysRevD.9.3471}{\emph{Phys. Rev. D} {\bfseries
  9} (1974) 3471}.

\bibitem{MITbag1}
A.~Chodos, R.L.~Jaffe, K.~Johnson and C.B.~Thorn, \emph{{Baryon structure in
  the bag theory}}, \href{https://doi.org/10.1103/PhysRevD.10.2599}{\emph{Phys.
  Rev. D} {\bfseries 10} (1974) 2599}.

\bibitem{MITbag2}
T.A.~DeGrand, R.L.~Jaffe, K.~Johnson and J.E.~Kiskis, \emph{{Masses and other
  parameters of the light hadrons}},
  \href{https://doi.org/10.1103/PhysRevD.12.2060}{\emph{Phys. Rev. D}
  {\bfseries 12} (1975) 2060}.

\bibitem{MITbag3}
T.A.~DeGrand and R.L.~Jaffe, \emph{{Excited states of confined quarks}},
  \href{https://doi.org/10.1016/0003-4916(76)90069-5}{\emph{Annals Phys.}
  {\bfseries 100} (1976) 425}.

\bibitem{MITbag4}
T.A.~DeGrand, \emph{{Excited states of confined quarks. 2.}},
  \href{https://doi.org/10.1016/0003-4916(76)90021-X}{\emph{Annals Phys.}
  {\bfseries 101} (1976) 496}.

\bibitem{BS2004}
A.~Bernotas and V.~Simonis, \emph{{Towards the unified description of light and
  heavy hadrons in the bag model approach}},
  \href{https://doi.org/10.1016/j.nuclphysa.2004.05.017}{\emph{Nucl. Phys. A}
  {\bfseries 741} (2004) 179}
  [\href{https://arxiv.org/abs/hep-ph/0403094}{{\ttfamily hep-ph/0403094}}].

\bibitem{BS2008}
A.~Bernotas and V.~Simonis, \emph{{Heavy hadron spectroscopy and the bag
  model}}, \href{https://doi.org/10.3952/lithjphys.49110}{\emph{Lith. J. Phys.}
  {\bfseries 49} (2009) 19} [\href{https://arxiv.org/abs/0808.1220}{{\ttfamily
  0808.1220}}].

\bibitem{BS2012}
A.~Bernotas and V.~Simonis, \emph{{Magnetic moments of heavy baryons in the bag
  model reexamined}},  \href{https://arxiv.org/abs/1209.2900}{{\ttfamily
  1209.2900}}.

\bibitem{Copley1979}
L.A.~Copley, N.~Isgur and G.~Karl, \emph{{Charmed baryons in a quark model with
  hyperfine interactions}},
  \href{https://doi.org/10.1103/PhysRevD.20.768}{\emph{Phys. Rev. D} {\bfseries
  20} (1979) 768}.

\bibitem{Jaffe2004}
R.L.~Jaffe, \emph{{Exotica}},
  \href{https://doi.org/10.1016/j.physrep.2004.11.005}{\emph{Phys. Rept.}
  {\bfseries 409} (2005) 1}
  [\href{https://arxiv.org/abs/hep-ph/0409065}{{\ttfamily hep-ph/0409065}}].

\bibitem{CMLW2016}
B.~Chen, K.-W.~Wei, X.~Liu and T.~Matsuki, \emph{{Low-lying charmed and
  charmed-strange baryon states}},
  \href{https://doi.org/10.1140/epjc/s10052-017-4708-x}{\emph{Eur. Phys. J. C}
  {\bfseries 77} (2017) 154}
  [\href{https://arxiv.org/abs/1609.07967}{{\ttfamily 1609.07967}}].

\bibitem{GGR1975}
A.~De~Rujula, H.~Georgi and S.L.~Glashow, \emph{{Hadron masses in a gauge
  theory}}, \href{https://doi.org/10.1103/PhysRevD.12.147}{\emph{Phys. Rev. D}
  {\bfseries 12} (1975) 147}.

\bibitem{GR1981}
S.~Gasiorowicz and J.L.~Rosner, \emph{{Hadron spectra and quarks}},
  \href{https://doi.org/10.1119/1.12597}{\emph{Am. J. Phys.} {\bfseries 49}
  (1981) 954}.

\bibitem{KR2014}
M.~Karliner and J.L.~Rosner, \emph{{Baryons with two heavy quarks: masses,
  production, decays, and detection}},
  \href{https://doi.org/10.1103/PhysRevD.90.094007}{\emph{Phys. Rev. D}
  {\bfseries 90} (2014) 094007}
  [\href{https://arxiv.org/abs/1408.5877}{{\ttfamily 1408.5877}}].

\bibitem{BLS1980}
V.D.~Barger, J.P.~Leveille and P.M.~Stevenson, \emph{{Nonspectator quark
  interactions and the $\Lambda_c^+$ lifetime}},
  \href{https://doi.org/10.1103/PhysRevLett.44.226}{\emph{Phys. Rev. Lett.}
  {\bfseries 44} (1980) 226}.

\bibitem{Rosner1996}
J.L.~Rosner, \emph{{Enhancement of the $\Lambda_b$ decay rate}},
  \href{https://doi.org/10.1016/0370-2693(96)00352-8}{\emph{Phys. Lett. B}
  {\bfseries 379} (1996) 267}
  [\href{https://arxiv.org/abs/hep-ph/9602265}{{\ttfamily hep-ph/9602265}}].

\bibitem{SV1987}
M.B.~Voloshin and M.A.~Shifman, \emph{{On the annihilation constants of mesons
  consisting of a heavy and a light quark, and $B^0 - \overline{B}^{0}$
  oscillations}}, {\emph{Sov. J. Nucl. Phys.} {\bfseries 45} (1987) 292}.

\bibitem{PW1988}
H.D.~Politzer and M.B.~Wise, \emph{{Leading logarithms of heavy quark masses in
  processes with light and heavy quarks}},
  \href{https://doi.org/10.1016/0370-2693(88)90718-6}{\emph{Phys. Lett. B}
  {\bfseries 206} (1988) 681}.

\bibitem{Neubert1992}
M.~Neubert, \emph{{Symmetry breaking corrections to meson decay constants in
  the heavy quark effective theory}},
  \href{https://doi.org/10.1103/PhysRevD.46.1076}{\emph{Phys. Rev. D}
  {\bfseries 46} (1992) 1076}.

\bibitem{BGMS2004}
A.~Babi\'c, B.~Guberina, B.~Meli\'c and H.~\v{S}tefan\v {c}i\'c,
  \emph{{Cabibbo-suppressed decays of the $\Omega_c^0$-- Feedback to the
  $\Xi_c^+$ lifetime}},
  \href{https://doi.org/10.1103/PhysRevD.70.117501}{\emph{Phys. Rev. D}
  {\bfseries 70} (2004) 117501}
  [\href{https://arxiv.org/abs/hep-ph/0406183}{{\ttfamily hep-ph/0406183}}].

\bibitem{Cheng2021RevI}
H.-Y.~Cheng, \emph{{Charmed Baryon Physics Circa 2021}},
  \href{https://arxiv.org/abs/2109.01216}{{\ttfamily 2109.01216}}.

\bibitem{Cheng2021RevII}
H.-Y.~Cheng, \emph{{The strangest lifetime: a bizarre story of
  $\tau(\Omega_c^0)$}},
  \href{https://doi.org/10.1016/j.scib.2021.11.025}{\emph{Science Bulletin}
  {\bfseries 67} (2021) 445}
  [\href{https://arxiv.org/abs/2111.09566}{{\ttfamily 2111.09566}}].

\bibitem{EFG2011}
D.~Ebert, R.N.~Faustov and V.O.~Galkin, \emph{{Spectroscopy and Regge
  trajectories of heavy baryons in the relativistic quark-diquark picture}},
  \href{https://doi.org/10.1103/PhysRevD.84.014025}{\emph{Phys. Rev. D}
  {\bfseries 84} (2011) 014025}
  [\href{https://arxiv.org/abs/1105.0583}{{\ttfamily 1105.0583}}].

\bibitem{Shifman1994}
M.A.~Shifman, \emph{{Theory of preasymptotic effects in weak inclusive
  decays}},  in \emph{{Workshop on Continuous Advances in QCD}}, 2, 1994
  [\href{https://arxiv.org/abs/hep-ph/9405246}{{\ttfamily hep-ph/9405246}}].

\bibitem{BDS1994}
B.~Blok, R.D.~Dikeman and M.A.~Shifman, \emph{{Calculation of $1/m_c^3$ terms
  in the total semileptonic width of $D$ mesons}},
  \href{https://doi.org/10.1103/PhysRevD.51.6167}{\emph{Phys. Rev. D}
  {\bfseries 51} (1995) 6167}
  [\href{https://arxiv.org/abs/hep-ph/9410293}{{\ttfamily hep-ph/9410293}}].

\bibitem{BBBF2003}
S.~Bianco, F.L.~Fabbri, D.~Benson and I.~Bigi, \emph{{A Cicerone for the
  physics of charm}},
  \href{https://doi.org/10.1393/ncr/i2003-10003-1}{\emph{Riv. Nuovo Cim.}
  {\bfseries 26} (2003) 1}
  [\href{https://arxiv.org/abs/hep-ex/0309021}{{\ttfamily hep-ex/0309021}}].

\bibitem{Umeeda2021}
H.~Umeeda, \emph{{Revisiting quark-hadron duality for heavy meson non-leptonic
  decays in two-dimensional QCD}},
  \href{https://doi.org/10.1016/j.physletb.2021.136854}{\emph{Phys. Lett. B}
  {\bfseries 825} (2022) 136854}
  [\href{https://arxiv.org/abs/2111.01401}{{\ttfamily 2111.01401}}].

\bibitem{SchwartzComms}
A. Schwartz, private communication.

\bibitem{HPQCD2014II}
{\scshape HPQCD} collaboration, \emph{{$\Upsilon$ and $\Upsilon^{\prime}$
  leptonic widths, $a_{\mu}^b$ and $m_b$ from full lattice QCD}},
  \href{https://doi.org/10.1103/PhysRevD.91.074514}{\emph{Phys. Rev. D}
  {\bfseries 91} (2015) 074514}
  [\href{https://arxiv.org/abs/1408.5768}{{\ttfamily 1408.5768}}].

\bibitem{ETM2016}
{\scshape European Twisted Mass} collaboration, \emph{{Mass of the $b$ quark
  and $B$-meson decay constants from $N_f=2+1+1$ twisted-mass lattice QCD}},
  \href{https://doi.org/10.1103/PhysRevD.93.114505}{\emph{Phys. Rev. D}
  {\bfseries 93} (2016) 114505}
  [\href{https://arxiv.org/abs/1603.04306}{{\ttfamily 1603.04306}}].

\bibitem{GMS2017}
P.~Gambino, A.~Melis and S.~Simula, \emph{{Extraction of heavy-quark-expansion
  parameters from unquenched lattice data on pseudoscalar and vector
  heavy-light meson masses}},
  \href{https://doi.org/10.1103/PhysRevD.96.014511}{\emph{Phys. Rev. D}
  {\bfseries 96} (2017) 014511}
  [\href{https://arxiv.org/abs/1704.06105}{{\ttfamily 1704.06105}}].

\bibitem{ETM2014III}
{\scshape European Twisted Mass} collaboration, \emph{{Leptonic decay constants
  $f_{K},f_{D},$ and $f_{{D}_{s}}$ with $N_{f} = 2+1+1$ twisted-mass lattice
  QCD}}, \href{https://doi.org/10.1103/PhysRevD.91.054507}{\emph{Phys. Rev. D}
  {\bfseries 91} (2015) 054507}
  [\href{https://arxiv.org/abs/1411.7908}{{\ttfamily 1411.7908}}].

\bibitem{MILC2017}
{\scshape Fermilab Lattice, MILC} collaboration, \emph{{$B$- and $D$-meson
  leptonic decay constants from four-flavor lattice QCD}},
  \href{https://doi.org/10.1103/PhysRevD.98.074512}{\emph{Phys. Rev. D}
  {\bfseries 98} (2018) 074512}
  [\href{https://arxiv.org/abs/1712.09262}{{\ttfamily 1712.09262}}].

\end{thebibliography}\endgroup
\end{document}